\documentclass[notitlepage,letterpaper,natbib,aps,prd,onecolumn,amsmath,amsfonts,nofootinbib,preprintnumbers,superscriptaddress,secnumarabic,groupedaddress]{revtex4-1}
\pdfoutput=1
\usepackage{amssymb,amsmath,latexsym,mathrsfs}
\usepackage{url}
\usepackage{enumitem}
\usepackage{graphicx}
\usepackage[usenames,dvipsnames]{color}
\usepackage[breaklinks,colorlinks,urlcolor=blue,citecolor=blue,linkcolor=magenta]{hyperref}

\usepackage{multirow}
\usepackage{float}
\usepackage{cases}
\usepackage{hhline}

\makeatletter
\renewcommand\tableofcontents{%
    \@starttoc{toc}%
}
\makeatother

\usepackage{titlesec}

\titleformat{\section}
  {\normalfont\fontsize{12}{12}\bfseries}{\centering \thesection}{1em}{}

\titleformat{\subsection}
  {\normalfont\fontsize{11}{11}\bfseries}{\centering \thesubsection}{0.5em}{}

\titlespacing*{\subsection}
  {0pt}{1\baselineskip}{0.5\baselineskip}

\titlespacing*{\section}
  {0pt}{1.6\baselineskip}{0.9\baselineskip}

\titleformat{\subsubsection}
  {\normalfont\fontsize{10}{10}\bfseries}{\centering \thesubsubsection}{1em}{}
\titlespacing*{\subsection}
  {0pt}{0.9\baselineskip}{0.4\baselineskip}
  
\usepackage{dsfont}

\begin{document}
\title{Refined Bounds on MeV-scale Thermal Dark Sectors from BBN and the CMB}

\preprint{KCL-2019-75}
\author{Nashwan Sabti$^\mathds{S}$}

\author{James Alvey$^\mathds{A}$}

\author{Miguel Escudero$^\mathds{E}$}

\author{Malcolm Fairbairn$^\mathds{F}$}

\author{Diego Blas$^\mathds{B}$}
\affiliation{\vspace{8pt}Theoretical Particle Physics and Cosmology Group\\ 
King's College London, Department of Physics, Strand, London WC2R 2LS, UK}

\def\thefootnote{$\mathds{S}$\hspace{0.4pt}}\footnotetext{\href{mailto:nashwan.sabti@kcl.ac.uk}{nashwan.sabti@kcl.ac.uk}}
\def\thefootnote{$\mathds{A}$\hspace{-0.5pt}}\footnotetext{\href{mailto:james.alvey@kcl.ac.uk}{james.alvey@kcl.ac.uk}}
\def\thefootnote{$\mathds{E}$}\footnotetext{\href{mailto:miguel.escudero@kcl.ac.uk}{miguel.escudero@kcl.ac.uk}}
\def\thefootnote{$\mathds{F}$}\footnotetext{\href{mailto:malcolm.fairbairn@kcl.ac.uk}{malcolm.fairbairn@kcl.ac.uk}}
\def\thefootnote{$\mathds{B}$}\footnotetext{\href{mailto:diego.blas@kcl.ac.uk}{diego.blas@kcl.ac.uk}}
\setcounter{footnote}{0}
\def\thefootnote{\arabic{footnote}}
\begin{abstract}
\noindent New light states thermally coupled to the Standard Model plasma alter the expansion history of the Universe and impact the synthesis of the primordial light elements. In this work, we carry out an exhaustive and precise analysis of the implications of MeV-scale BSM particles in Big Bang Nucleosynthesis (BBN) and for Cosmic Microwave Background (CMB) observations. We find that BBN observations set a lower bound on the thermal dark matter mass of $m_\chi > 0.4\,\text{MeV}$ at $2\sigma$. This bound is independent of the spin and number of internal degrees of freedom of the particle, of the annihilation being s-wave or p-wave, and of the annihilation final state. Furthermore, we show that current BBN plus CMB observations constrain purely electrophilic and neutrinophilic BSM species to have a mass, $m_\chi > 3.7\,\text{MeV}$ at $2\sigma$. We explore the reach of future BBN measurements and show that upcoming CMB missions should improve the bounds on light BSM thermal states to $m_\chi > (10-15)\,\text{MeV}$. Finally, we demonstrate that very light BSM species thermally coupled to the SM plasma are highly disfavoured by current cosmological observations. 
\end{abstract}

\maketitle
\vspace{5pt}
{
  \hypersetup{linkcolor=blue}
    \vspace{5pt}
    \setlength\parskip{0.5pt}
  \hrule
  \tableofcontents
  \vspace{10pt}
  \hrule
}
\vspace{10pt}
\section{Introduction}\label{sec:intro}

The nature of dark matter is still unknown and thermal relics associated with the electroweak scale are under increasing pressure from the LHC as well as from direct and indirect detection experiments~\cite{Aaboud:2017phn,Sirunyan:2018xlo,Aprile:2018dbl,Akerib:2016vxi,Cui:2017nnn,Fermi-LAT:2016uux,Aghanim:2018eyx,Escudero:2016gzx,Arcadi:2017kky,Roszkowski:2017nbc,Athron:2017kgt,Arcadi:2019lka,Blanco:2019hah}. Motivated by the fact that direct detection experiments are considerably less sensitive to sub-GeV dark matter particles, attention has naturally turned to lower mass alternatives~\cite{Essig:2013lka,Alexander:2016aln,Battaglieri:2017aum,Beacham:2019nyx}. 

From a theoretical perspective, MeV-scale thermal dark matter candidates were shown to be viable some years ago \cite{Boehm:2003hm,Boehm:2003bt,Feng:2008ya} and, since then, a large number of MeV-scale dark matter models have appeared in the literature, see e.g.~\cite{Boehm:2006mi,Farzan:2009ji,Farzan:2011ck,Batell:2017cmf,Ballett:2019cqp,Lamprea:2019qet,Blennow:2019fhy,Krnjaic:2015mbs,Bondarenko:2019vrb,Hochberg:2014dra,Agrawal:2014ufa,Kamada:2018zxi,Knapen:2017xzo,Hall:2009bx,Chu:2011be,Hambye:2019dwd,Dvorkin:2019zdi,Evans:2019vxr}. Experimentally, with the aim of testing as many scenarios as possible \cite{Bertone:2018xtm}, a complementary program has been developed to test the possible existence of MeV-scale dark matter particles and potential companions Beyond the Standard Model (BSM)~\cite{Essig:2013lka,Alexander:2016aln,Battaglieri:2017aum,Beacham:2019nyx}. Light dark matter particles and their potential mediators with the dark sector have been searched for at particle colliders~\cite{Borodatchenkova:2005ct,Batell:2009yf,Fox:2011fx,Essig:2013vha,Lees:2014xha,Babusci:2015zda,Lees:2017lec,Aaij:2017rft}, beam dump experiments~\cite{Bjorken:2009mm,Batell:2009di,Andreas:2012mt}, neutrino experiments~\cite{PalomaresRuiz:2007eu,Harnik:2012ni,deNiverville:2012ij,Batell:2014yra,Klop:2018ltd,Kelly:2019wow}, neutrino telescopes~\cite{Kamada:2015era,Arguelles:2017atb,Alvey:2019jzx}, as well as in direct~\cite{Essig:2015cda,Lee:2015qva,Derenzo:2016fse,Essig:2017kqs,Agnese:2018col,Agnes:2018oej,Abramoff:2019dfb,Aprile:2019xxb} and indirect~\cite{Slatyer:2015jla,Essig:2013goa,Bartels:2017dpb} dark matter detection experiments. Searches for light BSM species are not only carried out in terrestrial experiments, but a variety of astrophysical~\cite{Raffelt:1996wa,Dreiner:2013mua,Chang:2018rso,DeRocco:2019jti,Farzan:2002wx,Heurtier:2016otg} and cosmological~\cite{Xu:2018efh,Campo:2017nwh,Berlin:2018sjs,Wilkinson:2014ksa,Bertoni:2014mva,Vogel:2013raa,Escudero:2019gzq,Dolgov:2013una} constraints have been also derived on states with masses at the MeV scale. Up to now, however, all these searches have been unsuccessful. Future, ongoing and planned experiments are expected to cut into relevant regions of parameter space and perhaps yield a signal~\cite{Essig:2013lka,Alexander:2016aln,Battaglieri:2017aum,Beacham:2019nyx,Kou:2018nap,Ariga:2019ufm,Alekhin:2015byh,Chou:2016lxi,Akesson:2018vlm}.

Big Bang Nucleosynthesis (BBN) has been widely used as a probe of new physics~\cite{Sarkar:1995dd,Iocco:2008va,Pospelov:2010hj}. BBN occurred when the Universe was about three minutes old, in the temperature range $10\,\text{keV}  \lesssim T \lesssim 1\,\text{MeV}$, and therefore represents a key stage of the Universe that new states at the MeV scale can affect. Given the excellent agreement between observations and the Standard Model (SM) prediction of the primordial light nuclei abundances~\cite{pdg}, strong constraints can be set on the masses and properties of new light particles. Similarly, the agreement of Cosmic Microwave Background (CMB) observations with a vanilla $\Lambda$CDM Universe~\cite{Aghanim:2018eyx} can be used to set strong constraints on light new physics.

In this work, we perform an exhaustive and robust analysis of the cosmological implications of MeV-scale particles that are thermally coupled to electrons, neutrinos or both in the early Universe. This has been studied in the past by a number of groups~\cite{Kolb:1986nf,Serpico:2004nm,Boehm:2013jpa,Nollett:2013pwa,Nollett:2014lwa,Boehm:2012gr,Ho:2012ug,Wilkinson:2016gsy,Depta:2019lbe,Escudero:2018mvt}, but here we update and upgrade the constraints by:
\begin{itemize}[leftmargin=0.5cm,itemsep=0pt]
\item Using up-to-date measurements of the primordial element abundances \cite{pdg} and Planck 2018 CMB observations \cite{Aghanim:2018eyx}.
\item Accurately accounting for the early Universe evolution in the presence of MeV-scale states following \cite{Escudero:2018mvt,Escudero:2019new}.
\item Using the state-of-the-art Big Bang Nucleosynthesis code \texttt{PRIMAT} \cite{Pitrou:2018cgg}, which outputs the most accurate theoretical predictions for the helium and deuterium abundances to date. \texttt{PRIMAT} accounts for a variety of effects, such as up-to-date nuclear reaction rates, finite temperature corrections, incomplete neutrino decoupling and several other effects relevant to the proton-to-neutron conversion rates.
\item Performing a pure BBN analysis on light MeV-scale states. Namely, we set a bound on the masses of different species by using only the primordial helium and deuterium abundances and by marginalizing over any possible value of the baryon energy density.
\end{itemize}
While we are mostly interested in constraining MeV-scale dark matter particles, our analysis applies more generally to any additional BSM particles that are in thermal equilibrium with the SM during BBN and with a mass in the MeV range. This includes dark matter particles and mediators with the dark sector. In practice, this equilibrium should be maintained at temperatures below that of neutrino decoupling $T_\nu^{\rm dec} \sim 2\,\text{MeV}$~\cite{Dolgov:2002wy}. The requirement of being in thermal contact at least prior to neutrino decoupling restricts the couplings and masses of the particles we are able to constrain. For the case of weakly-interacting, stable, thermal BSM particles (WIMPs), it is well known that annihilation interactions with the SM plasma will decouple from chemical equilibrium at $T \sim m/20$~\cite{Kolb:1990vq}. Hence, our analysis will apply to thermal WIMPs with $m \lesssim 20\, T_\nu ^{\rm dec} \sim 40\,\text{MeV}$. In the case of unstable particles that decay into SM species, they will be in thermal equilibrium with the SM plasma during BBN, provided that their lifetime is $\tau \lesssim 0.1 \,\text{s}$. In other words, the lifetime should be shorter than the age of the Universe at the time of neutrino decoupling.

Our analysis will constrain particles that efficiently annihilate or decay into neutrinos or electrons/photons prior and during neutrino decoupling. This is the case if the interaction rate is larger than the expansion rate $H$ at neutrino decoupling:
\begin{align}
    \Gamma \gtrsim H|_{T = T_\nu^{\rm dec}} \simeq 10 \,\text{s}^{-1}.
\end{align}
Throughout the text we will denote a generic BSM particle as $\chi$. For a particle that annihilates into SM species, the rate is $\Gamma \sim n\left<\sigma v \right> \sim g_\chi^2 g_{\rm SM}^2T^3/(16 \pi m_\chi^2)$ and our analysis will generally be sensitive to
\begin{align}\label{eq:Annihilation}
\text{Stable Particles with} \qquad  \,\,\,\,\,
    \sqrt{g_\chi g_{\rm SM}} \gtrsim  2\times 10^{-5} \sqrt{\frac{m_\chi}{\text{MeV}}} \qquad\qquad\qquad \,\,\,\,\,\,\,\,\,\, \text{and} \qquad m_\chi \lesssim 20\,\text{MeV}\,,
\end{align}
where $g_\chi$ and $g_{\rm SM}$ are coupling constants. If the $\chi$ particle is unstable, then $\Gamma \sim g_{\rm SM}^2m_\chi /(4\pi) \, K_{1}(m_\chi/T_{\nu}^{\rm dec})$ -- $K_{1}$ being a modified Bessel function of the first kind -- and our analysis will constrain
\begin{align}\label{eq:Decay}
\text{Unstable Particles with} \qquad 
    g_{\rm SM} \gtrsim  5\times 10^{-10} \sqrt{\frac{m_\chi}{\text{MeV}}}\left(1+ \sqrt{\frac{\text{MeV}}{m_\chi}} \right) \qquad \text{and} \qquad m_\chi \lesssim 20\,\text{MeV}.
\end{align}
In summary, the bounds we derive in this paper will generically constrain MeV-scale particles coupled to the SM bath with couplings $ \gtrsim 10^{-5}$ if they are stable\footnote{This would cover well the case of thermal dark matter, for which $\sqrt{g_\chi g_{\rm SM}} \sim 10^{-3} \sqrt{\frac{m_\chi}{10\,\text{MeV}}} \sqrt[4]{\frac{\left<\sigma v\right>}{3\times 10^{-26}\,\text{cm}^3/s} }$.} and with couplings as small as $10^{-9}$ if they are unstable. Note that our bounds will apply even if these unstable BSM particles decay into other BSM states, under the condition that they possess couplings $\gtrsim 10^{-9}$ to SM particles. 

This paper is organized as follows: In Section~\ref{sec:earlyUniverse}, we describe our modelling of the early Universe evolution with light BSM species coupled to the SM plasma at temperatures $1\,\text{keV} < T < 30\,\text{MeV}$. In Section~\ref{sec:current_data_analyis}, we outline the cosmological data and statistical procedure used to set constraints on the masses and properties of particles in thermal equilibrium with neutrinos and/or electrons in the early Universe. In Section~\ref{sec:results}, we set a lower bound on the mass of purely electrophilic and neutrinophilic BSM species in thermal equilibrium with the SM plasma from a combination of BBN and CMB data. In Section~\ref{sec:results_mixed}, we set constraints on BSM species that efficiently interact with both neutrinos and electrons/photons. In Section~\ref{sec:future}, we forecast future BBN and CMB constraints. In Section~\ref{sec:discussion}, we discuss the robustness of our bounds, how they are modified in the presence of additional species, mention certain particle physics scenarios for which they represent the most stringent constraints, and compare with previous literature. Finally, in Section~\ref{sec:conclusions}, we present our conclusions. Details regarding the modification to the BBN code, comparison to previous literature, complete sets of results, and our CMB forecasting methodology can be found in the Appendices~\ref{app:App}.

\section{Cosmology with Light WIMPS}\label{sec:earlyUniverse}
The implications of light, thermally coupled BSM particles with the SM plasma at temperatures $T \sim 1 \,\text{MeV}$ are twofold~\cite{Kolb:1986nf}: \textit{i)} they contribute to the expansion rate of the early Universe and \textit{ii)} they release entropy into the plasma.
In addition, if the new particles interact with both neutrinos and electrons/photons, they would efficiently delay the process of neutrino decoupling~\cite{Serpico:2004nm,Escudero:2018mvt}.

\subsection{Temperature Evolution and Universe's Expansion}\label{sec:earlyUniverse_method}

In order to accurately account for such effects we follow~\cite{Escudero:2018mvt} and assume that all relevant species can be described by thermal distributions and characterized by a temperature $T_i$. We then calculate the evolution of neutrino decoupling in terms of the temperature of the neutrinos and electromagnetic components of the plasma. When a neutrinophilic or electrophilic BSM particle is present, the differential equations governing the evolution of $T_\nu$ and $T_\gamma$ read:
\begin{subnumcases}{\hspace{-2.8cm}\text{Neutrinophilic }}
\label{eq:dTgamdt_DM_nu}
    \!\! \,\, \frac{dT_\nu}{dt} = -\frac{ 12 H \rho_\nu +3 H( \rho_{\chi} + p_{\chi})  - 3 \frac{\delta \rho_{\nu}}{\delta t}}{3 \, \frac{\partial \rho_\nu}{\partial T_\nu } +  \frac{\partial \rho_{\chi}}{\partial T_\nu } }\,, \\
    \!\! \,\, \frac{dT_{\gamma}}{dt}  =- \frac{  4 H \rho_{\gamma} + 3 H \left( \rho_{e} + p_{e}\right) + 3 H \, T_\gamma \frac{dP_\text{int}}{dT_\gamma}+3 \frac{\delta \rho_{\nu}}{\delta t}  }{ \frac{\partial \rho_{\gamma}}{\partial T_\gamma} + \frac{\partial \rho_e}{\partial T_\gamma} +T_\gamma \frac{d^2 P_\text{int}}{dT_\gamma^2} }\,,
\end{subnumcases}
\begin{subnumcases}{\text{Electrophilic }}
    \!\! \,\, \frac{dT_\nu}{dt} = -\frac{ 12  H \rho_\nu  -  3 \frac{\delta \rho_{\nu}}{\delta t}}{3 \, \frac{\partial \rho_\nu}{\partial T_\nu }}\,, \\
    \!\! \,\, \frac{dT_{\gamma}}{dt}  =- \frac{  4 H \rho_{\gamma} + 3 H \left( \rho_{e} + p_{e}\right) +  3 H \left( \rho_{\chi} + p_{\chi}\right) + 3 H \, T_\gamma \frac{dP_\text{int}}{dT_\gamma}+ 3\frac{\delta \rho_{\nu}}{\delta t}  }{ \frac{\partial \rho_{\gamma}}{\partial T_\gamma} + \frac{\partial \rho_e}{\partial T_\gamma} + \frac{\partial \rho_{\chi}}{\partial T_\gamma} +T_\gamma \frac{d^2 P_\text{int}}{dT_\gamma^2} } \ , \label{eq:dTgamdt_DM_e}
\end{subnumcases}

where $\rho_i$ and $p_i$ correspond to the energy density and pressure of a given particle respectively, $H = \sqrt{(8\pi/3)\,\sum_i \rho_i/M_{\rm Pl}^2}$ is the Hubble parameter, $M_{\rm Pl} = 1.22\times 10^{19}\,\text{GeV}$ the Planck mass, and $P_{\rm int}$ and its derivatives account for finite temperature corrections. The reader is referred to~\cite{Escudero:2018mvt} for further details. Here, $\delta \rho_{\nu} /\delta t$ corresponds to the energy exchange rate between neutrinos and electrons. Accounting for Fermi-Dirac statistics in the rates and setting $m_e = 0$, it reads~\cite{Escudero:2019new}:
\begin{align}\label{eq:energyrates_nu_SM}
& \left. \frac{\delta \rho_{\nu}}{\delta t}  \right|_{\rm SM} = \frac{G_F^2}{\pi^5} \left( 1 - \frac{4}{3} s_W^2 + 8 s_W^4 \right) \times  \left[ 32 \, f_a^{\rm FD} \,  \left( T_\gamma^9-T_{\nu}^9  \right) +  56 \,f_s^{\rm FD} \,   T_\gamma^4 \, T_{\nu}^4 \, \left( T_\gamma - T_{\nu}\right)\right] \,,
\end{align}
where $s_\mathrm{W}^2 = 0.223$~\cite{pdg}, $G_\mathrm{F}$ is Fermi's constant, $f_a^{\rm FD} = 0.884$, $f_s^{\rm FD} = 0.829$, and we account for the electron mass as in~\cite{Escudero:2019new}. 

We solve Equations~\eqref{eq:dTgamdt_DM_nu} -- \eqref{eq:dTgamdt_DM_e} for $1\,\text{keV} < T_\gamma < 30\,\text{MeV}$. We start the integration at $t_0 = 1/(2 H)|_{T = 30\,\text{MeV}}$ for which we use as an initial condition $T_\gamma = T_\nu = 30 \,\text{MeV}$, since for such high temperatures SM neutrino-electron interactions are highly efficient. By solving this set of differential equations, we find all the key background evolution quantities as a function of time, scale factor and temperature. In addition, we evaluate the number of effective relativistic degrees of freedom $N_{\rm eff}$ as relevant for CMB observations,
\begin{align}\label{eq:Neff}
N_{\rm eff} \equiv \frac{8}{7}\left(\frac{11}{4} \right)^{4/3} \left( \frac{\rho_\text{rad}-\rho_\gamma}{\rho_\gamma}\right) = 3 \left(\frac{11}{4} \right)^{4/3} \left(\frac{T_\nu}{T_\gamma}\right)^4 \, ,
\end{align} 
where in the last step we have assumed that $\rho_\text{rad} = \rho_\nu + \rho_\gamma$. By solving this system of equations in the SM we find $N_{\rm eff}^{\rm SM} = 3.046$~\cite{Escudero:2019new}, a result that is in perfect agreement with state-of-the-art calculations~\cite{Mangano:2005cc,deSalas:2016ztq}. 

Finally, if the $\chi$ particle interacts with both electrons and neutrinos, an additional energy exchange between the latter two particles should be included. Considering a stable particle and neglecting scattering interactions, this reads~\cite{Escudero:2018mvt}:
\begin{align}\label{eq:energyrates_WIMP}
 \left. \frac{\delta \rho_{\nu}}{\delta t}  \right|_{\chi} =  \frac{g_\chi^2 m_{\rm \chi}^5}{4\pi^4} \,\left( \left< \sigma v \right>_{\chi \chi \to \bar{\nu}\nu} \left[T_\nu^2  \, K_2^2\left[\frac{m_{\chi}}{T_{\nu}} \right] -  T_\chi^2 \, K_2^2\left[\frac{m_{\chi}}{T_{\chi}} \right] \right] - \left< \sigma v \right>_{\chi \chi \to e^+ e^-} \left[T_\chi^2  \, K_2^2\left[\frac{m_{\chi}}{T_{\chi}} \right] -  T_\gamma^2 \, K_2^2\left[\frac{m_{\chi}}{T_{\gamma}} \right] \right] \right)  \,.
\end{align}
For a thermal WIMP $\left< \sigma v \right>_{\chi \chi \to e^+ e^-} + \left< \sigma v \right>_{\chi \chi \to \bar{\nu} \nu} = \left< \sigma v \right>_{\rm WIMP} \simeq 3 \times 10^{-26}\,\text{cm}^3/\text{s} $~\cite{Steigman:2012nb}. If the particle annihilates predominantly to electrons, i.e. $\left< \sigma v \right>_{\chi \chi \to e^+ e^-} > \left< \sigma v \right>_{\chi \chi \to \bar{\nu} \nu}$, then $T_\chi = T_\gamma $ and vice versa for neutrinos. In scenarios where we consider stable particles interacting with both electrons and neutrinos we shall fix the total annihilation cross section to $\left< \sigma v \right>_{\rm WIMP}$.

\subsection{Primordial Nucleosynthesis in the Presence of Thermal BSM Particles}\label{sec:earlyUniverse_method}

MeV-scale thermal relics affect the synthesis of the light elements, see e.g.~\cite{Kolb:1986nf,Serpico:2004nm,Boehm:2013jpa,Nollett:2013pwa,Nollett:2014lwa}. In this work, we have modified the publicly available state-of-the-art BBN code \texttt{PRIMAT}~\cite{Pitrou:2018cgg} to accommodate for the presence of light BSM particles in thermal equilibrium with the SM plasma. This is done by computing the background cosmology externally using \texttt{NUDEC\_BSM} \cite{Escudero:2018mvt,Escudero:2019new} and then passing on the relevant parameters\footnote{This includes the evolution as a function of time and scale factor of $T_\gamma$, $T_\nu$, $H$ and the residual entropy transfer between neutrinos and electrons as parametrized by $\mathcal{N}$ in \texttt{PRIMAT} \cite{Pitrou:2018cgg}.} to the section of \texttt{PRIMAT} that takes care of the nuclear reaction rates and the time evolution of nuclei abundances. We have explicitly verified that the differences in the primordial element abundances between the default version of \texttt{PRIMAT} and with the SM evolution as calculated in \cite{Escudero:2018mvt,Escudero:2019new} are below $0.1\,\%$, and hence one order of magnitude smaller than current observational errors. Our results agree quantitatively and qualitatively with previous studies~\cite{Kolb:1986nf,Serpico:2004nm,Boehm:2013jpa,Nollett:2013pwa,Nollett:2014lwa}\footnote{We agree particularly well with \cite{Nollett:2013pwa,Nollett:2014lwa}, see Appendix~\ref{app:ComparisonsLiterature}. } modulo differences we attribute to updated nuclear reaction rates and the fact that we account for non-instantaneous neutrino decoupling, see Appendix~\ref{app:ComparisonsLiterature} for details.

\begin{figure}[t]
    \centering
    \includegraphics[width=0.45\textwidth]{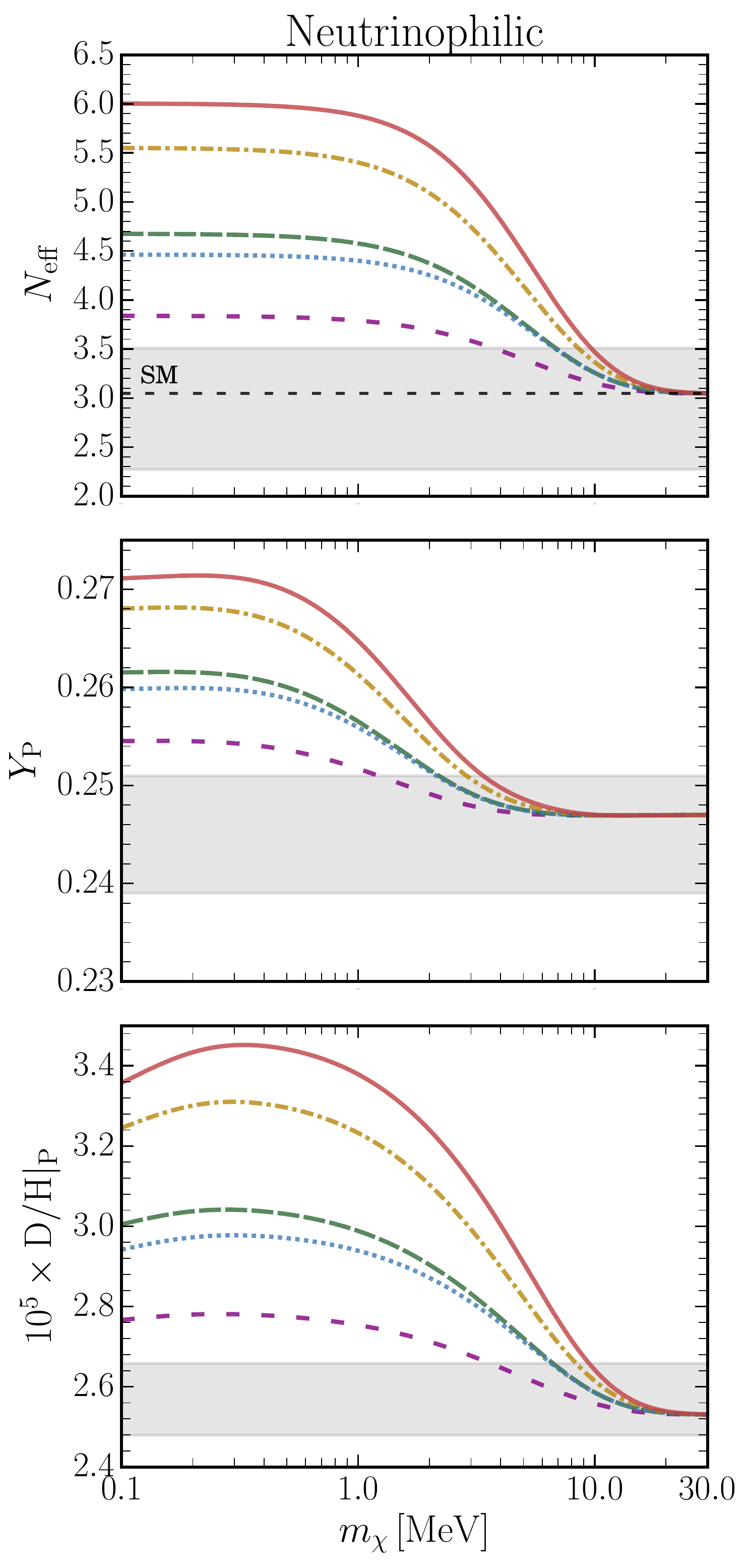} \qquad
    \includegraphics[width=0.45\textwidth]{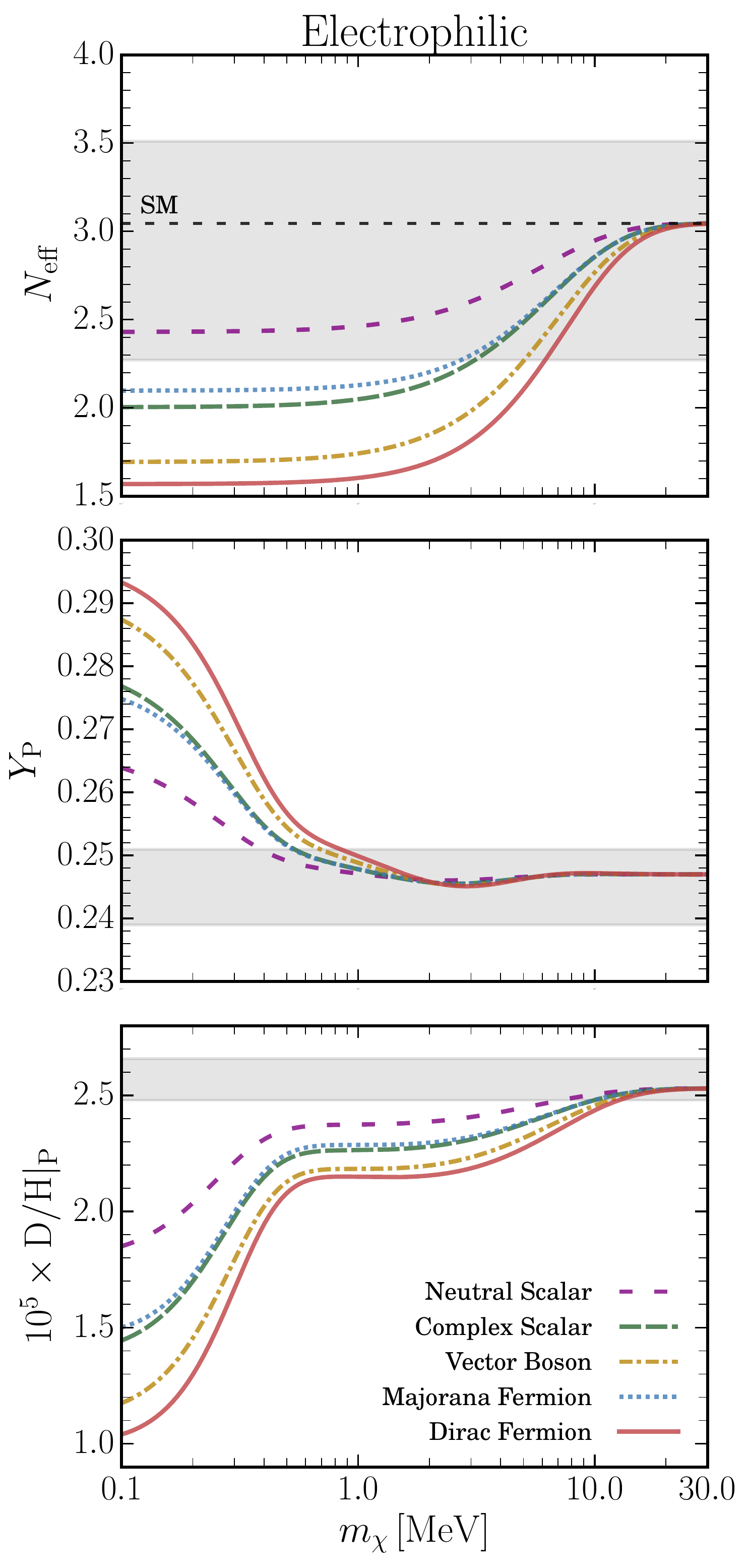}\vspace{-0.2cm}
    \caption{Cosmological impact of light BSM particles in thermal equilibrium with the SM plasma as a function of their mass $m_\chi$. The \textit{left/right panel} corresponds to neutrinophilic/electrophilic particles. \textit{Upper panels:} The number of effective relativistic neutrino species $N_{\rm eff}$ as relevant for CMB observations. \textit{Middle panels:} Primordial helium abundance $Y_{\rm P}$. \textit{Lower panels:} Primordial deuterium abundance ${\rm D/H}|_{\rm P}$. The $Y_{\rm P}$ and ${\rm D/H}|_{\rm P}$ predictions are computed with $\Omega_\mathrm{b} h^2 = 0.021875$ and $\tau_n = 879.5\,\text{s}$. The grey contours correspond to the mean $\pm \, 2\sigma$ measurements that enter our BBN and Planck data analyses, see Eqs.~\eqref{eq:chiBBN},~\eqref{eq:chi2_CMB}  and~\eqref{eq:CovarianceForm}.}
    \label{fig:Cosmoimply}
\end{figure}

\subsection{Cosmological Implications}\label{sec:earlyUniverse_method}
In this section we review the main cosmological implications of light BSM particles in thermal equilibrium with the Standard Model plasma during neutrino decoupling and Big Bang Nucleosynthesis. The reader is referred to~\cite{Kolb:1986nf,Serpico:2004nm,Boehm:2013jpa,Nollett:2013pwa,Nollett:2014lwa} for previous and complementary discussions of the impact of such particles on BBN and the CMB and to~\cite{Sarkar:1995dd,Iocco:2008va,Pospelov:2010hj} for reviews on the role of BBN as a probe of physics beyond the Standard Model. 

\vspace{0.1cm}
\textbf{Cosmic Microwave Background -- Modifications to $\boldsymbol{N_{\rm eff}}$} 

Once neutrinos decouple from the SM plasma at $T_\nu^{\rm dec} \lesssim 2\,\text{MeV}$, a light, neutrinophilic BSM particle of mass $m_\chi \lesssim 20 \,\text{MeV}$ will annihilate/decay into neutrinos, which results in $\rho_\nu > \rho_\nu^{\rm SM}$ or equivalently $N_{\rm eff} >N_{\rm eff}^{\rm SM}$. Analogously, an electrophilic particle will dump energy into the electromagnetic sector of the plasma at temperatures $T < T_\nu^{\rm dec}$ and yield $N_{\rm eff} < N_{\rm eff}^{\rm SM}$. In the upper panel of Figure~\ref{fig:Cosmoimply}, we display the corresponding value of $N_{\rm eff}$ for neutrinophilic and electrophilic BSM particles in thermal equilibrium with the SM plasma as a function of their mass. The grey contours correspond to the $\pm\,2\sigma$ measurements by Planck (see below). With the sole exception of an electrophilic scalar particle, it is clear that regardless of what the spin and number of internal degrees of freedom of a given species are, Planck would set a lower bound on its mass.

\vspace{0.1cm}
\textbf{Big Bang Nucleosynthesis}

\noindent Light BSM particles in thermal equilibrium during BBN lead to three main effects on the cosmological evolution that impact the synthesis of the primordial elements. Firstly, additional species present in the early Universe will alter the expansion rate and therefore also the temperature-to-time relation. This is important because they will modify the time at which various weak and nuclear processes freeze-out, in particular the proton-to-neutron conversion and $p + n \leftrightarrow  D + \gamma$. Secondly, the presence of these particles can change the evolution of the neutrino-to-photon temperature ratio. This has relevant implications because this ratio enters the proton-to-neutron conversion rates. Finally, light species thermally coupled to the electromagnetic sector of the plasma will release entropy after nucleosynthesis and therefore dilute the number density of all nuclei for a given primordial baryon-to-photon ratio. The impact of the particles considered in this work on the primordial abundances of helium and deuterium is depicted in the middle and lower panels of Figure~\ref{fig:Cosmoimply} and can be suitably categorised into three mass regions:
\begin{enumerate}[leftmargin=0.7cm,itemsep=0pt]
    \item[\textbf{A.}] $\boldsymbol{m_\chi \lesssim 0.05 \, \mathrm{MeV}}$ \\ Very light neutrinophilic particles simply contribute to the expansion rate of the Universe during BBN, and do not alter the baryon-to-photon ratio. As such, they simply shorten the timescales on which the weak and nuclear processes freeze out, increasing both $Y_{\mathrm{P}}$ -- which is approximately proportional to the neutron-to-proton ratio at $T \sim 0.07 \, \mathrm{MeV}$ \cite{Sarkar:1995dd,Iocco:2008va,Pospelov:2010hj} -- and $\mathrm{D}/\mathrm{H}|_{\mathrm{P}}$ relative to the SM. Whilst very light electrophilic particles contribute to the expansion rate, which increases the value of $Y_{\mathrm{P}}$, they also release substantial amounts of entropy into the electromagnetic plasma \emph{after} nucleosynthesis. This acts to dilute number of baryons per photon and hence leads to a smaller value of $\mathrm{D}/\mathrm{H}|_{\mathrm{P}}$ than in the SM.
    
    \item[\textbf{B.}] $\boldsymbol{0.5 \, \mathrm{MeV} \lesssim m_\chi \lesssim 10 \, \mathrm{MeV}}$ \\ Electrophilic particles in this region lead to a smaller energy density of the Universe during nucleosynthesis and again to entropy release. This leads to lower values for both $Y_{\mathrm{P}}$ and $\mathrm{D}/\mathrm{H}|_{\mathrm{P}}$ as is seen in the lower two panels of Figure \ref{fig:Cosmoimply}. For neutrinophilic particles on the other hand, there is a larger energy density than in the SM, and hence a larger expansion rate. This leads to larger values for both $Y_{\mathrm{P}}$ and $\mathrm{D}/\mathrm{H}|_{\mathrm{P}}$ for a given $\Omega_{\rm b} h^2$.
    
    \item[\textbf{C.}] $\boldsymbol{m_\chi \gtrsim 30 \, \mathrm{MeV}}$ \\ In this region of masses the energy density of the particles at the time of nucleosynthesis is negligible since their number density is Boltzmann suppressed. As such, one recovers the SM predictions for the primordial element abundances.
\end{enumerate}

Of course, the abundances of other light elements like ${}^3\text{He}$ or ${}^{7}\text{Li}$ are also affected by the presence of light thermal BSM particles. However, such abundances are not typically used as cosmological probes~\cite{pdg}, and therefore we do not use them in this work. The interested reader is referred to Appendix~\ref{app:cosmo_imp_other} for the effect of different types of thermal BSM particles on these abundances. 

\vspace{0.1cm}
\textbf{Expectations}

From Figure~\ref{fig:Cosmoimply} we notice that thermal species of mass $m_\chi \lesssim 0.4-3\,\text{MeV}$ will be ruled out by current $Y_{\rm P}$ measurements. Note that $Y_{\rm P}$ is only logarithmically dependent upon the baryon energy density $\Omega_\mathrm{b}h^2$. One might also expect current ${\rm D/H}|_{\rm P}$ measurements to set stringent constraints on the mass of various BSM particles; about $m_\chi \lesssim 3-10\,\text{MeV}$. However, note that the ${\rm D/H}|_{\rm P}$ predictions are shown for a fixed value of $\Omega_\mathrm{b} h^2$ and ${\rm D/H}|_{\rm P}$ is strongly sensitive to $\propto \Omega_\mathrm{b} h^2$, thus deuterium measurements will only yield constraints provided that $\Omega_\mathrm{b} h^2$ is inferred from CMB observations or in conjunction with $Y_{\rm P}$ measurements.

\vspace{-0.05 cm}
\section{Cosmological Data and Analysis}\label{sec:current_data_analyis}
\vspace{-0.05 cm}

In order to set constraints on the masses of various BSM particles, we perform very conservative analyses using the latest determinations of the primordial element abundances and CMB observations by the Planck satellite as described below. Table~\ref{tab:analysis_summary} provides a summary of the main data sets used in each analysis.

\begin{table}[t]
\begin{center}
{\def\arraystretch{0.9}
\begin{tabular}{lll}
\hline\hline
\textbf{Analysis}  $\qquad$            	   & \textbf{Cosmological Data} $ \,\,\,$	& \textbf{Description} \\ \hline\hline
\multirow{2}{*}{BBN}    &  \multirow{2}{*}{($Y_{\rm P}$,\,\,${\rm D/H}|_{\rm P}$)} & Mean values and error bars as recommended by the PDG. \\
 & & Theoretical uncertainties in the predictions are accounted for. \\ \hline

\multirow{2}{*}{BBN+$\Omega_\mathrm{b} h^2$}   & \multirow{2}{*}{$(Y_{\rm P}$,\,\,${\rm D/H}|_{\rm P},\,\Omega_\mathrm{b} h^2$)} & Same as BBN but with $\Omega_\mathrm{b} h^2 = 0.02225 \pm 0.00066$ from CMB observations. \\
 & & This represents a conservative and model independent range for $\Omega_\mathrm{b} h^2$. \\ \hline

\multirow{2}{*}{Planck}  & \multirow{2}{*}{$(\Omega_\mathrm{b} h^2,\,N_{\rm eff},\,Y_{\rm P})$} & From the Planck2018-TTTEEE+lowE analysis. \\
 & & Assumes $\Lambda$CDM + varying $N_{\rm eff}$ and $Y_{\rm P}$. \\ \hline

\multirow{2}{*}{Planck+$H_0$}  & \multirow{2}{*}{$(\Omega_\mathrm{b} h^2,\,N_{\rm eff},\,Y_{\rm P})$} & From the Planck2018-TTTEEE+lowE+lensing+BAO+$H_0$ analysis. \\
 & & Assumes $\Lambda$CDM + varying $N_{\rm eff}$ and $Y_{\rm P}$. \\ \hline

\multirow{2}{*}{Planck+BBN} & $(\Omega_\mathrm{b} h^2,\,N_{\rm eff},\,Y_{\rm P}) +$ & Joint constraint from Planck 2018 CMB observations and $Y_{\rm P}$ and ${\rm D/H}|_{\rm P}$\\
 & $(Y_{\rm P}$,\,\,${\rm D/H}|_{\rm P})$ & determinations as recommended by the PDG. \\   \hline \hline
\end{tabular}
}
\end{center}\vspace{-0.3cm}
\caption{Summary of the different baseline analyses carried out in this work in order to constrain light BSM particles in thermal equilibrium with the SM plasma during BBN. For each analysis, a likelihood is computed on a grid of $(\Omega_{\rm b} h^2,\,m_\chi)$. }\label{tab:analysis_summary}
\end{table}

\subsection{Big Bang Nucleosynthesis}\label{sec:BBN_data}

We use the PDG recommended means and error bars for the observed primordial abundances of helium and deuterium, which at $1\sigma$ read \cite{pdg}:
\begin{align}
Y_{\rm P} 	            &= 0.245 \pm 0.003 \label{eq:Yp}\, , \\
{\rm D/H}|_{\rm P}   	&= (2.569  \pm 0.027) \times 10^{-5} \label{eq:DH} \,.
\end{align}
These values are based on the analyses/measurements of~\cite{2017RMxAC..49..181P,Aver:2015iza,Izotov:2014fga} and~\cite{Cooke:2016rky,Balashev:2015hoe,2018MNRAS.477.5536Z,Riemer-Sorensen:2017pey} for helium and deuterium respectively. 
In addition to the observational uncertainties in $Y_{\rm P}$ and ${\rm D/H}|_{\rm P}$, we account for theoretical uncertainties in the predicted abundances arising from uncertainties in the neutron lifetime\footnote{Since we account for the uncertainty in the neutron lifetime, in all analyses presented in this work, we shall fix the neutron lifetime to the default value in \texttt{PRIMAT}: $\tau_n =  879.5\,\text{s}$. This value is compatible with the PDG within $1\sigma$, $\tau_n = 880.2\pm 1\,\text{s}$~\cite{pdg}. Choosing $\tau_n = 880.2\,\text{s}$ will not alter any of the results presented in this study.} and various nuclear reaction rates. These are given by~\cite{Pitrou:2018cgg}:
\begin{align}
\sigma(Y_{\rm P})^{\rm Theo} 	        		&= 0.00017 \,, \\
\sigma({\rm D/H}|_{\rm P} )^{\rm Theo}       	&= 0.036 \times 10^{-5} \label{eq:sigmaDH} \, .
\end{align}
It is clear that while the theoretical uncertainty on the $Y_{\rm P}$ prediction is negligible as compared to current observational errors, the ${\rm D/H}|_{\rm P} $ one is not. Earlier references than \texttt{PRIMAT}~\cite{Pitrou:2018cgg} found $\sigma({\rm D/H}|_{\rm P} )^{\rm Theo}  = 0.05\times 10^{-5}$~\cite{Coc:2015bhi}. We have explicitly checked that our conclusions are not altered if we use this latter value from~\cite{Coc:2015bhi}. Moreover, it is known that $\sigma({\rm D/H}|_{\rm P} )^{\rm Theo} $ depends slightly upon the value of the baryon energy density~\cite{pdg}. We have explicitly checked that such dependence leads to $\sigma({\rm D/H}|_{\rm P} )^{\rm Theo} < 0.05 \times 10^{-5}$~\cite{Pitrou:2018cgg} and therefore does not impact any of our conclusions either. 


Assuming Gaussian statistics and combining in quadrature the observational and theoretical errors, we define the following effective BBN $\chi^2$: 
\begin{align}\label{eq:chiBBN}
\chi_{\rm BBN}^2 = \frac{\left[Y_{\rm P} - Y_{\rm P}^{\rm Obs}\right]^2}{\sigma_{Y_{\rm P}}^2|{}^{\rm Theo} + \sigma_{Y_{\rm P}}^2|{}^{\rm Obs}} + \frac{\left[{\rm D/H}|_{\rm P} - {\rm D/H}|_{\rm P}^{\rm Obs}\right]^2}{\sigma_{{\rm D/H}|_{\rm P}}^2|{}^{\rm Theo} + \sigma_{{\rm D/H}|_{\rm P}}^2|{}^{\rm Obs}} \, ,
\end{align}
which we will use to quantify deviations from the observed primordial abundances due to the presence of the new particles in the thermal bath.

\subsection{Cosmic Microwave Background: Planck 2018}\label{sec:CMB_data}
CMB observations measure very precisely three parameters that are relevant for our analysis: $\Omega_\mathrm{b} h^2,\,N_{\rm eff},\,Y_{\rm P}$. The baryon abundance $\Omega_\mathrm{b} h^2$ is one of the 6 parameters in $\Lambda$CDM and Planck reports measurements on $\Omega_\mathrm{b} h^2$ with greater than $1\%$ accuracy. $N_{\rm eff}$ represents one of the most important cosmological parameters and the current accuracy by the Planck satellite on this parameter is $\mathcal{O}(10\,\%)$. $Y_{\rm P}$ is also constrained by CMB observations, albeit with error bars that are typically a factor of 6 - 7 larger than those inferred from blue compact galaxies \cite{pdg}, see Equation~\eqref{eq:Yp}.

In this work, we use the latest CMB observations by the Planck satellite to set constraints on the masses and interactions of various BSM particles. Since the disagreement between local~\cite{Riess:2019cxk} and CMB determinations of the Hubble constant~\cite{Aghanim:2018eyx} could potentially be attributed to additional contributions to $N_{\rm eff}$~\cite{Bernal:2016gxb,Verde:2019ivm}, we consider two data sets: \textit{i)} in which we consider the 2018 Planck baseline TTTEEE+lowE analysis and \textit{ii)} where we combine Planck CMB data with Baryon Acoustic Oscillation (BAO) measurements~\cite{Beutler:2011hx,Ross:2014qpa,Alam:2016hwk} and the local measurement of $H_0$ as reported by the SH0ES collaboration~\cite{Riess:2019cxk}. We shall call the former data set Planck and the latter Planck+$H_0$.  

We build a Gaussian likelihood for the relevant parameters:
\begin{align}\label{eq:chi2_CMB}
\chi_{\rm CMB}^2  = \left(\Theta - \Theta_{\rm Obs}\right)^T \,  \Sigma_{\rm CMB}^{-1} \,\left(\Theta - \Theta_{\rm Obs}\right)\, , \qquad \text{with} \qquad \Sigma_{\rm CMB} = \,
 \left(
\begin{array}{ccc}
\sigma_1^2 & \sigma_1 \sigma_2 \rho_{12} & \sigma_1 \sigma_3 \rho_{13}  \\
\sigma_1 \sigma_2 \rho_{12}     & \sigma_2^2 & \sigma_2 \sigma_3 \rho_{23} \\
\sigma_1 \sigma_3 \rho_{13}     & \sigma_2 \sigma_3 \rho_{23}        & \sigma_3^2 \\
\end{array}
\right) \,, 
\end{align}
where $\Theta = (\Omega_\mathrm{b} h^2,\,N_{\rm eff},\,Y_{\rm P})$ and

\vspace{8pt}
\begin{minipage}{0.49\textwidth}
\centering \textit{Planck 2018}
\begin{align}\label{eq:CovarianceForm}
(\Omega_\mathrm{b} h^2,\,N_{\rm eff},\,Y_{\rm P})|_{\rm Obs} &= (0.02225,\, 2.89,\, 0.246),\nonumber  \\
(\sigma_{1},\,\sigma_{2},\,\sigma_{3}) &= (0.00022,\, 0.31,\, 0.018),  \\
(\rho_{12},\,\rho_{13},\,\rho_{23}) &= ( 0.40,\, 0.18,\, -0.69) \,, \nonumber
\end{align}
\end{minipage}
\begin{minipage}{0.45\textwidth}
\centering \textit{Planck 2018+BAO+$H_0$}
\begin{align}
(\Omega_\mathrm{b} h^2,\,N_{\rm eff},\,Y_{\rm P})|_{\rm Obs} &= (0.02345,\, 3.36,\, 0.249), \nonumber  \\
(\sigma_{1},\,\sigma_{2},\,\sigma_{3}) &= (0.00025,\, 0.25,\, 0.020),      \\
(\rho_{12},\,\rho_{13},\,\rho_{23}) &= (0.011,\, 0.50,\, -0.64) \,  \nonumber,
\end{align}
\end{minipage}
\vspace{0.2cm}

 where the covariance matrix for the Planck 2018 analysis has been extracted from the Planck database~\cite{Aghanim:2018eyx,Aghanim:2019ame}. The covariance matrix for the Planck 2018+BAO+$H_0$ analysis was obtained by running a Markov-Chain-Monte-Carlo analysis using \texttt{CLASS}~\cite{Blas:2011rf,Lesgourgues:2011re} and \texttt{Monte Python}~\cite{Brinckmann:2018cvx,Audren:2012wb} with Planck 2018 data~\cite{Aghanim:2018eyx,Aghanim:2019ame}, various BAO measurements~\cite{Beutler:2011hx,Ross:2014qpa,Alam:2016hwk} and by including a Gaussian likelihood on $H_0$ from the results of~\cite{Riess:2019cxk}. Clearly, the main implication of including local measurements of $H_0$ in the fit is the upward shift on the reconstructed value of $N_{\rm eff}$ from $2.89$ to $3.36$.

\subsection{BBN+CMB Data Combinations}\label{sec:Statistics}
Combining measurements of the primordial element abundances and CMB observations proves useful in constraining light thermal species coupled to the SM plasma.

In this work, we will combine BBN+CMB data in two ways: \textit{i)} by constructing a joint $\chi^2$ that is obtained by summing the individual Planck and BBN $\chi^2$'s as defined in Eqs.~\eqref{eq:chiBBN} and~\eqref{eq:chi2_CMB} (labelled BBN+Planck across the paper) and \textit{ii)} by adding to $\chi^2_{\rm BBN}$ a measurement of $\Omega_\mathrm{b} h^2 = 0.02225 \pm 0.00066$, which is to be regarded as a cosmological model-independent Planck determination of the baryon energy density\footnote{The value $\Omega_\mathrm{b} h^2 = 0.02225 \pm 0.00066$ has an error $4.4$ times larger than the one associated with $\Lambda\text{CDM}$ using Planck 2018 observations~\cite{Aghanim:2018eyx}, and furthermore it covers well the inferred value of $\Omega_\mathrm{b} h^2$ in a well-motivated 12-parameter extension of $\Lambda$CDM using different data sets~\cite{DiValentino:2016hlg,DiValentino:2017zyq}.} (we shall call this analysis BBN+$\Omega_\mathrm{b} h^2$). See Appendix~\ref{app:Omegab} for details.

\subsection{Statistical Assessment}\label{sec:Statistics}

For each of the scenarios considered, the quantities $\chi^2_\mathrm{BBN}$ and $\chi^2_\mathrm{CMB}$ are computed on a grid of $(\Omega_\mathrm{b} h^2,\,m_\chi)$ and subsequently marginalized over $\Omega_\mathrm{b} h^2$. Then, by comparing the marginalized 1-D $\chi^2 (m_\chi)$ with the minimum $\chi^2_{\rm min}$, we consider a scenario to be ruled out at $2\sigma$ when $\Delta \chi^2 \equiv \chi^2 - \chi^2_\mathrm{min} = 4$. The statistical compatibility of each ${\chi}^2_\mathrm{min}$ is estimated by computing its p-value, which is found to be acceptable in all cases presented in this work. This is as expected, given that BBN predictions and CMB observations are compatible with each other within the Standard Model.

\begin{figure}[t]
    \centering
    \hspace*{-0.2 cm}\includegraphics[width=0.45\textwidth]{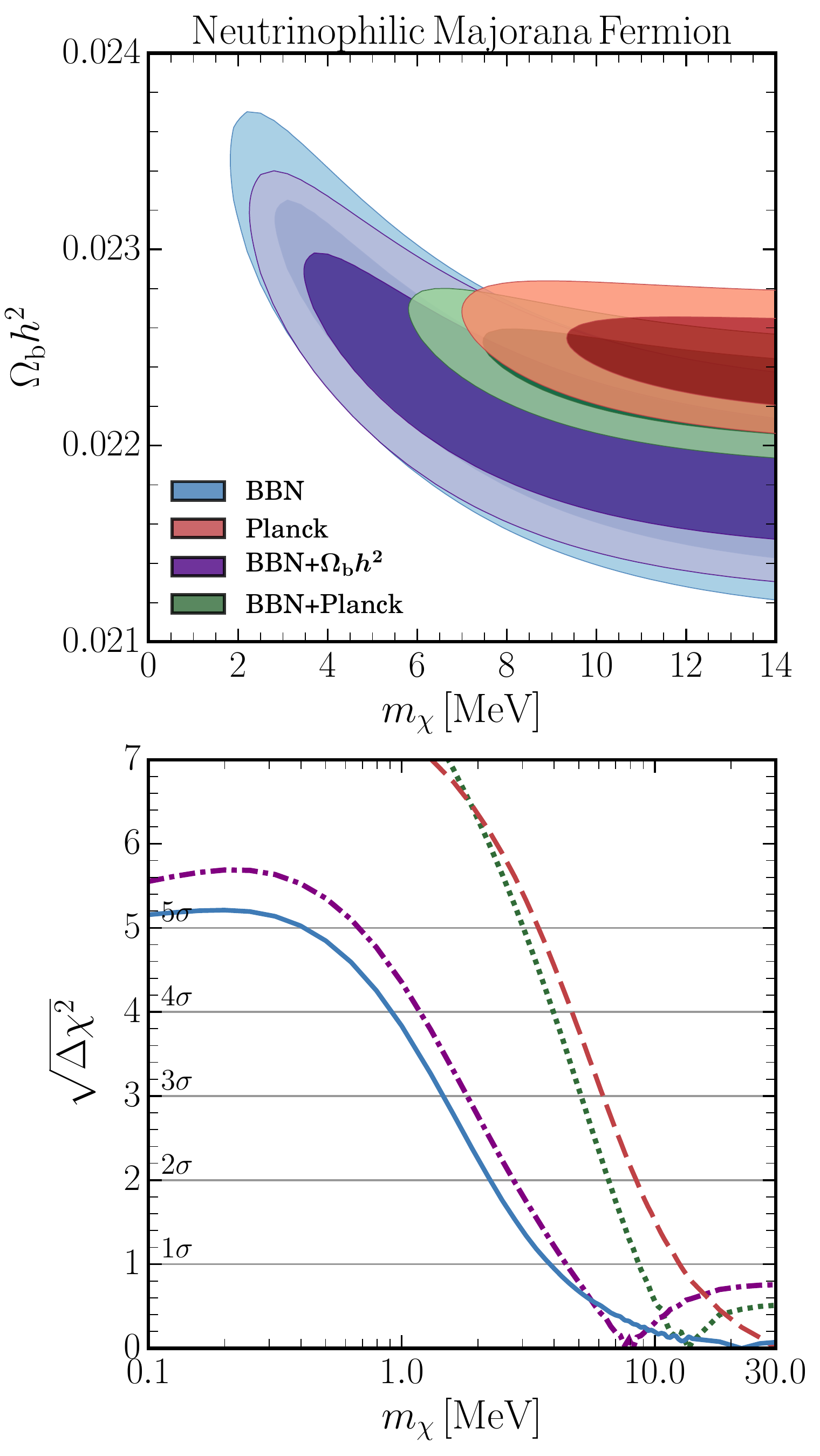} \quad
    \includegraphics[width=0.45\textwidth]{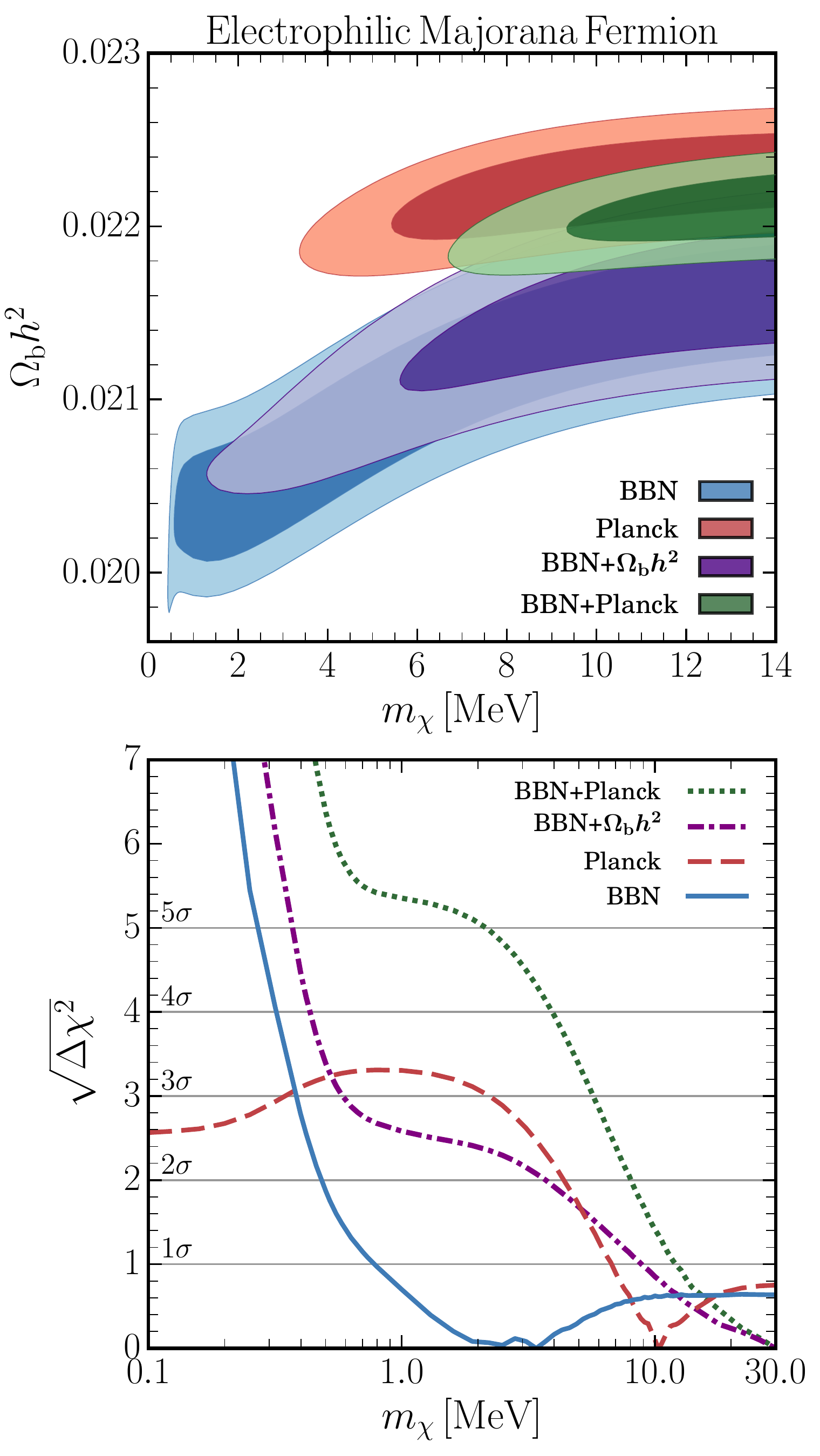}\vspace{-0.3cm}
    \caption{\emph{Upper Panels:} Contour plots showing the $1\sigma$ and $2\sigma$ confidence intervals in the $(\Omega_\mathrm{b} h^2$,\,$m_\chi)$ plane for a Majorana fermion with mass $m_\chi$ in thermal equilibrium with the SM plasma. \emph{Lower Panels:} Marginalized $\Delta \chi^2$ as a function of $m_\chi$. Solid lines correspond to BBN, dashed to Planck CMB 2018 observations, dash-dotted to the combination BBN+$\Omega_{\mathrm{b}}h^2$ and dotted to BBN+Planck. The \textit{left/right panel} corresponds to neutrinophilic/electrophilic particles.}
    \label{fig:Current_1-2D}
\end{figure}

\section{Current Cosmological Constraints on Purely Electrophilic or Neutrinophilic Species}\label{sec:results}
\vspace{-0.2cm}

\begin{table*}[t]
\begin{center}
{\def\arraystretch{1.35}
\begin{tabular}{l|lc|ccccc|cc}
\hline\hline
\multirow{2}{*}{\textbf{Type}$\,$}	& \multicolumn{2}{c|}{\textbf{BSM Particle}}   	 &  \multicolumn{5}{c|}{$\,\, $ \textbf{Current Constraints} $\,\,$}  &  \multicolumn{2}{c}{$\,\, $ \textbf{Forecasted Constraints} $\, $} \\
   &	 \multirow{1}{*}{Particle}    	 &   \multirow{1}{*}{g-Spin}  &   
   \multicolumn{1}{c}{$\,\,\,$ BBN $\,\,\,$}  &
   \multicolumn{1}{c}{BBN+$\Omega_\mathrm{b} h^2$}  &
   \multicolumn{1}{c}{$\,\,\,$ Planck $\,\,\,$}  &
   \multicolumn{1}{c}{Planck+$H_0$ $\,$}  &   \multicolumn{1}{c|}{BBN+Planck$\,\,\,$} &   \multicolumn{1}{c}{$\,$ Simons Obs.}  &   \multicolumn{1}{c}{$\,$ CMB-S4 $\,$} \\ 
  \hline \hline
\parbox[t]{8mm}{\multirow{5}{*}{\rotatebox[origin=c]{90}{\textbf{Neutrinophilic}}}}   

& Majorana   & 2-F  & 2.2 &  2.8  & 8.4 & 4.9 & 6.6  & 12.5   & 13.5  \\ \cline{2-10} 

& Dirac  & 4-F  & 3.7 &  5.4  & 11.3 & 8.0 & 9.4  & 15.3  & 16.2    \\ \cline{2-10} 

& Scalar   & 1-B  & 1.2 & 1.3  & 5.6 & 1.6 & 3.7  & 9.8  & 10.7\\ \cline{2-10} 

& Complex Scalar  & 2-B  & 2.3 & 2.9  & 8.5 & 5.1 & 6.7  & 12.5  & 13.5\\ \cline{2-10} 

& Vector  & 3-B  & 3.1 & 4.4  & 10.1 & 6.8 & 8.3  & 14.1  & 15.1   \\ \cline{2-10} 
\hline	

  \parbox[t]{8mm}{\multirow{5}{*}{\rotatebox[origin=c]{90}{\textbf{Electrophilic}}}}   

&Majorana & 2-F  &  0.5 & 3.7 & 4.4 & 9.2 & 8.0  & 12.2  & 13.2\\ \cline{2-10} 

&Dirac   & 4-F  &  0.7 & 7.0 & 7.4 & 12.0 & 10.9  & 14.9  & 15.9\\  \cline{2-10} 

&Scalar  & 1-B  & 0.4 & 0.6  & $\>\>2.4^{*}$ & 6.4 & 5.2  & 9.4  & 10.5\\  \cline{2-10} 

&Complex Scalar   & 2-B  &  0.5 & 4.0 &  4.6 & 9.2 & 8.1  & 12.2  & 13.2\\  \cline{2-10} 

&Vector    & 3-B  & 0.6 & 5.8  & 6.3 & 10.9 & 9.8  & 13.8  & 14.8   \\ 
\hline \hline
\end{tabular}
}
\end{center}
\vspace{-0.3cm}
\caption{Lower bounds at 95.4\% CL on the masses of various thermal BSM particles in MeV. The columns correspond to constraints using data from various sources as detailed in Sections~\ref{sec:current_data_analyis} and~\ref{sec:future} for current and forecasted constraints respectively. The rows correspond to BSM particles with a different number of internal degrees of freedom $g$ and spin (F: fermion, B: boson). The upper/lower parts of the table correspond to purely neutrinophilic/electrophilic particles. $^{*}$This bound is only at 86\% CL.}\label{tab:DMbounds}
\vspace{-0.5cm}
\end{table*}

Using different sets of cosmological observations, we set stringent constraints on the mass of various BSM species that are thermally coupled to the SM plasma. In Table~\ref{tab:analysis_summary} we provide a summary of the data sets used in each analysis, and in Table~\ref{tab:DMbounds} we report the $95.4\%$ CL lower bounds on the mass of such BSM species.

In order to illustrate the extent to which cosmological observations constrain the masses of light thermally coupled BSM species, we depict in the upper frames of Figure~\ref{fig:Current_1-2D} the $1\sigma$ and $2\sigma$ confidence intervals in the $(\Omega_\mathrm{b}h^2$, $m_\chi)$ plane for our four baseline analyses for the case of a Majorana fermion. We observe degeneracies between $\Omega_\mathrm{b}h^2$ and $m_\chi$ in some regions of parameter space for the BBN analysis, but thanks to the precision with which the primordial element abundances are measured and the CMB is observed, a lower bound on $m_\chi$ can be set. From the lower panels of Figure~\ref{fig:Current_1-2D} we see that neutrinophilic BSM states with $m_\chi \lesssim 1$ MeV are strongly disfavoured by BBN. In the case of electrophilic states, the same holds, albeit for relatively lighter BSM particles with $m_\chi \lesssim 0.3$ MeV. These results show that current cosmological observations set very stringent constraints on light species in thermal equilibrium during the time of BBN. In addition, the constraints derived from BBN are independent of the assumed cosmological model. In particular, BBN disfavours thermal particles with $m_\chi < 0.1$ MeV at more than $5\sigma$ -- with the sole exception of a neutrinophilic neutral scalar that is disfavoured at $3.3\sigma$. This is done only by using the observed values of $Y_{\rm P}$ and ${\rm D/H}|_{\rm P}$. The reader is referred to Figure~\ref{fig:Multiplot} in Appendix~\ref{app:FullResults} for the $\Delta \chi^2 (m_\chi)$ of each scenario considered in this study.

\textbf{BBN and BBN}$\boldsymbol{+\Omega_\mathrm{b}h^2}$

From Table~\ref{tab:DMbounds} we observe that from the current determinations of the primordial helium and deuterium abundances (BBN) alone we are able to place a lower bound on the mass of $m_\chi >0.4$ MeV at 95.4\% CL. This bound is independent of the spin, number of internal degrees of freedom of the species at hand and also of whether the particle interacts only with neutrinos or electrons/photons. We notice that the bounds for neutrinophilic species, $m_\chi > (1.2-3.7)\,\text{MeV}$, are stronger as compared with the bounds for electrophilic species, $m_\chi > (0.4-0.7)\,\text{MeV}$. When very conservative information about the value of $\Omega_\mathrm{b}h^2$ from CMB observations is included (BBN+$\Omega_\mathrm{b}h^2$), the bounds get slightly stronger to the level of $m_\chi > (1.3-4.4)\,\text{MeV}$ for neutrinophilic species and $m_\chi > (0.6-7.0)\,\text{MeV}$ for electrophilic ones.\\\\
\textbf{Planck} 

From the Planck column in Table~\ref{tab:DMbounds} one can clearly see that Planck typically sets more restrictive constraints than BBN. For neutrinophilic relics $m_\chi > (5.6-11.3)\,\text{MeV}$, while for electrophilic relics $m_\chi > (4.4-7.4)\,\text{MeV}$. The sole exception to this is an electrophilic scalar boson that cannot be constrained at $2\sigma$ from Planck CMB observations (as can be seen from Figure~\ref{fig:Cosmoimply}). Nonetheless, we find that a lower bound of $m_\chi > 2.4$ MeV at 86\% CL can still be set.\\\\
\textbf{Planck}$\boldsymbol{+H_0}$

Planck constraints are based solely on CMB observations. However, the actual value of $N_{\rm eff}$ may be different if local determinations of the Hubble constant are taken into account as discussed in Section \ref{sec:current_data_analyis}. We find that when local measurements of $H_0$, BAO data and Planck CMB observations are considered, the bounds for neutrinophilic relics are relaxed as compared to Planck data alone, while the bounds for electrophilic relics become stronger. This is because the inclusion of the local determination of $H_0$ results in a higher mean value of $N_\mathrm{eff}$, which leads to a preference for neutrinophilic relics that generally contribute to $N_\mathrm{eff} >N_\mathrm{eff}^{\rm SM}$. Still, this data combination rules out thermal BSM particles of $m_\chi > 1.6\,\text{MeV}$ at 95.4\% CL.\\\\
\textbf{BBN+Planck}

Finally, when $Y_{\rm P}$ and  ${\rm D/H}|_{\rm P}$ data are combined with Planck CMB observations we find that the constraints for neutrinophilic species are slightly relaxed as compared to Planck alone, yielding $m_\chi > (3.7-9.4)\,\text{MeV}$, while the bounds get stronger for electrophilic relics, yielding $m_\chi > (5.2-10.9)\,\text{MeV}$. This is a mere result of a slight $\sim 0.9\sigma $ tension between the $\Omega_\mathrm{b}h^2$ that is inferred from BBN and CMB observations~\cite{Pitrou:2018cgg}. Note that from the lower panels of Figure~\ref{fig:Current_1-2D}, Plank+BBN data strongly disfavours very light BSM thermal species. \\\\
\textbf{Summary} 

We have set strong constraints from a combination of cosmological measurements, including the primordial helium and deuterium abundances and CMB observations by the Planck satellite. For the combination of BBN+Planck data, we find that the mass of purely electrophilic and neutrinophilic BSM species in thermal equilibrium with the SM plasma -- independently of their spin and number of degrees of freedom -- should satisfy $m_\chi > 3.7\,\text{MeV}$ at 95.4\% CL.

\section{Current Cosmological Constraints on Generic WIMPs}\label{sec:results_mixed}

\begin{figure}[b]
    \centering
    \includegraphics[width=0.95\textwidth]{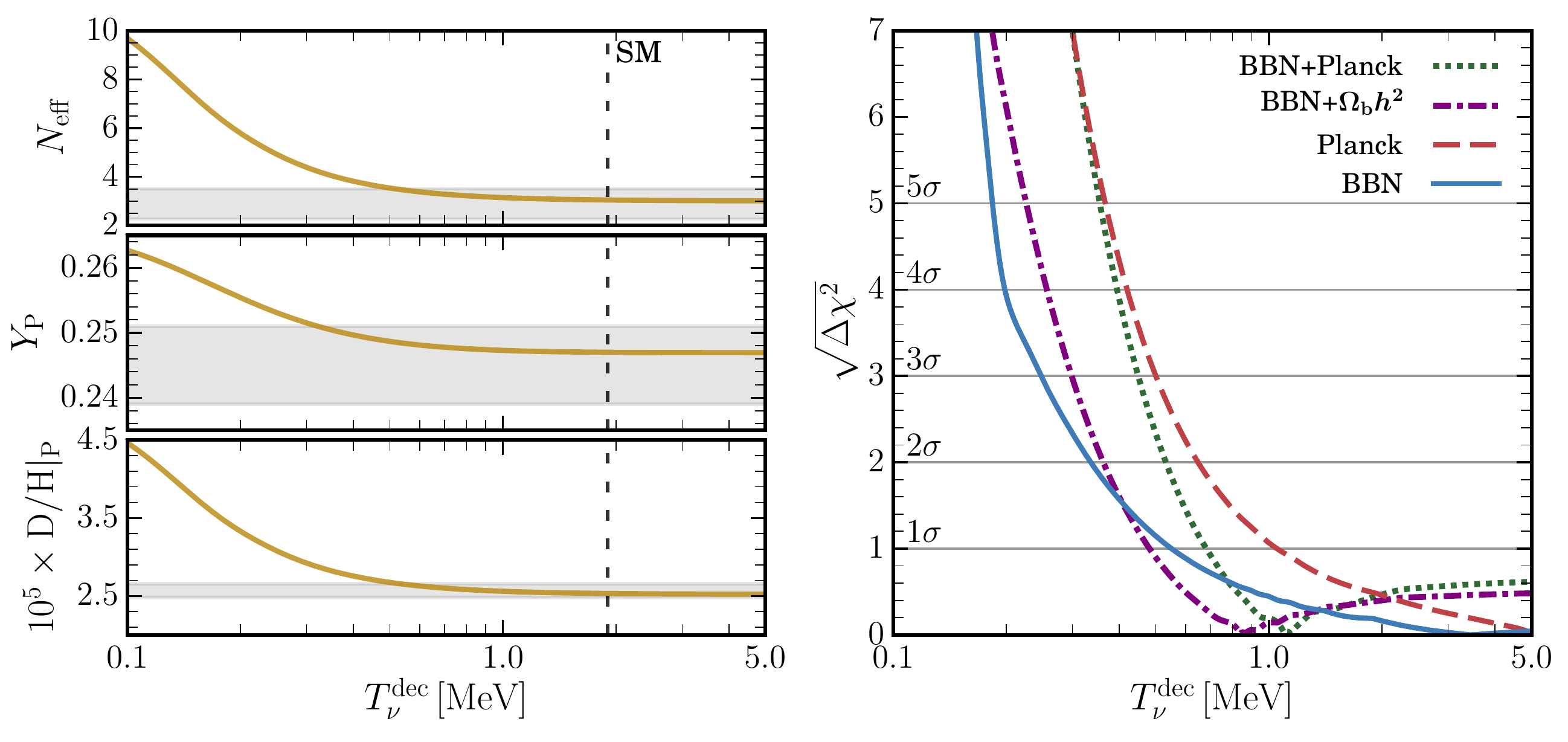}\vspace{-0.4cm}
    \caption{\textit{Left panels:} Impact of a non-standard neutrino decoupling temperature on the cosmological observables $N_{\mathrm{eff}}$, $Y_{\mathrm{P}}$ and $\mathrm{D}/\mathrm{H}|_{\mathrm{P}}$. The $Y_{\rm P}$ and ${\rm D/H}|_{\rm P}$ predictions are computed with $\Omega_\mathrm{b} h^2 = 0.021875$ and $\tau_n = 879.5\,\text{s}$. The grey contours correspond to the mean $\pm \, 2\sigma$ measurements that enter our BBN and Planck data analysis, see Eqs.~\eqref{eq:chiBBN},~\eqref{eq:chi2_CMB}  and~\eqref{eq:CovarianceForm}. \textit{Right panel:} Marginalized $\Delta \chi^2$ as a function of the neutrino decoupling temperature. Solid lines correspond to BBN constraints, dashed to Planck CMB observations, dash-dotted to the combination BBN+$\Omega_{\mathrm{b}}h^2$ and dotted to BBN+Planck.}
    \label{fig:Cosmo_imp_Tnudec}
\end{figure}

Up to this point we have restricted our analysis to particles that interact solely with neutrinos or electrons/photons. In this section, we explore the cosmological constraints that can be placed on particles which interact with both neutrinos and electrons. First, we provide a brief overview of the phenomenology of this scenario and then we describe the constraints we find on such particles from BBN and CMB observations. The reader is referred to Table \ref{tab:BRBounds} for a suite of constraints on thermal WIMPs that annihilate at a different rate to electrons/photons and neutrinos, to Table \ref{tab:Tnudec_bounds} for lower bounds on the neutrino decoupling temperature $T_{\nu}^{\rm dec}$, and to Figure \ref{fig:Cosmo_imp_Tnudec} for the cosmological implications of a non-standard $T_{\nu}^{\rm dec}$.

\subsection{Overview of Cosmological Implications}

The key consequence of BSM particles in thermal equilibrium with the SM plasma through interactions with both electrons and neutrinos is that they efficiently act as to delay the process of neutrino decoupling. A reduction of the neutrino decoupling temperature as compared to the SM case, $T_{\nu} ^{\rm dec}|_{\rm SM}$, implies that:\vspace{-0.1cm}
\begin{itemize}[leftmargin=0.7cm,itemsep=-1pt]
    \item The entropy released by electron-positron annihilation at $T\lesssim m_e$ is shared among both photons and neutrinos, yielding $T_\nu > T_\nu^{\rm SM}$ and $T_\gamma < T_\gamma ^{\rm SM}$ and therefore implying $N_{\rm eff} > N_{\rm eff}^{\rm SM}$.
    \item The number density of baryons is enhanced as compared with the SM case, as a result of the smaller number density of photons after electron-positron annihilation. \item A higher expansion rate of the Universe for $T \lesssim m_e$. 
\end{itemize}

Both stable and unstable particles can efficiently delay the process of neutrino decoupling. For example, in the case of a thermal WIMP that annihilates equally to neutrinos and electrons, the energy will be efficiently transferred between the electromagnetic and neutrino sectors of the plasma as a result of $e^+e^- \leftrightarrow \chi \chi  \leftrightarrow \bar{\nu} \nu$ processes\footnote{For regions of parameter space in which $m_\chi<m_e$, it is understood that the annihilation proceeds into $\gamma \gamma$ and not into $e^+e^-$ by pure kinematics.} until $T/m_\chi \sim 1/10$ \cite{Escudero:2018mvt}. If one considers $m_\chi = 2$ MeV, neutrino decoupling will occur at $T_{\nu}^{\rm dec} \sim 0.2$ MeV and hence at a temperature that is much lower than the SM one of $T_{\nu} ^{\rm dec}|_{\rm SM} \simeq 1.9$ MeV.

Unstable particles that interact with both neutrinos and electrons would delay the process of neutrino decoupling, typically for even smaller temperatures than WIMPs. For example, consider a light $U(1)_{B-L}$ gauge boson $Z'$ that decays at a similar rate to neutrinos and electrons. By comparing the inverse decay rate to the Hubble parameter one can show that via $e^+e^- \leftrightarrow Z' \leftrightarrow \bar{\nu}\nu$ interactions the process of neutrino decoupling will be delayed until
\begin{align}\label{eq:dec_invdecay}
\frac{m_{Z'}}{T} \simeq \log\left[\frac{ \Gamma_{Z'} M_{\rm Pl}}{1.66\,\sqrt{g_\star} m_{Z'}^2}\right] + \frac{7}{2} \log\left[\frac{m_{Z'}}{T}\right]\, ,
\end{align}
where $\Gamma_{Z'} \simeq g_{B-L}^2 m_{Z'}/(4\pi)$ is the decay width of the $Z'$ and $g_{B-L}$ is the gauge coupling of the corresponding $U(1)_{B-L}$ symmetry. Taking representative numbers, we can see that this particle will delay the process of neutrino decoupling until
\begin{align}\label{eq:nudec_decay}
\frac{m_{Z'}}{T} \simeq 30 + \log\left[ \left(\frac{g_{B-L}}{10^{-6}}\right)^2 \frac{5\,\text{MeV}}{m_{Z'}}\right] + \frac{7}{2} \log\left[\frac{m_{Z'}}{30\,T}\right]\, ,
\end{align}
and hence a $Z'$ with $m_{Z'} \simeq 5$ MeV and $g_{B-L} \simeq 10^{-6}$ will delay neutrino decoupling until $T_{\nu}^{\rm dec} \sim 0.2$ MeV; a temperature that is one order of magnitude smaller than the SM one.

\begin{table}[b]
\begin{center}
{\def\arraystretch{1.3}
\begin{tabular}{c|ccccc|cc}
\hline\hline
\multirow{2}{*}{\shortstack{\textbf{Temperature of}\\ \textbf{Neutrino Decoupling}}}	&  \multicolumn{5}{c|}{$\,\, $ \textbf{Current Constraints} $\,\,$}  &  \multicolumn{2}{c}{$\,\, $ \textbf{Forecasted Constraints} $\, $} \\
  &   
   \multicolumn{1}{c}{$\,\,\,$ BBN $\,\,\,$}  &
   \multicolumn{1}{c}{BBN+$\Omega_\mathrm{b} h^2$}  &
   \multicolumn{1}{c}{$\,\,\,$ Planck $\,\,\,$}  &
   \multicolumn{1}{c}{Planck+$H_0$ $\,$}  &   \multicolumn{1}{c|}{BBN+Planck$\,\,\,$} &   \multicolumn{1}{c}{$\,$ Simons Obs.}  &   \multicolumn{1}{c}{$\,$ CMB-S4 $\,$} \\ 
  \hline \hline
 $T_{\nu} ^{\rm dec}$    & 0.34 &  0.36  & 0.65 & 0.43 & 0.53  & 1.0   & 1.1 \\
\hline \hline
\end{tabular}
}
\end{center}
\vspace{-0.3cm}
\caption{Lower bounds at 95.4\% CL on the neutrino decoupling temperature in MeV.}\label{tab:Tnudec_bounds}
\end{table}

\newpage
\subsection{Constraints}

In this section we follow two avenues to set constraints on particles that interact with both neutrinos and electrons: \textit{i)} we set constraints on the masses of various types of WIMPs that annihilate with a thermal cross-section, but with varying final state annihilation ratios to electrons and neutrinos, and \textit{ii)} we set constraints on the neutrino decoupling temperature $T_{\nu}^{\rm dec}$ that can be mapped into many extensions of the SM with light interacting species. 

We consider all possible combinations of spins and final state annihilation ratios of electrons to neutrinos, ranging from $e: \nu = 10^{6}:1 $ to $e: \nu = 1:10^{6} $. All the resulting lower bounds of the WIMP mass at 95.4\% CL can be found in Table~\ref{tab:BRBounds} (see Table~\ref{tab:MajoranaBRBounds} for the particular case of a Majorana dark matter fermion). The three main conclusions that can be inferred from this analysis are:
\begin{itemize}
    \item Current BBN observations bound the thermal dark matter mass to be  $m_\chi > 0.4\,\text{MeV}$ at $2\sigma$. This can be seen from Table~\ref{tab:BRBounds}. This constraint is independent of the spin or the number of internal degrees of freedom of the given WIMP, as well as of the annihilation channel to SM species. 
    \item For any $e:\nu$ annihilation ratio and given type of WIMP, with the exception of a neutral scalar particle, Planck CMB observations set a $2\sigma$ lower bound on $m_\chi$. Similarly, Planck+BBN bound set $m_\chi > 0.8\,\text{MeV}$ a $2\sigma$. These bounds are independent upon the spin of the particle and whether the annihilation is s-wave or p-wave. 
    \item WIMPs with an annihilation ratio $e:\nu \sim 10^4: 1$ are particularly elusive to BBN and CMB observations; only BSM particles of $m_\chi \lesssim 1.3\,\text{MeV}$ can be constrained at present. This results from the fact that on one hand, such species preferably dump entropy into the electromagnetic sector of the plasma, while on the other hand, delayed neutrino decoupling causes that entropy to be shared with neutrinos. This yields $N_{\rm eff} \simeq N_{\rm eff}^{\rm SM}$. 
\end{itemize}

In addition, we set a lower bound on the temperature at which neutrinos decouple. In order to do so, we do not include any BSM species in the evolution, but solve for neutrino decoupling as in the Standard Model but with a modified Fermi constant $G_{\rm F}'$ in the neutrino-electron rates as in Equation~\eqref{eq:energyrates_nu_SM}, such that the temperature of neutrino decoupling is altered with respect to the SM, $T_{\nu}^{\rm dec} = 1.9\,\text{MeV} \, ({G_{\rm F}}/{G_{\rm F}'})^{2/3}$. For this computation we neglect the electron mass and evaluate the relevant $\chi^2$ on a grid of $(T_{\nu}^{\rm dec},\,\Omega_\mathrm{b}h^2)$. The results of such analysis are presented in Table~\ref{tab:Tnudec_bounds}.

We find that $T_{\nu}^{\rm dec} > 0.34\,\text{MeV}$ at 95.4\% CL from BBN observations. Planck CMB observations are more stringent than BBN and set $T_{\nu}^{\rm dec} > (0.43-0.65)\,\text{MeV}$ depending on whether local measurements of $H_0$ are included or not. In addition, from the right panel of Figure \ref{fig:Cosmo_imp_Tnudec}, we can appreciate that $ T_{\nu}^{\rm dec} \lesssim 0.2\,\text{MeV}$ are highly disfavoured by current cosmological observations. Finally, joint BBN+Planck CMB observations constrain $T_{\nu}^{\rm dec} > 0.63\,\text{MeV}$ and we expect next generation of CMB observations to be able to test $T_{\nu}^{\rm dec} \lesssim 1\,\text{MeV}$.

The bound on the neutrino decoupling temperature is generic and can be directly mapped into a constraint on the mass and couplings of various BSM species interacting both with neutrinos and electrons. The phenomenology of such type of scenarios has been studied in detail in the context of a very light $U(1)_{\mu-\tau}$ gauge boson \cite{Escudero:2019gzq}, but from Equation \eqref{eq:nudec_decay} we can directly map a bound on $T_{\nu}^{\rm dec}> 0.3\,\text{MeV}$ into a constraint on other scenarios. For example, by considering the conservative bound $T_{\nu}^{\rm dec}> 0.3\,\text{MeV}$ in the case of a light $U(1)_{B-L}$ gauge boson, from Equation \eqref{eq:nudec_decay} we can appreciate that gauge bosons of $m_{Z'} \lesssim 10 \, \,\text{MeV}$ with $g_{B-L} \gtrsim  5 \times 10^{-7} \sqrt{10\,\text{MeV}/m_{Z'}}$ are excluded by cosmological observations. Note that this constraint is independent of the branching fraction of such $Z'$ to invisible particles.

\begin{table}[t]
\begin{center}
{\def\arraystretch{1.3}
\begin{tabular}{c|l|p{0.7cm}|p{0.7cm}|p{0.7cm}|p{0.7cm}|p{0.7cm}|p{0.7cm}|p{0.7cm}|p{0.7cm}|p{0.7cm}|p{0.7cm}|p{0.7cm}|p{2cm}|p{0.7cm}}
\hline\hline
\multirow{2}{*}{\textbf{Type}$\,$} & \multirow{2}{*}{\vspace*{-0.1 cm}\hspace{0.44 cm}\textbf{Probe}} & \multicolumn{13}{c}{$\boldsymbol{e :}  \boldsymbol{\nu}$ \textbf{Annihilation Ratio}}\\
  &                       & \multicolumn{1}{c}{1:$10^6$} & \multicolumn{1}{c}{1:$10^5$} & \multicolumn{1}{c}{1:$10^4$} & \multicolumn{1}{c}{1:$10^3$} & \multicolumn{1}{c}{1:$10^2$} & \multicolumn{1}{c}{1:$10^1$} & \multicolumn{1}{c}{1:1} & \multicolumn{1}{c}{$10^1$:1} & \multicolumn{1}{c}{$10^2$:1} & \multicolumn{1}{c}{$10^3$:1} & \multicolumn{1}{c}{$10^4$:1} & \multicolumn{1}{c}{$10^5$:1} & \multicolumn{1}{c}{$10^6$:1} \\ \hline\hline
\parbox[t]{3mm}{\multirow{5}{*}{\rotatebox[origin=c]{90}{\hspace{-0.32 cm}\textbf{Majorana}}}} & BBN      & \hfil 2.2           &    \hfil 2.0        &  \hfil 1.8          & \hfil 1.8           & \hfil 2.0           &    \hfil 2.3        & \hfil 2.4     &  \hfil 2.2         &  \hfil 1.9           &  \hfil 1.6           &    \hfil 1.1         &  \hfil 0.7           & \hfil 0.5           \\
\rule{0pt}{3ex} & BBN+$\Omega_\mathrm{b}h^2$      & \hfil 2.6           &    \hfil 2.2        &   \hfil 1.8         & 
\hfil 1.9           & \hfil 2.1           &    \hfil 2.4        &  \hfil 2.6     &  \hfil 2.3          & \hfil 1.9           & \hfil 1.5           &    \hfil 1.1        & \hfil  0.9          &  \hfil 3.3          \\
\rule{0pt}{3ex} & Planck      &   \hfil 8.2         &    \hfil 7.6        &  \hfil 4.6          & \hfil  3.4          & \hfil 3.6           &   \hfil  4.1        & \hfil 4.5      &  \hfil 3.8       & \hfil 3.0           &   \hfil  2.1        &  \hfil 1.1          &  \hfil 0.3          & \hfil 4.1           \\
\rule{0pt}{3ex} & Planck$+H_0$      &  \hfil 4.7          &  \hfil 3.3          &     \hfil 2.0       &  \hfil   2.0        &   \hfil 2.3         & \hfil 2.7           & \hfil 3.0      & \hfil 3.2    &  \hfil 2.5      & \hfil  1.8          &     \hfil 0.9       &  \hfil 0.2; 0.4 - 8.4 &\hfil   8.9           \\
\rule{0pt}{3ex} & BBN+Planck      &    \hfil 6.4        &    \hfil 5.4        & \hfil 3.1           &  \hfil 2.7          &     \hfil 2.9       &  \hfil   3.4        &  \hfil 3.7     &\hfil  3.2           & \hfil 2.6           &    \hfil  1.9       &\hfil     1.1        &\hfil   7.0          &  \hfil 7.8          \\
 \hline\hline
\end{tabular}
}
\caption{Lower bounds at 95.4\% CL on the mass of a Majorana thermal dark matter particle in MeV. The rows correspond to constraints using data from various sources as detailed in Section~\ref{sec:current_data_analyis}. The columns correspond to the annihilation ratio between electrons/photons and neutrinos in the final state. Here we have assumed $\left<\sigma v\right> = \left<\sigma v\right>_{\rm WIMP}$. }
\label{tab:MajoranaBRBounds}
\end{center}
\end{table}

\section{Future Cosmological Constraints}\label{sec:future}

\subsection{Cosmic Microwave Background}\label{sec:future_CMB}
There are a number of proposed future CMB experiments that would provide an accurate determination of the relevant cosmological parameters for this study: $\Omega_\mathrm{b} h^2,\,N_{\rm eff},\,Y_{\rm P}$. Proposed experiments include satellite missions like PICO~\cite{Hanany:2019lle} or CORE~\cite{DiValentino:2016foa} and ground-based experiments like the Simons Observatory~\cite{Ade:2018sbj}, CMB-S4~\cite{Abazajian:2016yjj,Abazajian:2019eic} and CMB-HD~\cite{Sehgal:2019ewc}. In this section we consider the reach of the Simons Observatory\footnote{\url{https://simonsobservatory.org/}.}, because it is fully funded and expected to deliver measurements within the next few years, and that of CMB-S4\footnote{\url{https://cmb-s4.org/}.}, because it aims to reach a sub-percent determination of $N_{\rm eff}$.

We use the Fisher Matrix method to forecast the reach of CMB-S4 (see Appendix~\ref{app:CMBfisher} for details) and use the baseline covariance matrix from the Simons Observatory collaboration. In analogy with~\eqref{eq:chi2_CMB}, the relevant parameters read:

\begin{center}
\begin{minipage}{0.49\textwidth}
\centering \textit{Simons Observatory}
\begin{align}
(\Omega_\mathrm{b} h^2,\,N_{\rm eff},\,Y_{\rm P})|_{\rm Fiducial} &= (0.022360,\, 3.046,\, 0.2472) \, \nonumber, \\
(\sigma_{1},\,\sigma_{2},\,\sigma_{3}) &= (0.000073,\, \,\,\,0.11,\, 0.0066)\, , \nonumber \\
(\rho_{12},\,\rho_{13},\,\rho_{23}) &= (0.072,\, 0.33,\, -0.86) \, ,\nonumber
\end{align}
\end{minipage}
\begin{minipage}{0.49\textwidth}
\centering \textit{CMB-S4}
\begin{align}
(\Omega_\mathrm{b} h^2,\,N_{\rm eff},\,Y_{\rm P})|_{\rm Fiducial} &= (0.022360,\, 3.046,\, 0.2472) \, , \nonumber  \\
(\sigma_{1},\,\sigma_{2},\,\sigma_{3}) &= (0.000047,\, 0.081,\, 0.0043)\,  ,\nonumber\\ 
(\rho_{12},\,\rho_{13},\,\rho_{23}) &= (0.25,\, 0.22,\, -0.84) \, ,\nonumber
\end{align}
\end{minipage}
\end{center}

Note that the forecasted errors on $N_{\rm eff}$ look substantially larger than what is typically quoted in the literature and this is simply a result of the fact we also allow $Y_{\rm P}$ to vary.

\begin{figure*}[t]
    \centering
    \includegraphics[width=0.45\textwidth,height=200pt]{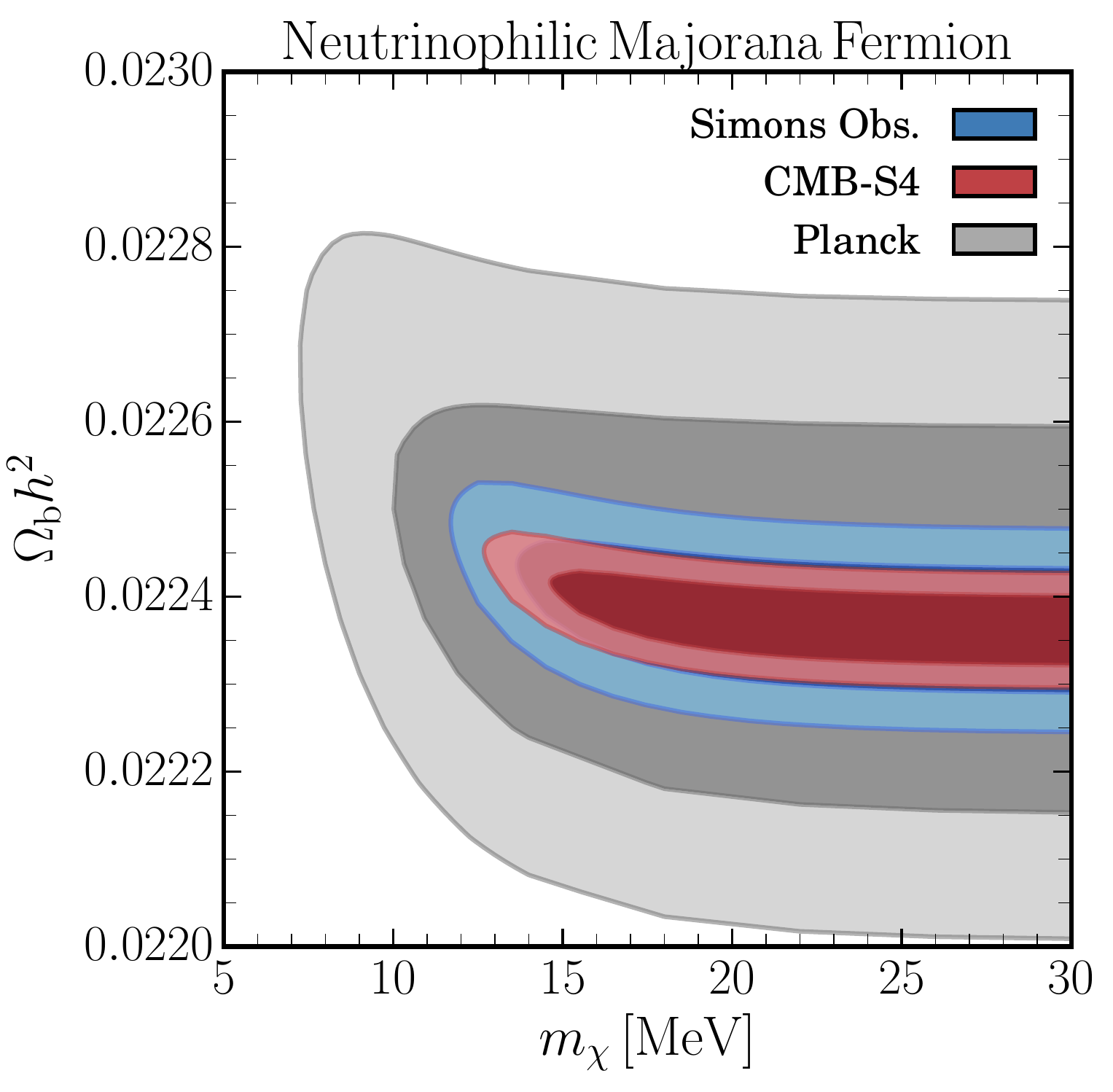} \qquad
    \includegraphics[width=0.45\textwidth,height=200pt]{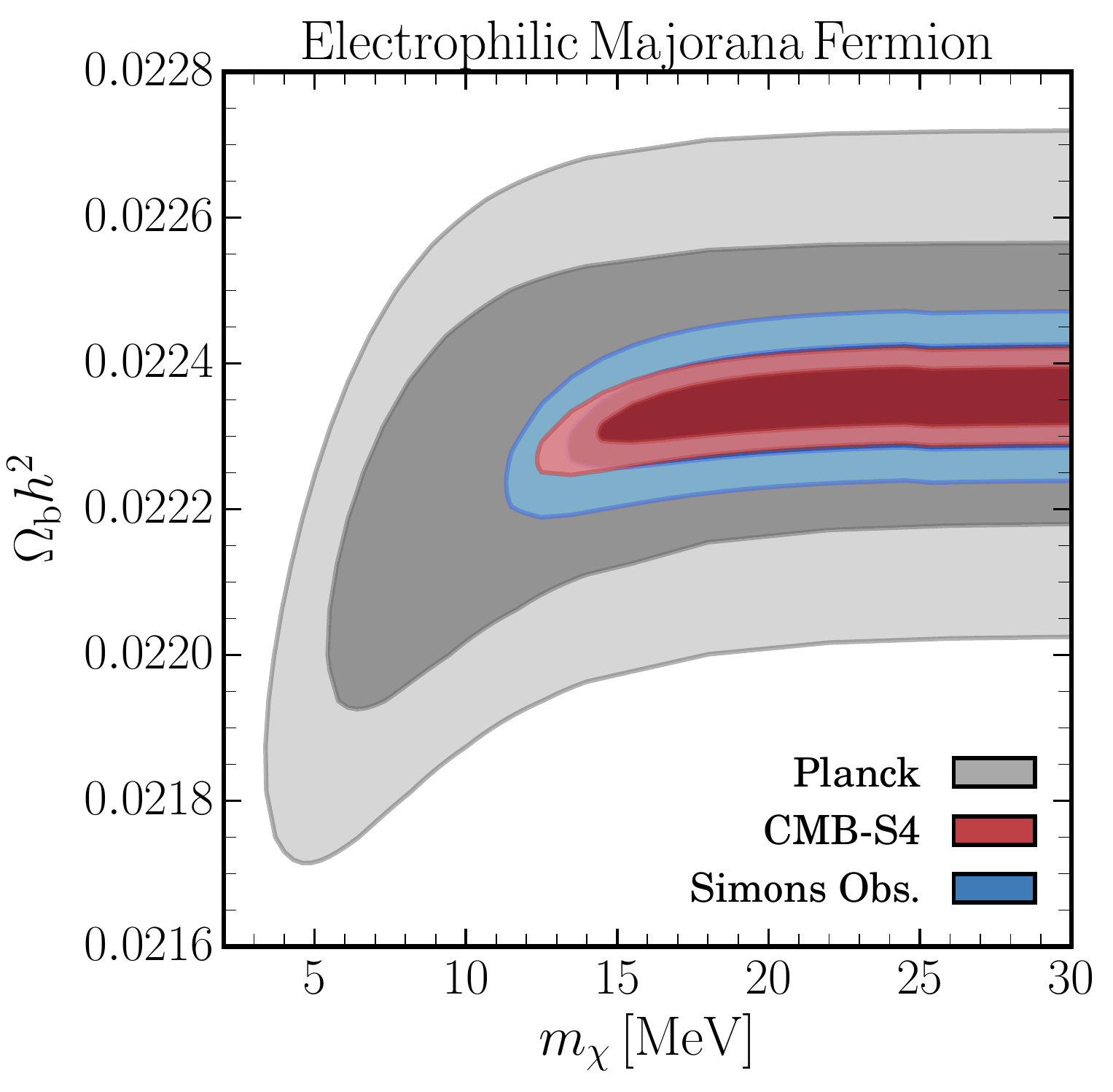}
    \caption{Contour plots showing the current and forecasted $1\sigma$ and $2\sigma$ confidence intervals in the $(\Omega_\mathrm{b} h^2$,\,$m_\chi)$ plane for a BSM Majorana fermion in thermal equilibrium with the SM plasma during BBN. The grey contours correspond to current constraints by Planck 2018, while the blue and red contours correspond to the expected reach of the Simons Observatory and CMB-S4 experiments respectively. \emph{Left}: Neutrinophilic. \emph{Right}: Electrophilic.}
    \label{fig:Future_1-2D}
\end{figure*}

Figure~\ref{fig:Future_1-2D} shows the reach of future CMB observations to $m_\chi$ and $\Omega_{\rm b}h^2$ for a Majorana fermion. From Table \ref{tab:analysis_summary}, we notice that future CMB experiments will be able to probe substantially heavier BSM states than current CMB observations. In particular, the Simons Observatory is expected to set a lower bound on the mass of light BSM species of $m_\chi > 9.4$ MeV, while CMB-S4 is expected to extend this bound to $m_\chi > 10.5$ MeV, both at 95.4\% CL.
 
\subsection{Big Bang Nucleosynthesis}\label{sec:future_BBN}

Forecasting the reach of future determinations of the light primordial element abundances is not straightforward. However, from an observational perspective, the precision with which the primordial deuterium abundance is measured, is expected to improve by an order of magnitude with upcoming 30 m telescope facilities~\cite{Cooke:2016rky,Grohs:2019cae}. From a theoretical perspective, the nuclear reaction rates that significantly contribute to the error budget in the theoretical prediction of ${\rm D/H}|_{\rm P}$ are expected to be measured with higher accuracy by the LUNA collaboration~\cite{Trezzi:2018qjs}. It is therefore feasible that, in the near future, a per mille determination of ${\rm D/H}|_{\rm P}$ could be achieved. Regarding $Y_{\rm P}$, while the situation is much less clear, it is still conceivable that $Y_{\rm P}$ could be narrowed down with greater than $1\,\%$ accuracy in the future~\cite{Grohs:2019cae}.

In order to account for many possible future scenarios, and in a similar spirit to \cite{Lague:2019yvs}, we estimate the reach of future measurements of $Y_{\rm P}$ and ${\rm D/H}|_{\rm P}$ to the mass of thermal BSM state by assuming that the measured values of $Y_{\rm P}$ and ${\rm D/H}|_{\rm P}$ correspond to the values as predicted by \texttt{PRIMAT} using $\Omega_\mathrm{b} h^2 = 0.02236$ and $\tau_n = 879.5$ s -- namely, $Y_{\rm P} = 0.2472$ and ${\rm D/H}|_{\rm P} = 2.439 \times 10^{-5}$ -- and by varying the joint theoretical + observational accuracy with which they are determined.

\begin{figure*}[t]
    \centering
    \includegraphics[width=0.45\textwidth]{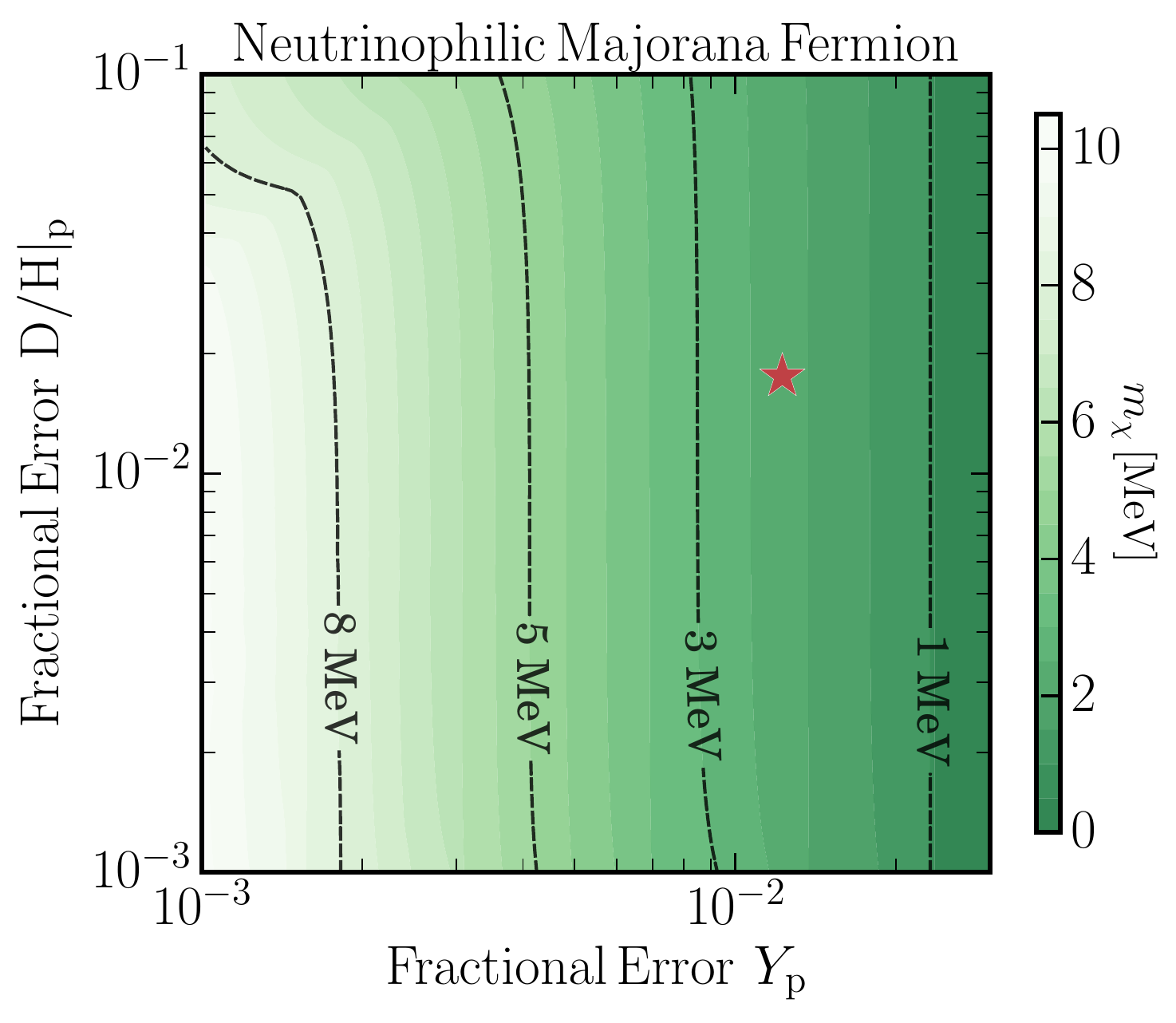} \qquad
    \includegraphics[width=0.465\textwidth]{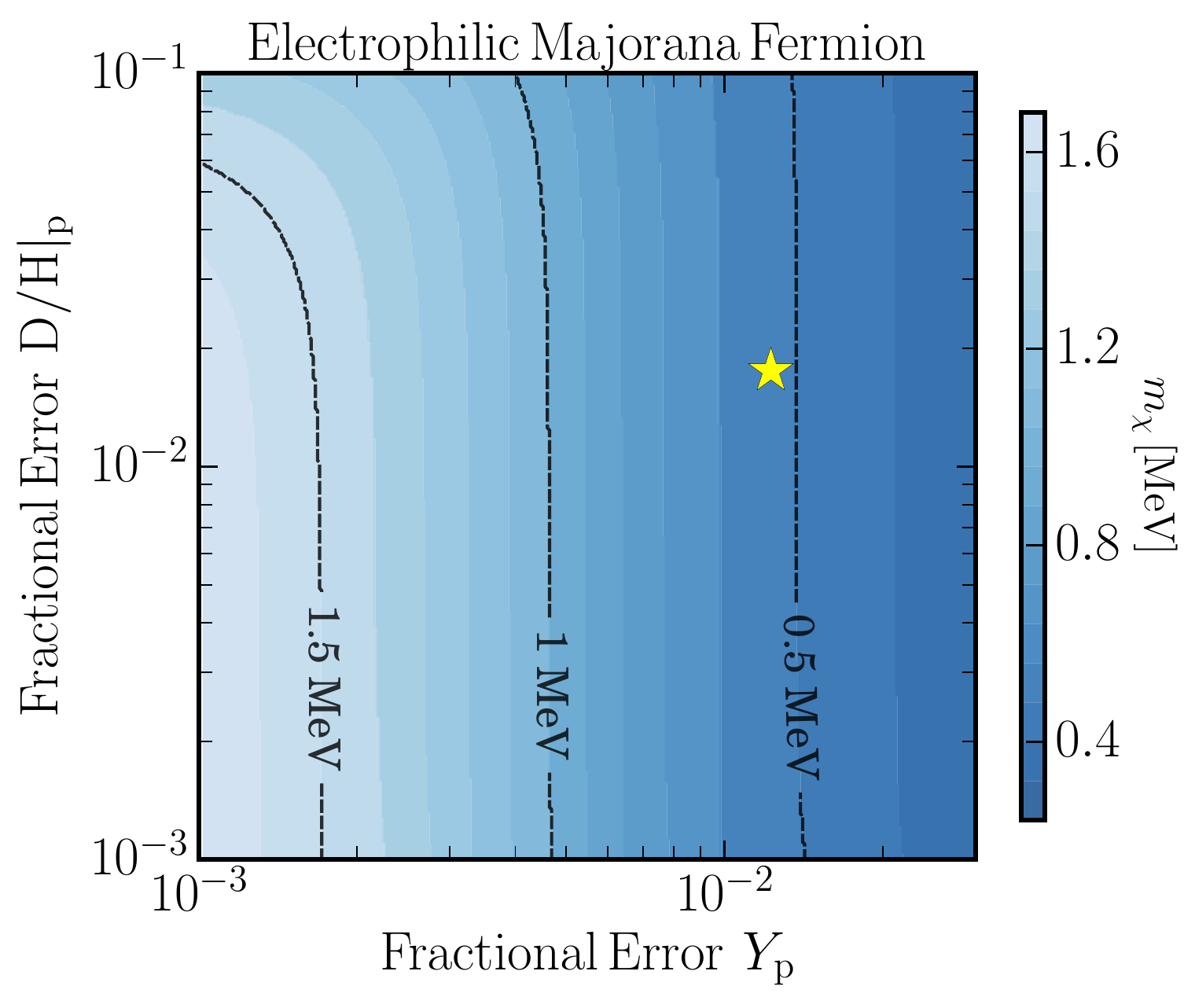} 
    \caption{Projected $2\sigma$ exclusion limits for the mass of a Majorana BSM particle in thermal equilibrium during BBN. Bounds are shown as a function of the fractional errors (joint theoretical and observational) in the primordial helium and deuterium abundances. These bounds are based on a BBN-only analysis. The red and yellow stars correspond to the current precision. \emph{Left}: Neutrinophilic. \emph{Right}: Electrophilic.}
    \label{fig:futureBBN}
\end{figure*}

In Figure~\ref{fig:futureBBN}, we show the forecasted $2\sigma$ lower bounds on the mass of a Majorana fermion in thermal equilibrium with the SM plasma as a function of the fractional error in $Y_{\rm P}$ and ${\rm D/H}|_{\rm P}$. It is clear that the bounds are largely driven by helium measurements, while ${\rm D/H}|_{\rm P}$ measurements are instead expected to provide accurate determinations of $\Omega_\mathrm{b}h^2$. As such, if a prior for $\Omega_\mathrm{b}h^2$ is provided from CMB observations, then ${\rm D/H}|_{\rm P}$ measurements do play an important role in constraining light BSM species in thermal equilibrium with the SM plasma. 
 
\section{Discussion}\label{sec:discussion}
In this section we comment on some of the implications of the constraints on thermal BSM species derived in this work. In particular, we discuss examples of theoretical scenarios in which these bounds are relevant, how the bounds are altered when additional BSM states are present and the robustness of our constraints with respect to non-standard expansion histories of the Universe. Finally, we provide a brief comparison with recent literature. 
\subsection{Particle Physics Scenarios} 

Our constraints on thermally coupled BSM species apply to various particle physics models, typically within the context of thermal dark matter. The bounds outlined in Table~\ref{tab:DMbounds} apply to both s-wave and p-wave annihilating thermal relics. Such bounds are particularly relevant for s-wave and p-wave dark matter particles annihilating to neutrinos~\cite{Boehm:2006mi,Farzan:2009ji,Farzan:2011ck,Batell:2017cmf,Ballett:2019cqp}, since they are difficult to test at neutrino experiments \cite{Kamada:2015era,Arguelles:2017atb,Alvey:2019jzx,Klop:2018ltd,Kelly:2019wow}. In addition, these bounds will  be relevant for p-wave annihilating relics to electrons and positrons, as in the case of Majorana/Dirac dark matter annihilating via dark photon/Higgs exchange \cite{Krnjaic:2015mbs,Alexander:2016aln,Battaglieri:2017aum,Beacham:2019nyx}. Furthermore, the bounds apply to species that need not to be the entirety of the dark matter, see e.g.~\cite{Berlin:2018sjs}, and have also been applied to scenarios in which dark matter particles interact with  quarks~\cite{Krnjaic:2019dzc}. 

The bounds derived for BSM species that annihilate into electrons and neutrinos with different ratios are, for instance, relevant for scenarios involving a gauging of SM global symmetries such as $U(1)_{L_\mu-L_\tau}$ \cite{Arcadi:2018tly,Kamada:2018zxi,Foldenauer:2018zrz} or $U(1)_{B-L}$ \cite{Okada:2010wd,Escudero:2018fwn}. Finally, these bounds also apply to asymmetric dark matter sectors interacting with SM species~\cite{Zurek:2013wia,Petraki:2013wwa}. Asymmetric dark matter set-ups require annihilation cross sections that are larger than for WIMPs, and as such thermal equilibrium in the early Universe is realised. Note, however, that these bounds do not apply to scenarios in which the given BSM species is never brought into thermal equilibrium, as in the case of freeze-in \cite{Hall:2009bx,Dvorkin:2019zdi}, or simply for significantly smaller couplings than those outlined in Equations \eqref{eq:Annihilation} and \eqref{eq:Decay}~\cite{Berlin:2017ftj}. For slightly smaller couplings than those in Equations \eqref{eq:Annihilation} and \eqref{eq:Decay}, BBN can still serve as a useful probe~\cite{Berlin:2019pbq}.

The bounds presented in this study do not only apply to dark matter particles, but also to unstable mediators. For example, the bounds constrain relevant parameter spaces for various neutrinophilic scalars and vector bosons, regardless of whether they are related to dark matter \cite{Blennow:2019fhy,Kelly:2019wow} or not \cite{Blinov:2019gcj}. Similarly, light dark Higgses or dark photons are also constrained. The constraints are particularly relevant for dark photons that decay into hidden sector species. Specifically, the bounds rule out MeV-scale dark photons that decay invisibly for kinetic mixing parameters in the range $10^{-7} \lesssim \epsilon \lesssim 10^{-5}$. This region of parameter space is mildly constrained from colliders, beam dump experiments and supernova cooling~\cite{Ilten:2018crw,Bauer:2018onh,Chang:2018rso}. 

\subsection{Modified Cosmological Histories} 

The bounds derived above were obtained assuming that only one particle alters the usual SM picture of a radiation dominated Universe between neutrino decoupling and recombination. Here we comment on how we expect the constraints to be altered when additional BSM species are present or non-standard thermal histories are considered.  

Typically, dark matter particles are accompanied with mediators of similar mass. The presence of two (or more) neutrinophilic/electrophilic particles in thermal equilibrium in the early Universe would result in stronger bounds on the individual masses of the particles as compared to those outlined in Table \ref{tab:DMbounds}. Another very plausible contribution to the energy density in the Universe during BBN and recombination is massless dark radiation adding to $\Delta N_{\rm eff}$. Massless dark radiation will contribute to the expansion rate of the Universe and hence lead to an enhancement of $Y_{\rm P}$ with respect to the SM prediction \cite{Sarkar:1995dd,Iocco:2008va,Pospelov:2010hj}. This is precisely the same effect as the light BSM particles we consider (see middle panels of Figure \ref{fig:Cosmoimply}) and hence the BBN bounds should simply strengthen in such a scenario. On the other hand, in the presence of non-interacting, free-streaming dark radiation the CMB constraints will be relaxed in the case of electrophilic particles \cite{Steigman:2013yua} but strengthen for neutrinophilic species. Similarly, perhaps a more exotic non-negligible primordial leptonic asymmetry could be present and modify nucleosynthesis and $N_{\rm eff}$ as relevant for CMB observations \cite{Dolgov:2002wy}. In such a scenario, we expect the bounds presented here to be modified \cite{Berezhiani:2012ru} but not substantially given the accuracy with which $N_{\rm eff}$ and the helium and deuterium abundances have now been measured.

One of the key assumptions to derive the bounds in this study was that the particles we consider must have been in thermal equilibrium. Since we know from both BBN and CMB observations that the Universe should have at least reached a temperature of $T > 1.8\,\text{MeV}$ \cite{deSalas:2015glj,Hasegawa:2019jsa}, the particles we consider will indeed have reached thermal equilibrium. 

Another assumption in order to derive these bounds is that the baryon-to-photon ratio remains constant between the end of BBN and recombination. This is well justified on the basis that late time electromagnetic energy injections are strongly constrained by BBN \cite{Kawasaki:2017bqm,Hufnagel:2018bjp,Forestell:2018txr} and CMB spectral distortions \cite{Hu:1992dc,Hu:1993gc}.

\subsection{Comparison with Previous Literature} 

The cosmological implications of MeV-scale thermal dark matter particles were highlighted a while ago in~\cite{Kolb:1986nf}. Since then, a number of groups~\cite{Kolb:1986nf,Serpico:2004nm,Boehm:2013jpa,Nollett:2013pwa,Nollett:2014lwa,Boehm:2012gr,Ho:2012ug,Wilkinson:2016gsy,Depta:2019lbe,Escudero:2018mvt} have used BBN and/or CMB observations to set constraints on the masses and properties of various thermally coupled species in the early Universe. 

One of the main differences between previous studies and the one presented here is the accuracy with which the primordial element abundances have been calculated. In particular, we account for non-instantaneous neutrino decoupling in the presence of light BSM particles~\cite{Escudero:2018mvt,Escudero:2019new} and we use the state-of-the-art BBN code \texttt{PRIMAT}~\cite{Pitrou:2018cgg}. With respect to previous CMB analyses, we find very similar results to those presented in~\cite{Escudero:2018mvt} that accounted for the same effects and used Planck 2018 data. Regarding BBN constraints, we can differentiate between two types of studies: some that fixed the baryon-to-photon ratio to be the best-fit from CMB observations at the time \cite{Serpico:2004nm,Boehm:2013jpa,Boehm:2012gr,Depta:2019lbe}, while others allowed $\Omega_{\rm b}h^2$ to vary and then fitted it to measurements of $Y_{\rm P}$ and ${\rm D/H}|_{\rm P}$ simultaneously with $m_\chi$~\cite{Nollett:2013pwa,Nollett:2014lwa,Wilkinson:2016gsy}. In this work, we marginalize over all possible values of $\Omega_{\rm b}h^2$. The comparison with each reference goes as follows: firstly, when comparing with~\cite{Nollett:2013pwa}, we find that the constraints presented in this work on purely electrophilic BSM states are a factor $1.5-2.5$ more stringent. Secondly, a direct comparison with~\cite{Nollett:2014lwa} is not possible since there are no bounds reported from a BBN only analysis. Thirdly, \cite{Wilkinson:2016gsy} did not found a BBN bound for a real neutrinophilic scalar boson at 95.4\% CL while we find $m_\chi > 1.2\,\text{MeV}$ at such CL. We believe that these differences with previous studies are largely driven by the use of more recent and precise determinations of $Y_{\rm P}$ and ${\rm D/H}|_{\rm P}$. 

Finally, our work represents the first exhaustive study of WIMPs that annihilate differently to both electrons and neutrinos. Furthermore, we provide bounds on the temperature at which neutrinos decouple that can be mapped into various relevant particle physics scenarios.

\section{Conclusions}\label{sec:conclusions}

MeV-scale BSM species in thermal equilibrium with the Standard Model plasma during Big Bang Nucleosynthesis have important cosmological consequences, as can be seen from Figure \ref{fig:Cosmoimply}. In this work, we have analyzed in detail and with precision the impact of such states on the synthesis of the primordial element abundances and CMB observations. To this end, we have modelled the early Universe evolution using the methods of \cite{Escudero:2018mvt,Escudero:2019new} and by modifying the state-of-the-art BBN code \texttt{PRIMAT}~\cite{Pitrou:2018cgg}. We have used a suite of cosmological observations, as summarized in Table~\ref{tab:analysis_summary}, to set constraints on the masses of various types of BSM states in thermal equilibrium with the SM plasma during BBN. We summarize the derived constraints in Table~\ref{tab:DMbounds} for purely electrophilic and neutrinophilic BSM states. In Tables~\ref{tab:MajoranaBRBounds} and~\ref{tab:BRBounds} we consider WIMPs that have different annihilation ratios to SM species. Finally, in Table~\ref{tab:Tnudec_bounds} we outline the lower bound on a non-standard neutrino decoupling temperature. The main conclusions that can be drawn from this study are:
\vspace{-0.4cm}
 
\begin{itemize}[leftmargin=0.6cm,itemsep=1pt] 
\item BBN observations set a lower bound on electrophilic/neutrinophilic thermal species of $m_\chi > 0.4/1.2 \,\text{MeV}$ at 95.4\% CL. This bound is independent of the spin and the number of internal degrees of freedom of the species at hand. In particular, any WIMP, irrespective of its annihilation being s-wave or p-wave and the annihilation final state, is bounded to have $m_\chi > 0.4\,\text{MeV}$ at 95.4\% CL.

\item Very light ($m_\chi < 0.1\,\text{MeV}$) thermal relics are highly disfavoured by current measurements of the primordial light elements (at more than $5\sigma$). The sole exception to this rule is a purely neutrinophilic neutral scalar state, which is nonetheless ruled out at $3.3\sigma$. 

\item BBN and CMB observations jointly constrain neutrinophilic and electrophilic thermal BSM states to have a mass $m_\chi > 3.7\,\text{MeV}$ at 95.4\% CL. This bound is independent of the spin or internal degrees of freedom of the given species and applies to both s-wave and p-wave annihilating dark matter relics, as well as to unstable dark sector mediators. Table \ref{tab:DMbounds} summarizes the constraints for various BSM states.

\item We argue that the bounds presented in this study are expected to be strengthened in the presence of additional species beyond those considered here. In addition, bounds based on BBN are largely insensitive to modifications of the assumed cosmological model.

\item We have set constraints on BSM particles with masses $m_\chi \lesssim 20 \,\text{MeV}$ that interact with both electrons and neutrinos. Such states efficiently delay the process of neutrino decoupling, which allows BBN to constrain the temperature of neutrino decoupling to be $T_{\nu} ^{\rm dec} > 0.34\,\text{MeV}$ at 95.4\% CL. Moreover, we find that decoupling temperatures $T_{\nu} ^{\rm dec} < 0.2 \,\text{MeV}$ are highly disfavoured ($> 5\sigma$) by BBN and/or CMB observations.

\item Future CMB experiments such as the Simons Observatory~\cite{Ade:2018sbj} and CMB-S4~\cite{Abazajian:2016yjj,Abazajian:2019eic}, will constrain generic thermal BSM particles of $m_\chi \lesssim (10-15)\,\text{MeV}$. Similarly, we highlighted the impact of future primordial helium and deuterium determinations to light BSM states in thermal equilibrium with the SM plasma during nucleosynthesis. 
\end{itemize}

To summarize, cosmology strongly constrains new physics at the MeV scale. Cosmological constraints are competitive with and complementary to those from  colliders, beam dump and neutrino experiments, (in)direct dark matter searches, as well as from astrophysical probes.

\begin{acknowledgements}
We are very happy to be able to thank Ren\'ee Hlo\v{z}ek, Subir Sarkar and Chris McCabe for valuable comments. We are grateful to Joel Meyers for providing us with the covariance matrix for the Simons Observatory forecast, and Sunny Vagnozzi and Erminia Calabrese for useful correspondence. We acknowledge the use of the public cosmological codes \texttt{PRIMAT}~\cite{Pitrou:2018cgg}, \texttt{CLASS}~\cite{Blas:2011rf,Lesgourgues:2011re} and \texttt{Monte Python}~\cite{Audren:2012wb,Brinckmann:2018cvx}. ME and MF are supported by the European Research Council under the European Union's Horizon 2020 program (ERC Grant Agreement No 648680 DARKHORIZONS). In addition, the work of MF was supported partly by the STFC Grant ST/P000258/1. JA is a recipient of an STFC quota studentship. NS is a recipient of a King's College London NMS Faculty Studentship.
\end{acknowledgements}

\cleardoublepage


\setcounter{secnumdepth}{1}
\bibliography{biblio}

\clearpage
\phantomsection
\section{Appendices}
\label{app:App}
\setcounter{secnumdepth}{2}
\renewcommand{\thesubsection}{\Alph{subsection}}
\setcounter{subsection}{0}

\subsection{Consistency Checks of Modified BBN Code}\label{app:ConsistencyChecks}\vspace{-0.2cm}
We checked whether our modifications do not significantly change the values of the primordial helium and deuterium abundances in Standard Model BBN compared to the base version of \texttt{PRIMAT}. Table \ref{tab:SBBNcheck} shows the relative difference in the output of the two codes and it is clear that the accuracy is better than $0.1\%$.
\begin{table}[h]
    \centering
    {\def\arraystretch{1.35}
    \begin{tabular}{c|c|c|c}
        \toprule
      $\,\,$   \textbf{Abundances} $\,\,$ & $\,\,$ \texttt{PRIMAT}$\,\,$ & $\,\,$ \textbf{Modified} \texttt{PRIMAT} $\,\,$ & $\,\,$ \textbf{Relative Difference (\%)} $\,\,$ \\
        \hline\hline
        $Y_\mathrm{p}$ & 0.24709 & 0.24717 & 0.03\\
        $10^5\times {\rm D/H}|_{\rm P}$ & 2.4592 & 2.4613 & 0.08 \\ \hline  \hline
    \end{tabular}}\vspace{-0.1cm}
    \caption{Primordial abundances as computed using \texttt{PRIMAT} and our modified version with $\Omega_\mathrm{b} h^2 = 0.02225$ and $\tau_n = 879.5$ s.}
    \label{tab:SBBNcheck}
\end{table} 

\begin{figure}[t]
    \centering
    \includegraphics[width=0.45\textwidth]{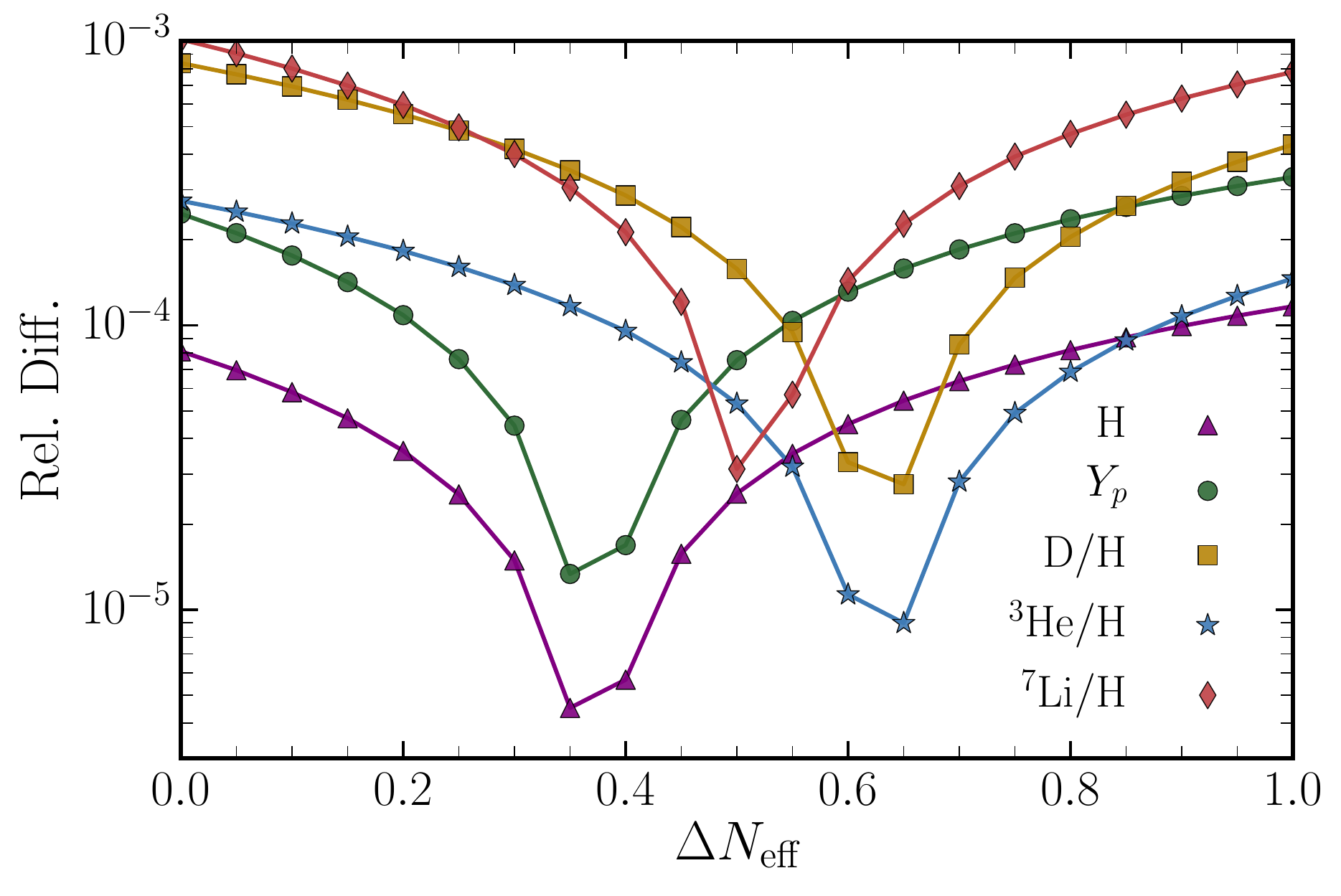}\vspace{-0.5cm}
    \caption{Relative difference in the primordial abundances between the default version of \texttt{PRIMAT} and our modified version of it as a function of $\Delta N_{\mathrm{eff}}$. Predictions are done using $\Omega_\mathrm{b} h^2 = 0.02225$ and $\tau_n = 879.5$ s.}
    \label{fig:CheckDeltaNeff}
\end{figure}

We also compared our modifications to \texttt{PRIMAT} when massless dark radiation is present, which we parametrize in terms of $\Delta N_{\rm eff}$. In \texttt{PRIMAT} this is done by increasing $N_\mathrm{eff}$ directly in the Friedmann equations while in our modified version of the code it is done by including the evolution of a non-interacting, relativistic component. The result is shown in Figure \ref{fig:CheckDeltaNeff}.
The test shows an accuracy better than $0.1\%$ for all relevant nuclides in the range $0 \leq \Delta N_{\rm eff} \leq 1$.

\subsection{Comparison with Previous Literature}
\begin{figure*}[t]
    \begin{center}
    \begin{tabular}{cc}
     \hspace{-0.5cm} \includegraphics[width=0.46\textwidth]{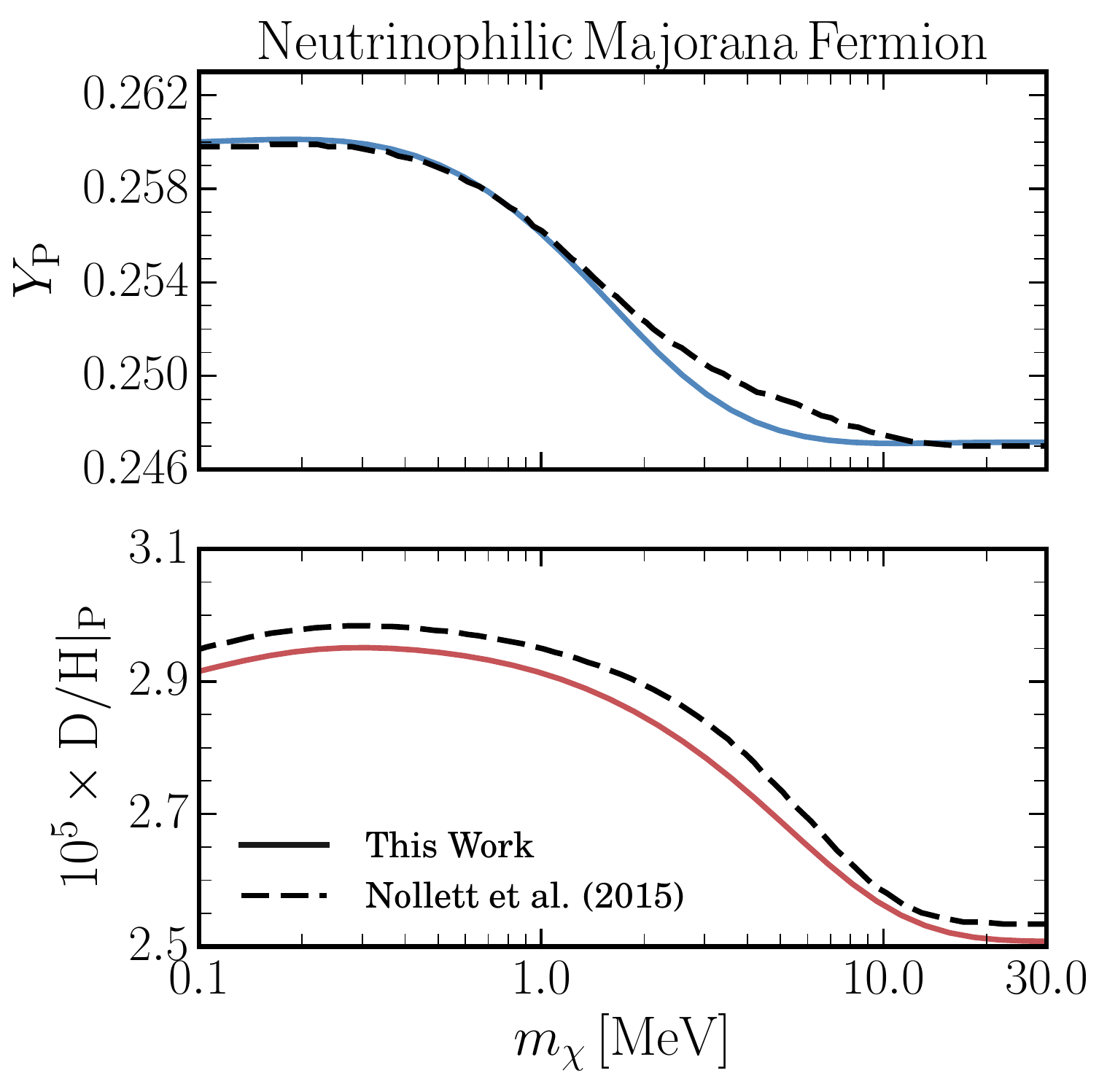}  \hspace{0.4cm} \includegraphics[width=0.45\textwidth]{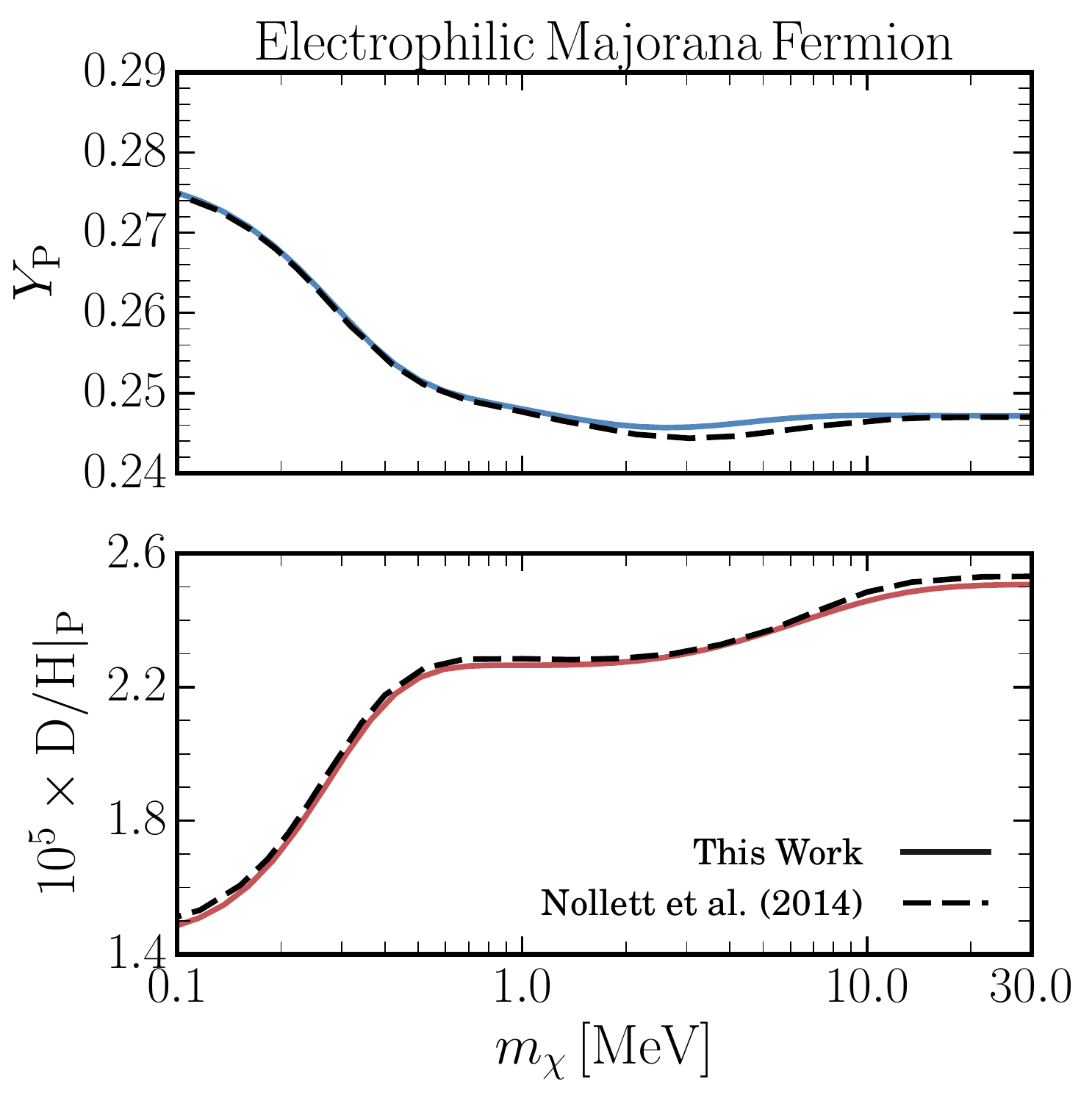}
      \end{tabular}
      \end{center}\vspace{-0.8cm}
    \caption{Comparison with previous literature for the primordial abundances $Y_{\mathrm{P}}$ and $\mathrm{D}/\mathrm{H}|_{\mathrm{P}}$ as a function of the mass of a Majorana BSM particle that couples exclusively to neutrinos (left panels) or electrons (right panels). The solid lines are from this work and the dashed lines from ~\cite{Nollett:2013pwa} and~\cite{Nollett:2014lwa}.}
    \label{fig:mccabe}
\end{figure*}

\vspace{-0.2cm} \label{app:ComparisonsLiterature}
In this appendix we make a direct comparison between our results and those reported in~\cite{Nollett:2013pwa} and~\cite{Nollett:2014lwa}. Refs~\cite{Nollett:2013pwa} and~\cite{Nollett:2014lwa} used a modified version of the Kawano code~\cite{Kawano:1988vh,Kawano:1992ua}. In Figure \ref{fig:mccabe}, we consider two cases: a Majorana fermion that is purely neutrinophilic or electrophilic. We compute the predictions for helium and deuterium using $\tau_\mathrm{n} =  880.1$ s and $\Omega_\mathrm{b}h^2 = 0.022 $ as in~\cite{Nollett:2013pwa} and~\cite{Nollett:2014lwa}. We observe a few small differences:
\begin{enumerate}[leftmargin=0.5cm,itemsep=0pt]\vspace{-0.1cm}
    \item Our predicted values of $\mathrm{D}/\mathrm{H}|_{\mathrm{P}}$ are smaller than those reported in~\cite{Nollett:2013pwa,Nollett:2014lwa}. Since the difference is WIMP mass independent, we attribute it to updated nuclear reaction rates in \texttt{PRIMAT}. 
    \item The predicted values of $Y_{\mathrm{P}}$ are slightly different for $ 1 \,\text{MeV} \lesssim   m_{\chi} \lesssim 15 \,\text{MeV}$. The reason is twofold:
    \begin{enumerate}[leftmargin=0.5cm,itemsep=0pt]\vspace{-0.1cm}
    \item \cite{Nollett:2013pwa,Nollett:2014lwa} considered that neutrinos decoupled instantaneously and tracked the temperature evolution by using entropy conservation, while we solve for the time evolution of neutrino decoupling. Imposing entropy conservation leads to a feature in the neutrino temperature evolution that affects both the Universe's expansion and the proton-to-neutron conversion rates, see Figure 2 of \cite{Escudero:2018mvt}.
    \item \cite{Nollett:2013pwa,Nollett:2014lwa} considered instantaneous neutrino decoupling at $T_\nu^{\rm dec} = 2.0\,\text{MeV}$, while an estimate based on the actual neutrino temperature time evolution yields $T_\nu^{\rm dec} = 1.91\,\text{MeV}$ \cite{Escudero:2018mvt}. Considering a smaller neutrino decoupling temperature leads to an impact on the proton-to-neutron rates and also reduces the impact of heavier BSM species in neutrino decoupling. 
    \end{enumerate}
\end{enumerate}\vspace{-0.1cm}
We have also compared our predictions of $Y_{\mathrm{P}}$ and $\mathrm{D}/\mathrm{H}|_{\mathrm{P}}$ with those reported in~\cite{Boehm:2013jpa} (which used  \texttt{PArthENoPEv1} \cite{Pisanti:2007hk}, see \cite{Consiglio:2017pot} for an updated version of the code). We find good overall agreement with~\cite{Boehm:2013jpa} and small differences similar to those we find when comparing to~\cite{Nollett:2013pwa,Nollett:2014lwa}. Note that~\cite{Wilkinson:2016gsy} provided updated bounds to those presented in~\cite{Boehm:2013jpa} although the predictions for $Y_{\mathrm{P}}$ and $\mathrm{D}/\mathrm{H}|_{\mathrm{P}}$ are not displayed in that reference.

\subsection{Conservative Range for the Baryon Density from CMB observations} \label{app:Omegab}

\begin{figure}[h]
    \centering
    \includegraphics[width=0.6\textwidth]{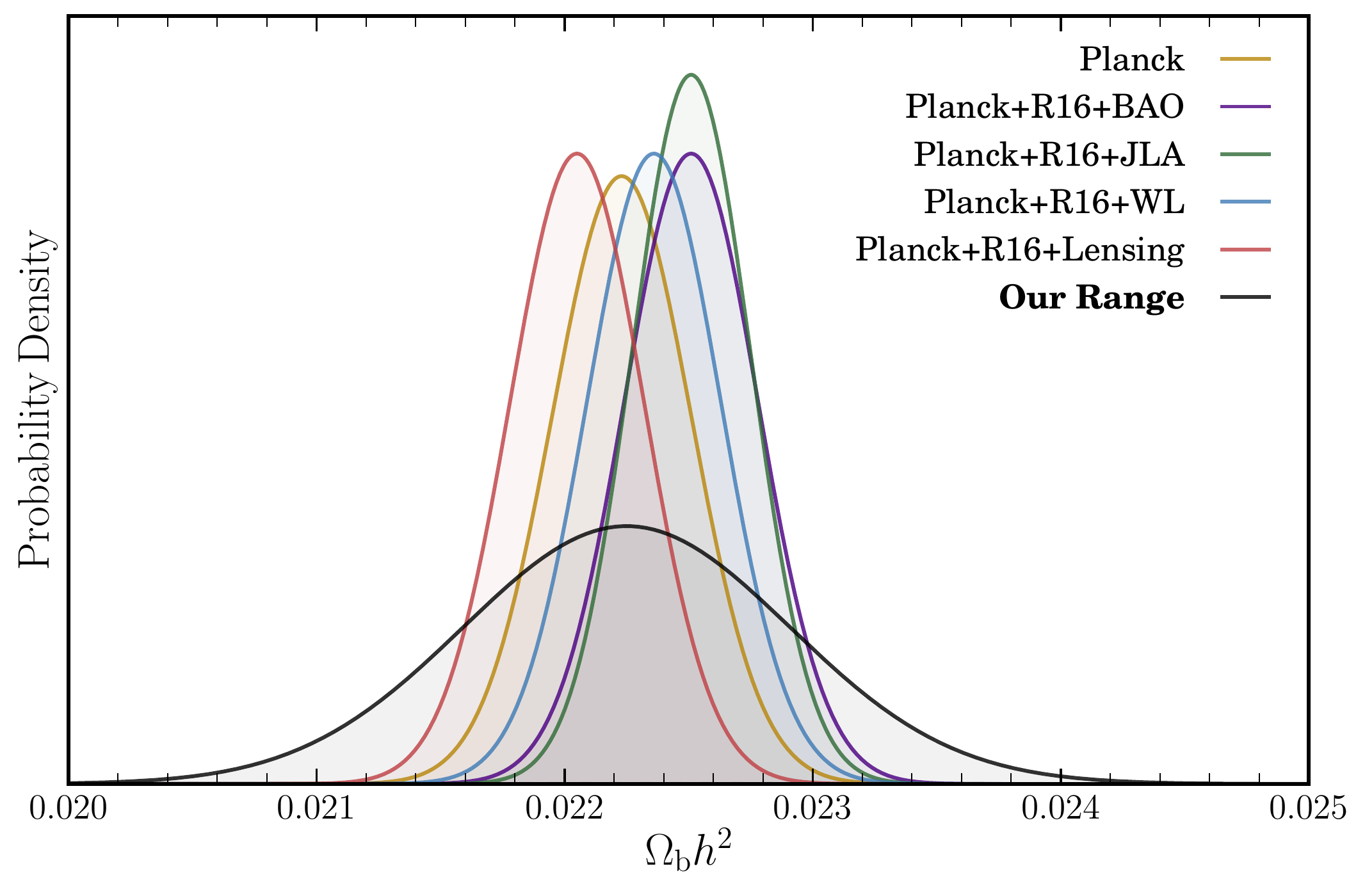}\vspace{-0.3cm}
    \caption{Illustration of the parameter range for the baryon density we consider in the BBN+$\Omega_{\mathrm{b}}h^2$ analysis as compared to the best-fit values and errors given in Table II of \cite{DiValentino:2017zyq}. The authors of \cite{DiValentino:2017zyq} infer $\Omega_{\mathrm{b}}h^2$ in a well-motivated 12-parameter extension to $\Lambda$CDM using the different data sets shown in the legend. Note in particular that our conservative range for the baryon density encompasses all derived central values and errors.}
    \label{fig:omegab}
\end{figure}

In the BBN+$\Omega_{\rm b}h^2$ analysis we consider $\Omega_\mathrm{b} h^2 = 0.02225 \pm 0.00066$ to be a conservative and cosmological model independent determination of the baryon energy density by current CMB observations. $\Omega_\mathrm{b} h^2 = 0.02225 \pm 0.00066$ has a $4.4$ times larger error bar than the one associated with $\Lambda\text{CDM}$ using Planck 2018 observations~\cite{Aghanim:2018eyx}, and furthermore, it covers well the inferred value of $\Omega_\mathrm{b} h^2$ in a well-motivated 12-parameter extensions of $\Lambda$CDM using different data sets~\cite{DiValentino:2016hlg,DiValentino:2017zyq}. In Figure \ref{fig:omegab}, one can appreciate that indeed the range with a central value of $\Omega_\mathrm{b} h^2 = 0.02225 \pm 0.00066$ covers very well the posterior distributions of $\Omega_\mathrm{b} h^2 $ of such a 12-parameter extension of $\Lambda$CDM including various data sets in conjuntion to Planck CMB observations. 

\newpage
\subsection{Constraints for all Scenarios}
\label{app:FullResults}

\begin{figure}[t]
    \centering
    \includegraphics[width=0.44\textwidth]{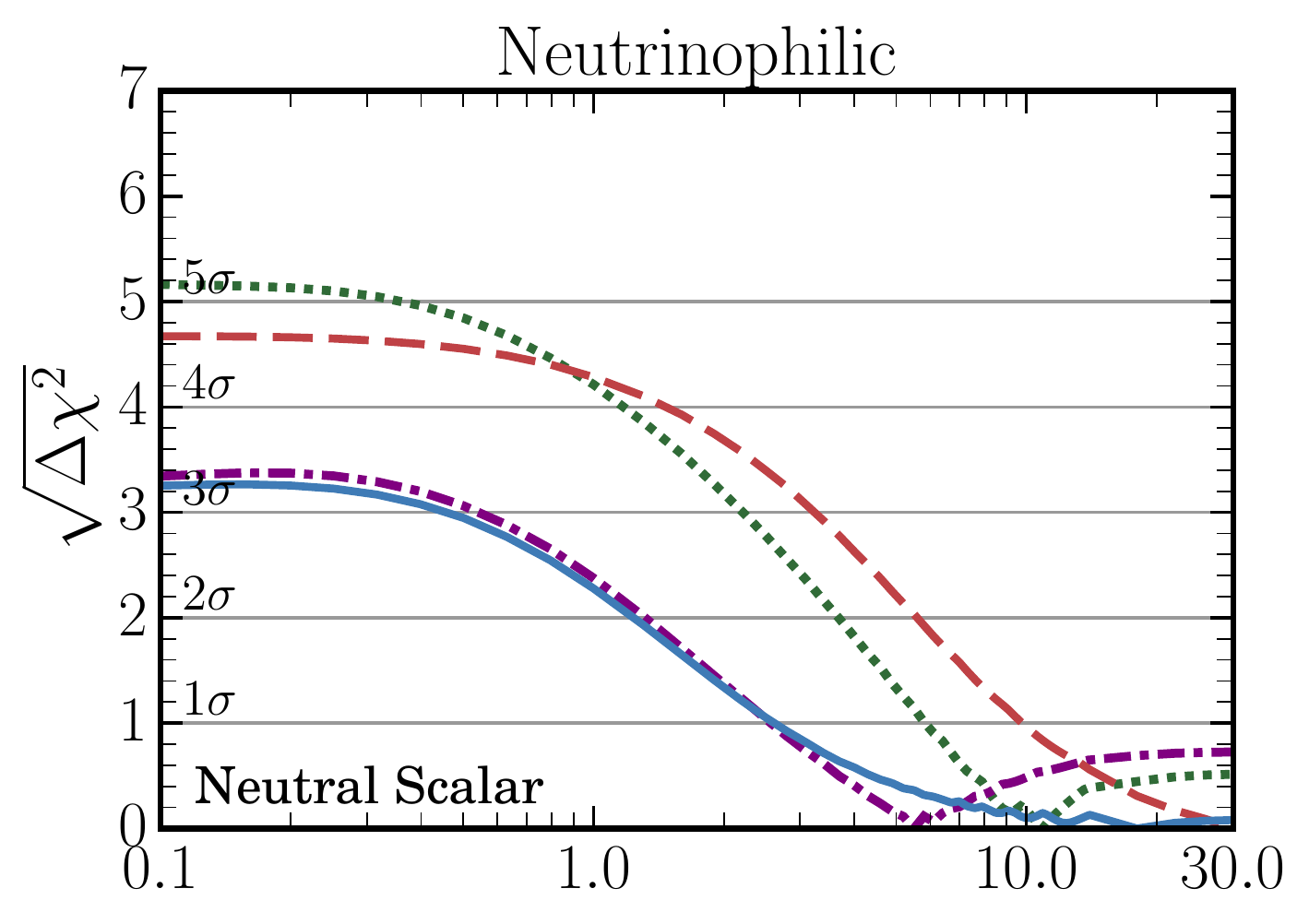} \quad \includegraphics[width=0.44\textwidth]{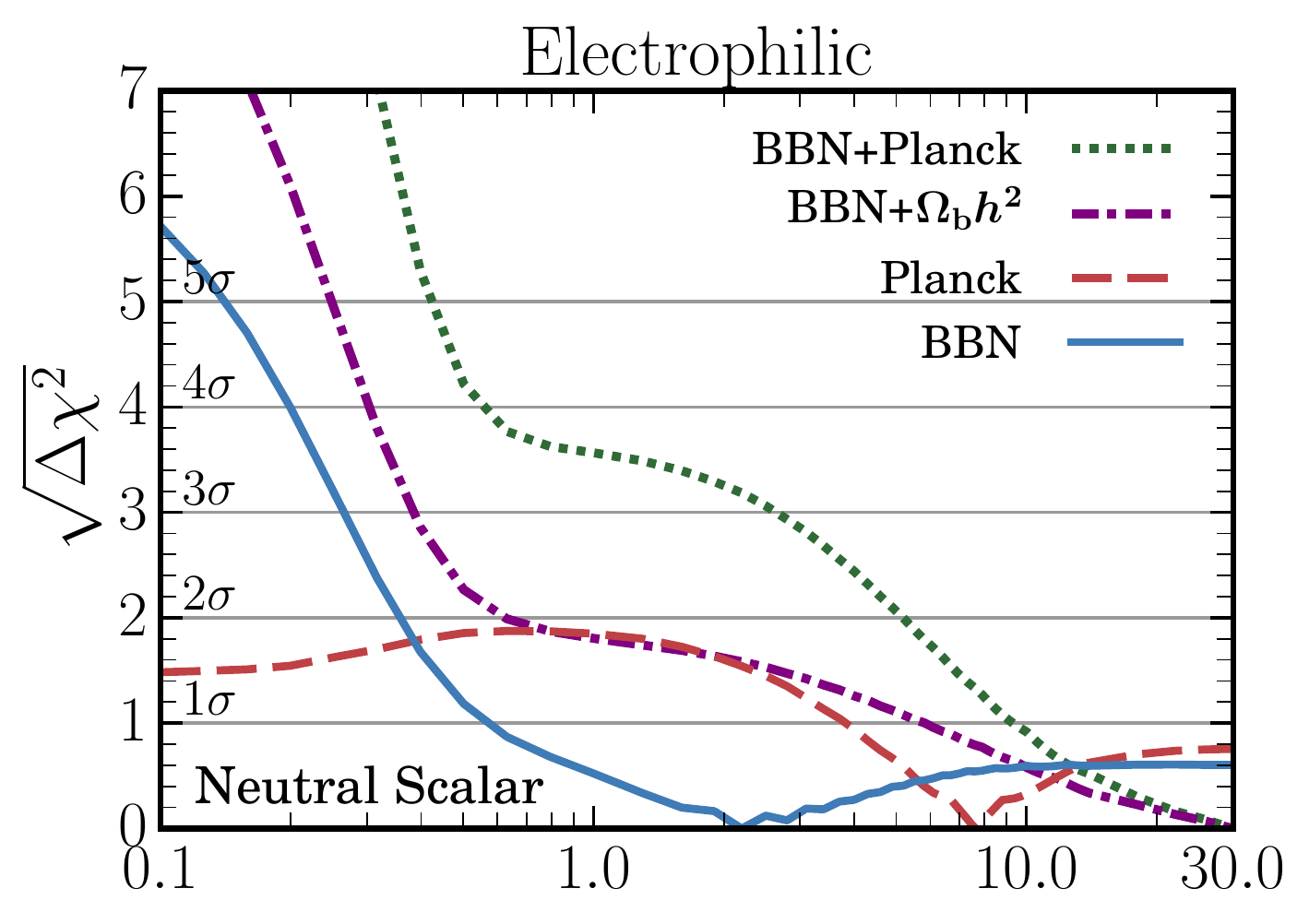} \\  \includegraphics[width=0.44\textwidth]{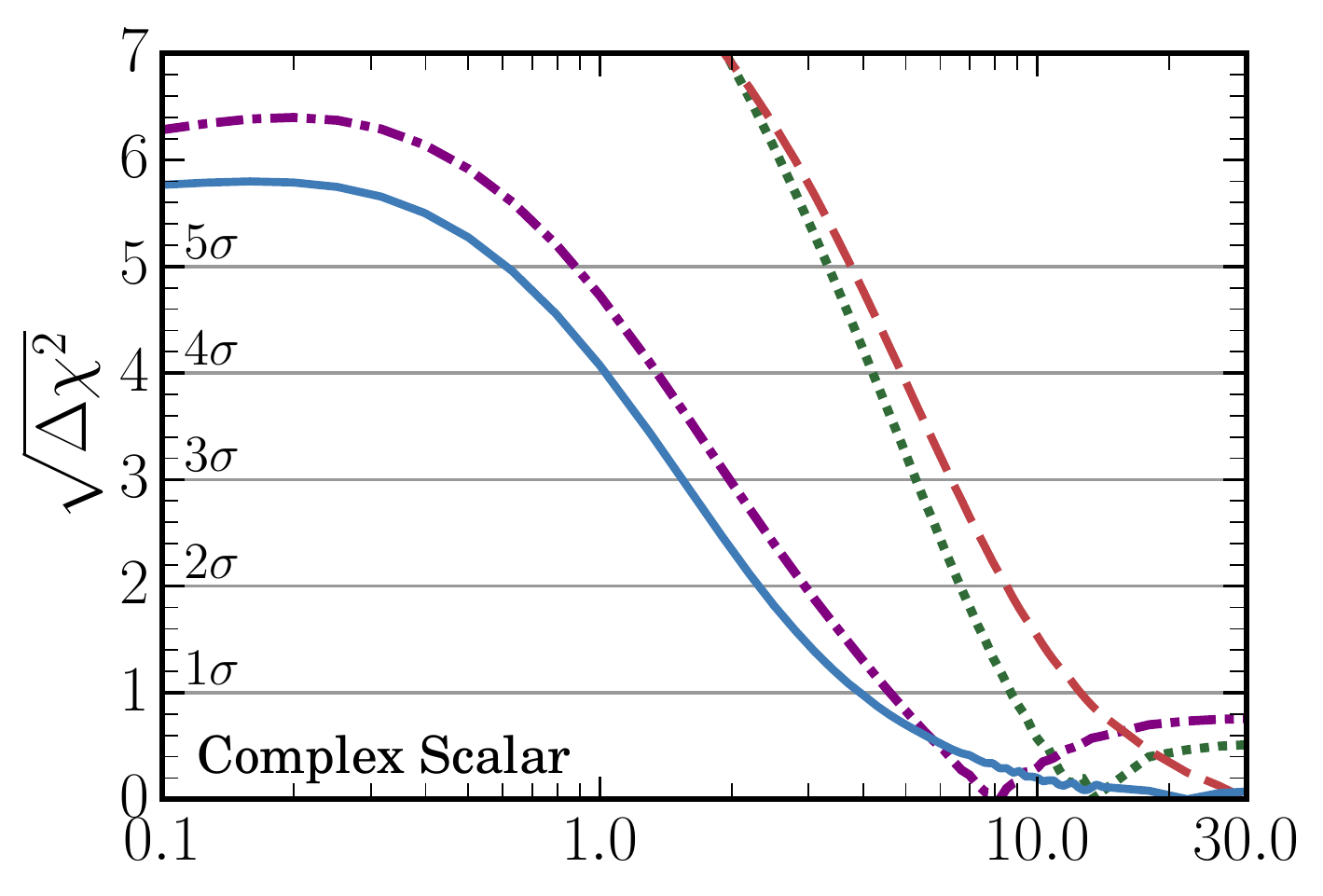} \quad \includegraphics[width=0.44\textwidth]{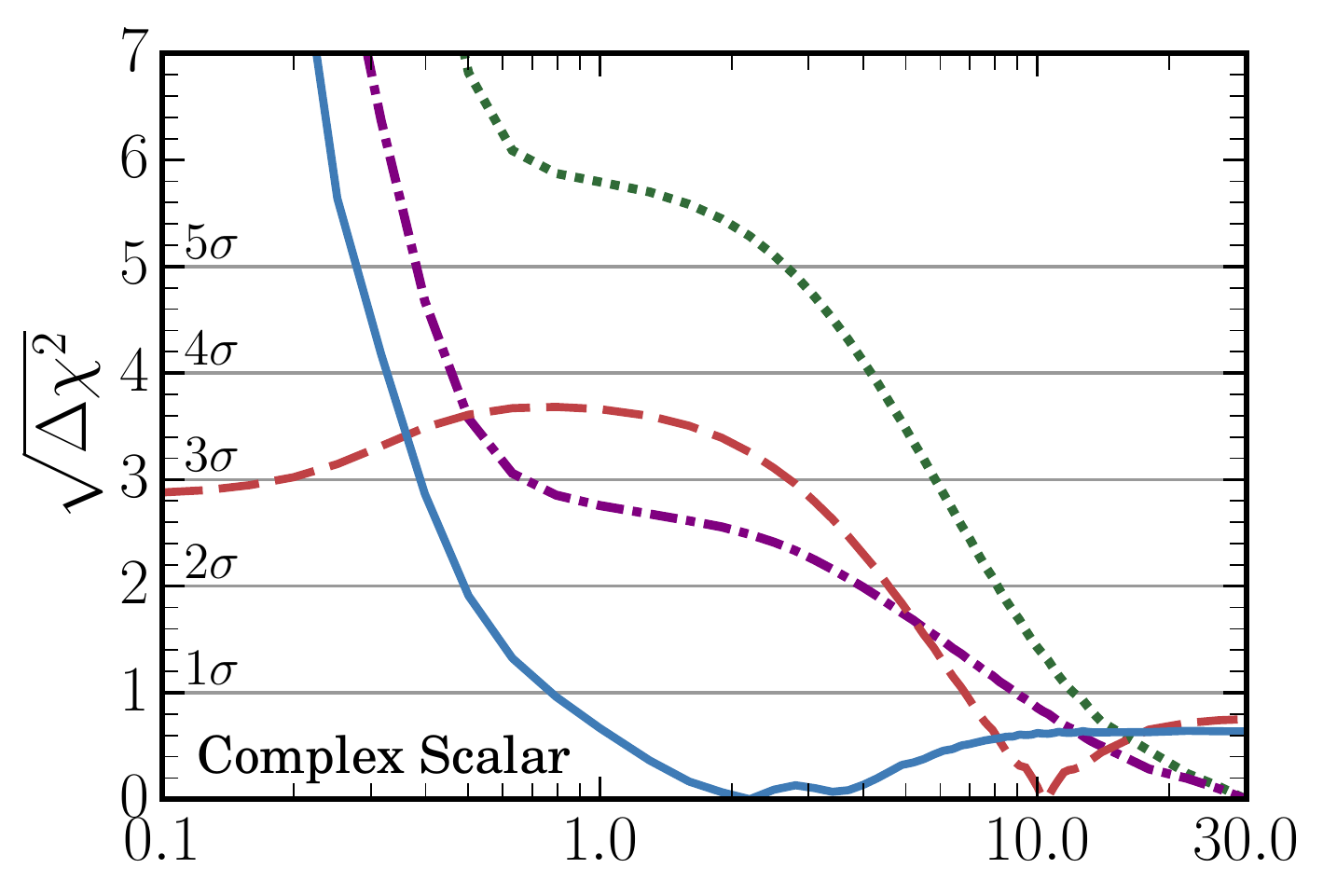} \\ \includegraphics[width=0.44\textwidth]{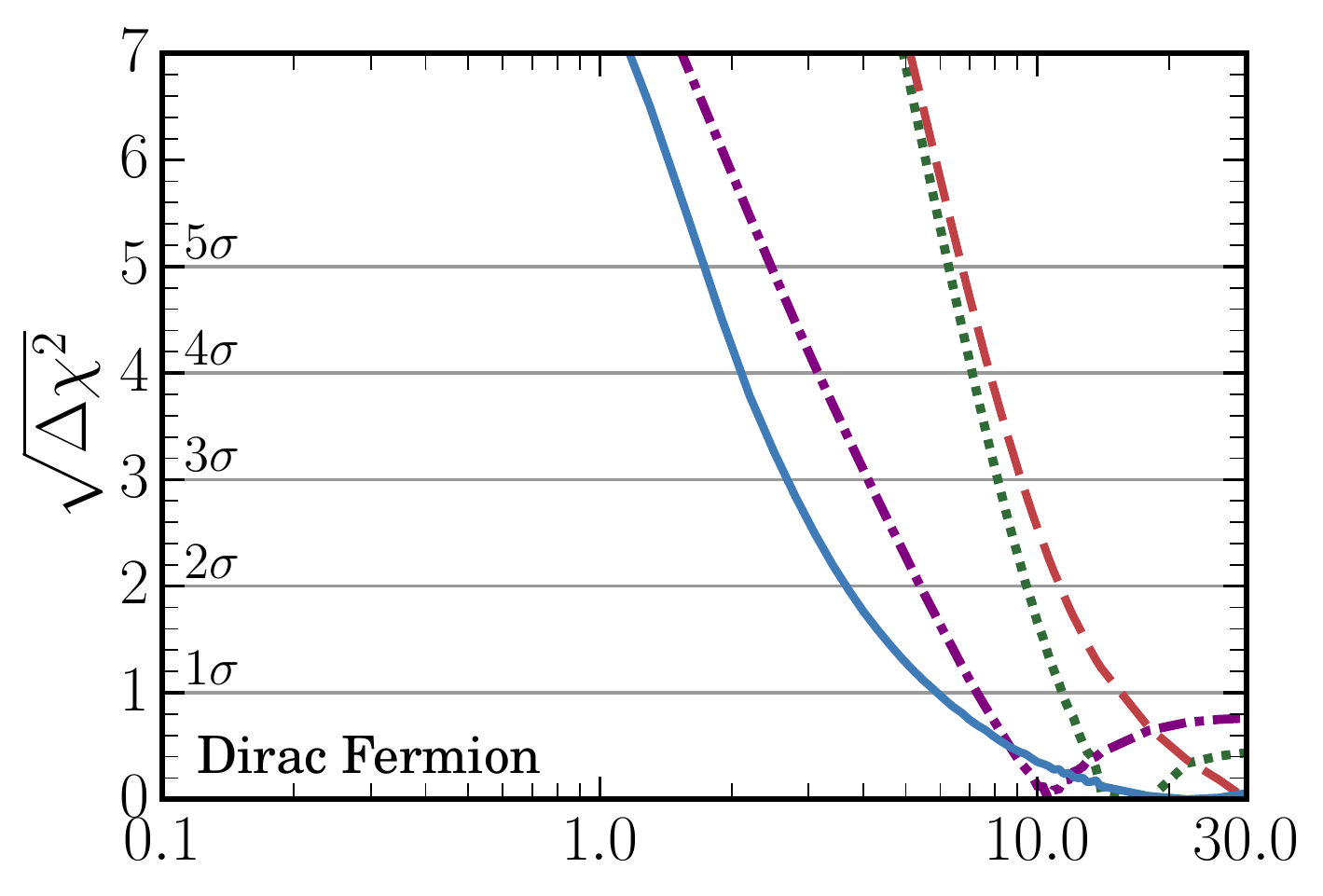} \quad \includegraphics[width=0.44\textwidth]{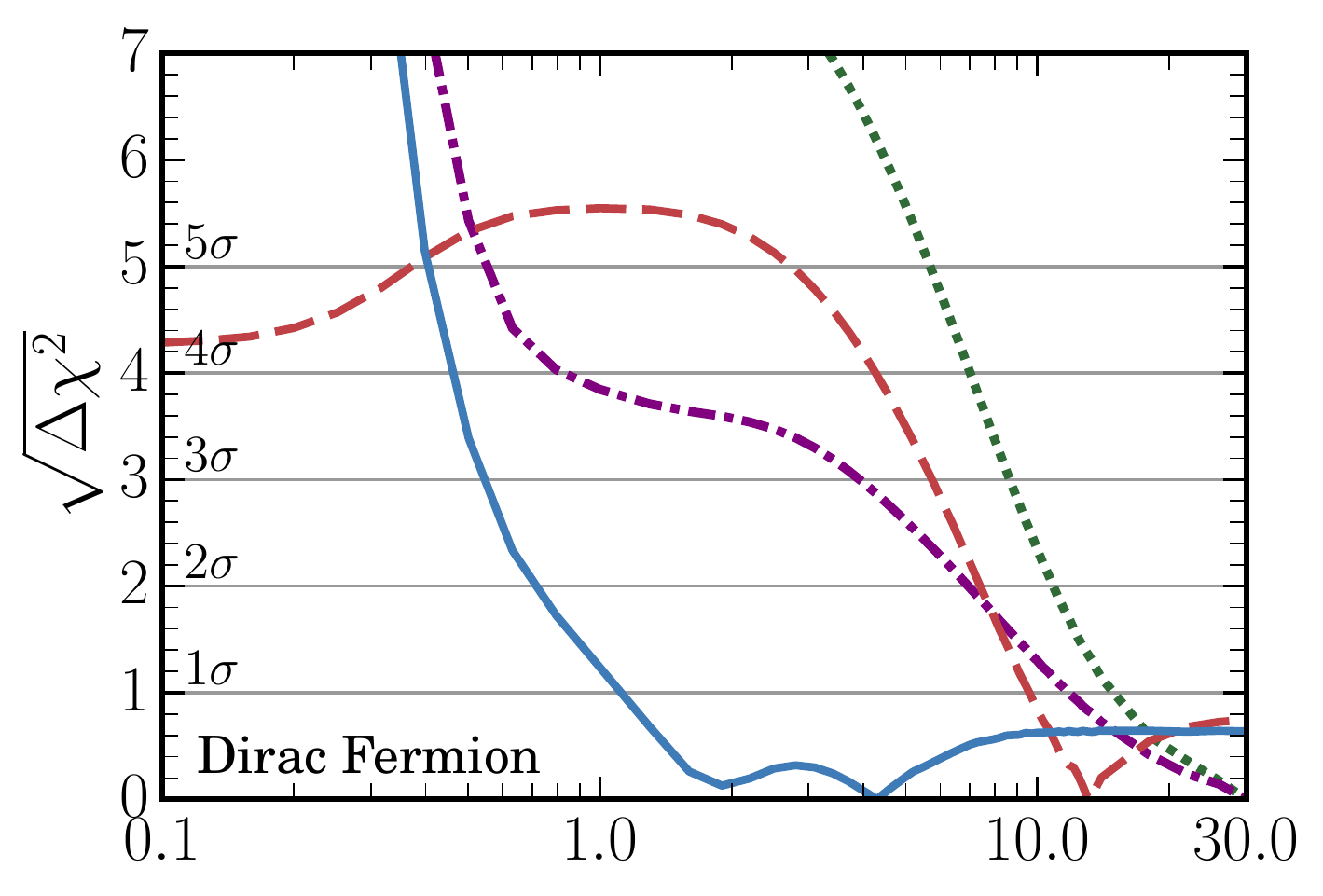} \\ 
    \includegraphics[width=0.44\textwidth]{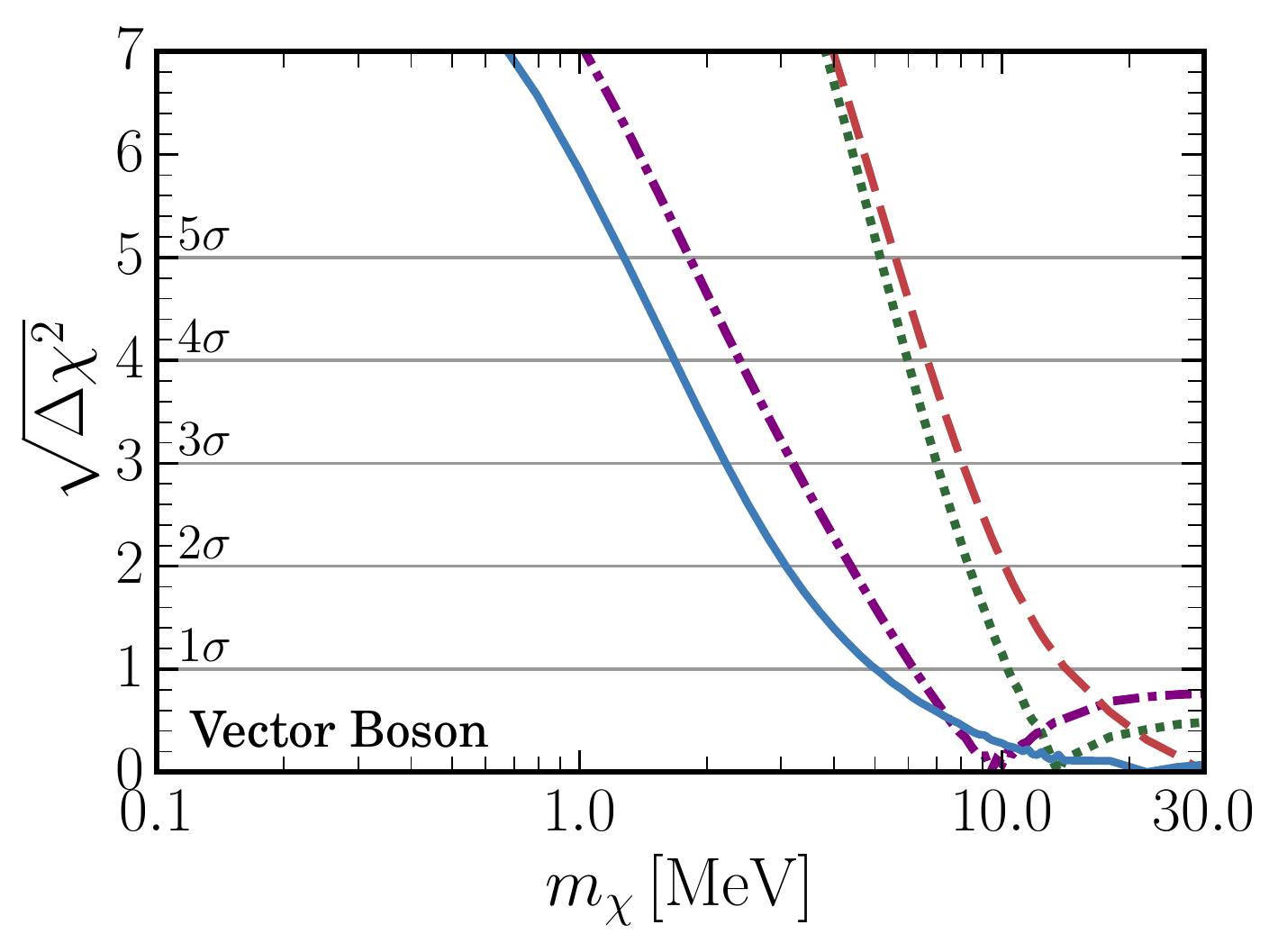} \quad \includegraphics[width=0.44\textwidth]{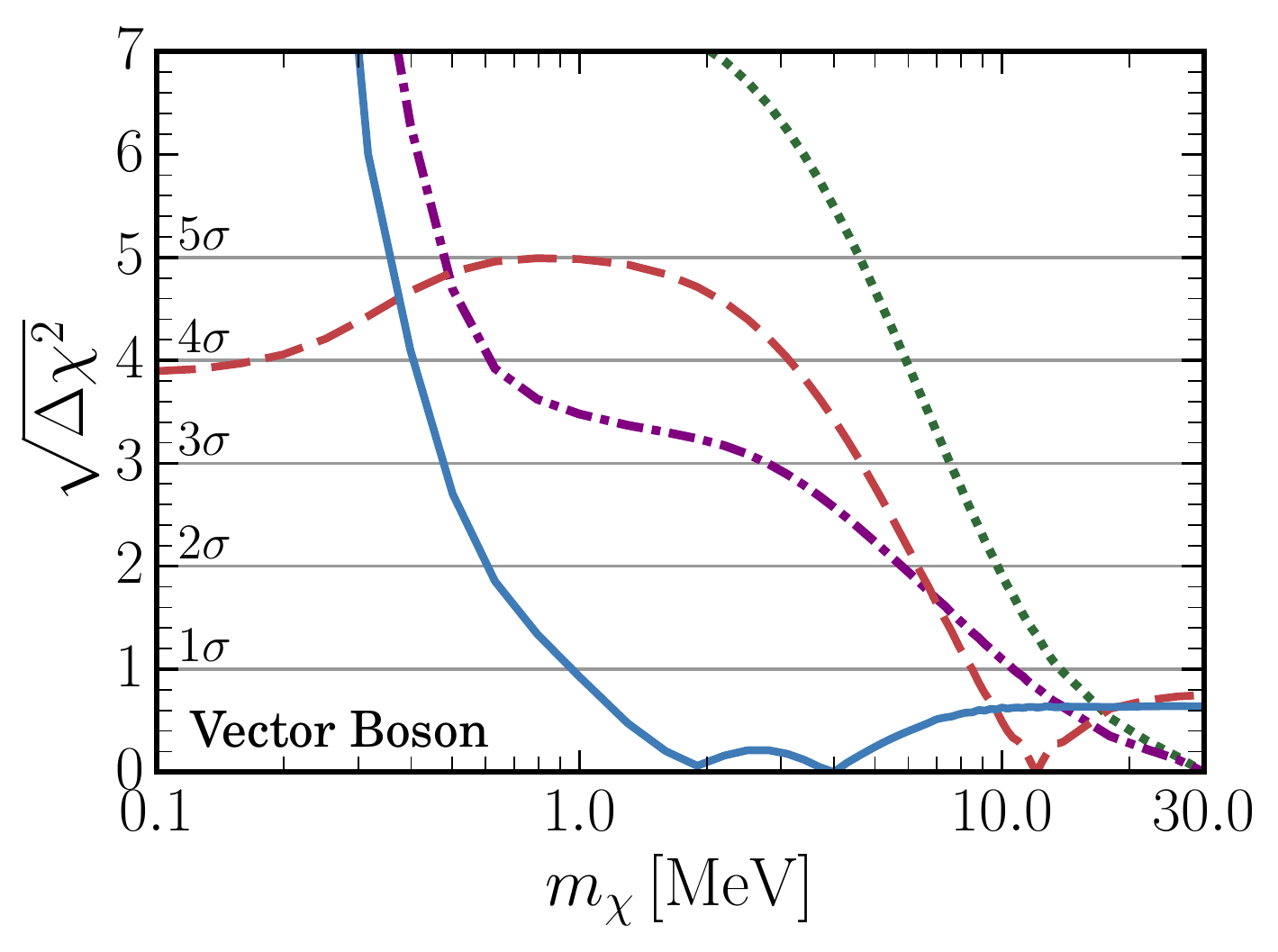}\vspace{-0.3cm}
    \caption{Additional set of marginalized $\Delta \chi^2$ as a function of the BSM particle mass $m_\chi$. Solid lines correspond to the BBN constraints, the dashed to CMB observations, the dash-dotted to the combination BBN+$\Omega_{\mathrm{b}}h^2$ and the dotted to BBN+CMB. The \textit{left/right panels} correspond to purely neutrinophilic/electrophilic particles.}
    \label{fig:Multiplot}
\end{figure}

In this appendix we display the marginalized $\chi^2(m_\chi)$ for neutrinophilic and electriphilic neutral scalars, complex scalars, Dirac fermions and vector bosons. They can be seen in Figure~\ref{fig:Multiplot}.

We also outline the bounds at 95.4\% CL for thermal WIMPs with different annihilation final states to electrons/photons and neutrinos. They are shown in Table~\ref{tab:BRBounds}.

\begin{table}[h]
{\def\arraystretch{1.3}
\begin{tabular}{c|l|p{0.7cm}|p{0.7cm}|p{0.7cm}|p{0.7cm}|p{0.7cm}|p{0.7cm}|p{0.7cm}|p{0.7cm}|p{0.7cm}|p{0.7cm}|p{2.08cm}|p{2.08cm}|p{0.7cm}}
\hline\hline
\multirow{2}{*}{\textbf{Probe}$\,$} & \multirow{2}{*}{\vspace*{-0.1 cm}\textbf{BSM Particle}} & \multicolumn{13}{c}{$\boldsymbol{e :}  \boldsymbol{\nu}$ \textbf{Annihilation Ratio}}\\
  &                       & \multicolumn{1}{c}{1:$10^6$} & \multicolumn{1}{c}{1:$10^5$} & \multicolumn{1}{c}{1:$10^4$} & \multicolumn{1}{c}{1:$10^3$} & \multicolumn{1}{c}{1:$10^2$} & \multicolumn{1}{c}{1:$10^1$} & \multicolumn{1}{c}{1:1} & \multicolumn{1}{c}{$10^1$:1} & \multicolumn{1}{c}{$10^2$:1} & \multicolumn{1}{c}{$10^3$:1} & \multicolumn{1}{c}{$10^4$:1} & \multicolumn{1}{c}{$10^5$:1} & \multicolumn{1}{c}{$10^6$:1} \\ \hline\hline
\parbox[t]{5mm}{\multirow{5}{*}{\rotatebox[origin=c]{90}{\textbf{BBN}}}} & Majorana      & \hfil 2.2           &    \hfil 2.0        &  \hfil 1.8          & \hfil 1.8           & \hfil 2.0           &    \hfil 2.3        & \hfil 2.4     &  \hfil 2.2         &  \hfil 1.9           &  \hfil 1.6           &    \hfil 1.1         &  \hfil 0.7           & \hfil 0.5            \\
                          & Dirac        & \hfil 3.3           &  \hfil 2.6          &  \hfil 2.2          &  \hfil 2.2          &  \hfil 2.3          &      \hfil 2.4      & \hfil 2.6      &  \hfil 2.4          &  \hfil 2.1          &  \hfil 1.9          &  \hfil 1.6          &   \hfil 1.1         &  \hfil 0.8          \\
                          & Scalar       & \hfil 1.2           & \hfil 1.2           & \hfil 1.3           & \hfil 1.5           & \hfil 1.7           & \hfil 2.0           &    \hfil 2.2   & \hfil 2.0           & \hfil 1.6           & \hfil 1.2           & \hfil 0.8           & \hfil 0.4           & \hfil 0.4           \\
                          & Complex Scalar & \hfil 2.2           & \hfil 2.0           & \hfil 1.8           & \hfil 1.8           & \hfil 2.0           & \hfil 2.2           & \hfil 2.4      & \hfil 2.2           & \hfil 1.9           & \hfil 1.6           & \hfil 1.1           & \hfil 0.7           & \hfil 0.5           \\
                          & Vector      & \hfil 2.9      & \hfil 2.3  & \hfil 2.0         & \hfil 2.0           & \hfil 2.2           & \hfil 2.4           & \hfil 2.5      & \hfil 2.3           &  \hfil 2.1          &  \hfil 1.8          &  \hfil 1.4          &  \hfil 0.9          &  \hfil 0.7          \\  \hline
\parbox[t]{5mm}{\multirow{5}{*}{\rotatebox[origin=c]{90}{\textbf{BBN+$\boldsymbol{\Omega_\mathrm{b}h^2}$}}}} & Majorana      & \hfil 2.6           &    \hfil 2.2        &   \hfil 1.8         & \hfil 1.9           & \hfil 2.1           &    \hfil 2.4        &  \hfil 2.6     &  \hfil 2.3          & \hfil 1.9           & \hfil 1.5           &    \hfil 1.1        & \hfil  0.9          &  \hfil 3.3            \\
                          & Dirac        &  \hfil 4.8          &      \hfil 3.0      &  \hfil 2.3          & \hfil 2.2           & \hfil 2.4           &  \hfil 2.6          &  \hfil 2.8     &  \hfil 2.5          &   \hfil 2.2         &  \hfil 1.9          &   \hfil 1.5         &   \hfil 1.7         &  \hfil 6.2          \\
                          & Scalar       & \hfil 1.3           & \hfil 1.3           & \hfil 1.3           & \hfil 1.5           & \hfil 1.8           & \hfil 2.1           &    \hfil 2.3   & \hfil 2.1           & \hfil 1.7           & \hfil 1.2           & \hfil 0.7           & \hfil 0.5           & \hfil 0.6           \\
                          & Complex Scalar & \hfil 2.8           & \hfil 2.2           & \hfil 1.8           & \hfil 1.9           & \hfil 2.1           & \hfil 2.4           & \hfil 2.6      & \hfil 2.3           & \hfil 1.9           & \hfil 1.6           & \hfil 1.1           & \hfil 1.0           & \hfil 3.5           \\
                          & Vector      & \hfil 4.0      & \hfil 2.7  & \hfil 2.1         & \hfil 2.1           & \hfil 2.3           & \hfil 2.5           & \hfil 2.7      & \hfil 2.4           &  \hfil 2.1          &  \hfil 1.7          &  \hfil 1.3          &  \hfil 1.5          &  \hfil 5.2          \\  \hline
\parbox[t]{5mm}{\multirow{5}{*}{\rotatebox[origin=c]{90}{\textbf{Planck}}}}     & Majorana     &   \hfil 8.2         &    \hfil 7.6        &  \hfil 4.6          & \hfil  3.4          & \hfil 3.6           &   \hfil  4.1        & \hfil 4.5      &  \hfil 3.8       & \hfil 3.0           &   \hfil  2.1        &  \hfil 1.1          &  \hfil 0.3          & \hfil 4.1 \\
                          & Dirac         &   \hfil 11.0         &    \hfil 10.1        &  \hfil 6.1          &  \hfil 4.0          & \hfil 4.0           &  \hfil 4.4          &  \hfil 4.8     &    \hfil 4.2        &    \hfil 3.4        &    \hfil 2.4        &    \hfil 1.3        &    \hfil 0.4; 1.2 - 4.1        &     \hfil 7.0       \\
                          & Scalar     & \hfil 5.4           & \hfil 5.1            & \hfil 3.3            & \hfil 2.8            & \hfil 3.1            & \hfil 3.7            & \hfil 4.1       & \hfil 3.5            & \hfil 2.7            & \hfil 1.8            & \hfil 0.8            & \hfil -           & \hfil -            \\
                          & Complex Scalar &   \hfil 8.3         & \hfil 7.6           & \hfil 4.6           & \hfil 3.4           & \hfil 3.6           & \hfil 4.1           & \hfil 4.5      & \hfil 3.8           & \hfil 3.0           & \hfil 2.1           & \hfil 1.1           & \hfil 0.2           & \hfil 4.3           \\
                          & Vector         & \hfil 9.8           & \hfil 8.9          & \hfil 5.3          & \hfil 3.7          & \hfil 3.8           & \hfil 4.3           & \hfil 4.7      & \hfil 4.1          & \hfil 3.2           &  \hfil 2.3          & \hfil 1.2          & \hfil 0.3; 1.2 - 3.4          & \hfil 6.0          \\
                        \hline
\parbox[t]{5mm}{\multirow{5}{*}{\rotatebox[origin=c]{90}{\textbf{Planck+}$\boldsymbol{H_0}$}}}     & Majorana     &  \hfil 4.7          &  \hfil 3.3          &     \hfil 2.0       &  \hfil   2.0        &   \hfil 2.3         & \hfil 2.7           & \hfil 3.0      & \hfil 3.2    &  \hfil 2.5      & \hfil  1.8          &     \hfil 0.9       &  \hfil 0.2; 0.4 - 8.4 &\hfil   8.9 \\
                          & Dirac         &   \hfil 7.6         & \hfil 5.4           &     \hfil 2.6       &     \hfil 2.4       & \hfil 2.6           &     \hfil 3.0       & \hfil 3.2      &  \hfil 3.5          &  \hfil 2.8          &  \hfil 2.1          &  \hfil 1.1; 3.5 - 5.9          &   \hfil 0.4; 0.6 - 10.5         &  \hfil 11.4          \\
                          & Scalar     & \hfil  1.6          & \hfil    1.4         & \hfil 1.4            & \hfil   1.6          & \hfil  2.0           & \hfil    2.4         & \hfil 2.7       & \hfil 2.9            & \hfil    2.2         & \hfil 1.5            & \hfil 0.7            & \hfil 0.3 - 5.8           & \hfil     6.2        \\
                          & Complex Scalar &   \hfil 4.9         & \hfil    3.5        & \hfil 2.0           & \hfil    2.0        & \hfil  2.3          & \hfil    2.7        & \hfil 3.0      & \hfil 3.2           & \hfil 2.5           & \hfil     1.8       & \hfil 0.9           & \hfil 0.2; 0.4 - 8.3            & \hfil    8.9        \\
                          & Vector         & \hfil  6.6          & \hfil    4.6       & \hfil 2.4          & \hfil 2.2          & \hfil     2.5       & \hfil 2.8           & \hfil 3.1      & \hfil    3.4       & \hfil  2.7          &  \hfil    1.9       & \hfil 1.1; 3.4 - 5.4          & \hfil 0.3; 0.5 - 9.7          & \hfil   10.3        \\
                        \hline
\parbox[t]{5mm}{\multirow{5}{*}{\rotatebox[origin=c]{90}{\textbf{BBN+Planck}}}} & Majorana      &    \hfil 6.4        &    \hfil 5.4        & \hfil 3.1           &  \hfil 2.7          &     \hfil 2.9       &  \hfil   3.4        &  \hfil 3.7     &\hfil  3.2           & \hfil 2.6           &    \hfil  1.9       &\hfil     1.1        &\hfil   7.0          &  \hfil 7.8                      \\
                          & Dirac        & \hfil 9.2           &  \hfil 7.7          &  \hfil 3.9          &  \hfil 3.1          &  \hfil 3.3          &  \hfil 3.6          & \hfil 3.9      & \hfil 3.5           &  \hfil 2.9          &  \hfil 2.2          &  \hfil 1.4          & \hfil 9.3           &  \hfil 10.4          \\
                          & Scalar       & \hfil 3.6           & \hfil 3.2           & \hfil 2.2           & \hfil 2.2            & \hfil 2.6           & \hfil 3.0           & \hfil 3.4      & \hfil 2.9           & \hfil 2.3           & \hfil 1.6           & \hfil 0.8           & \hfil 4.5           & \hfil 5.0           \\
                          & Complex Scalar & \hfil 6.5           & \hfil 5.5           & \hfil 3.1           & \hfil 2.7           & \hfil 2.9           & \hfil 3.3           & \hfil 3.7      & \hfil 3.2           & \hfil 2.6           & \hfil 1.8           & \hfil 1.1           & \hfil 7.1           & \hfil 7.8           \\
                          & Vector       &  \hfil 8.2          &  \hfil 6.9          &  \hfil 3.5          & \hfil 2.9           & \hfil 3.1           & \hfil 3.5           & \hfil 3.8      &  \hfil 3.4          &  \hfil 2.7          &  \hfil 2.0          &  \hfil 1.3          &  \hfil 8.5          &  \hfil 9.4            \\ \hline
\parbox[t]{5mm}{\multirow{5}{*}{\rotatebox[origin=c]{90}{\textbf{Simons Obs.}}}}     & Majorana     & \hfil 12.5  & \hfil 12.3 & \hfil 11.0 & \hfil 7.4 & \hfil 6.3 & \hfil 6.7 & \hfil 7.2 & \hfil 5.8 & \hfil 4.4 & \hfil 2.8 & \hfil 1.3; 1.9 - 10.3 & \hfil 11.9 & \hfil 12.2 \\
                          & Dirac         & \hfil 15.2  & \hfil 15.0 & \hfil 13.3 & \hfil 8.8 & \hfil 7.0 & \hfil 7.2 & \hfil 7.7 & \hfil 6.4 & \hfil 4.8 & \hfil 3.1 & \hfil 1.6; 2.2 - 12.4 & \hfil 14.6 & \hfil 14.9 \\
                          & Scalar & \hfil 9.7  & \hfil 9.6 & \hfil 8.7 & \hfil 6.1 & \hfil 5.5 & \hfil 6.0 & \hfil 6.6 & \hfil 5.4 & \hfil 4.0 & \hfil 2.4 & \hfil 1.0; 1.6 - 8.0 & \hfil 9.3 & \hfil 9.4 \\
                          & Complex Scalar & \hfil 12.5  & \hfil 12.3 & \hfil 11.1 & \hfil 7.5 & \hfil 6.3 & \hfil 6.7 & \hfil 7.2 & \hfil 5.9 & \hfil 4.4 & \hfil 2.8 & \hfil 1.3; 1.9 - 10.4 & \hfil 12.0 & \hfil 12.2 \\
                          & Vector         & \hfil 14.1  & \hfil 13.9 & \hfil 12.4 & \hfil 8.3 & \hfil 6.7 & \hfil 7.0 & \hfil 7.5 & \hfil 6.2 & \hfil 4.7 & \hfil 3.0 & \hfil 1.4; 2.1 - 11.6 & \hfil 13.5 & \hfil 13.8 \\
                        \hline
\parbox[t]{5mm}{\multirow{5}{*}{\rotatebox[origin=c]{90}{\textbf{CMB-S4}}}}     & Majorana     & \hfil 13.5 & \hfil 13.4 & \hfil 12.4 & \hfil 9.0 & \hfil 7.1 & \hfil 7.4 & \hfil 8.0 & \hfil 6.6 & \hfil 4.8 & \hfil 3.0 & \hfil 1.3; 1.8 - 11.8 & \hfil 13.0 & \hfil 13.2 \\
                          & Dirac         & \hfil 16.2  & \hfil 16.0 & \hfil 14.7 & \hfil 10.6 & \hfil 8.0 & \hfil 8.1 & \hfil 8.6 & \hfil 7.1 & \hfil 5.3 & \hfil 3.3 & \hfil 1.6; 2.1 - 14.0 & \hfil 15.6 & \hfil 15.9 \\
                          & Scalar & \hfil 10.7 & \hfil 10.7 & \hfil 10.0 & \hfil 7.4 & \hfil 6.3 & \hfil 6.7 & \hfil 7.3 & \hfil 5.9 & \hfil 4.3 & \hfil 2.5 & \hfil 1.0; 1.5 - 9.5 & \hfil 10.3 & \hfil 10.5 \\
                          & Complex Scalar & \hfil 13.5 & \hfil 13.3 & \hfil 12.4 & \hfil 9.0 & \hfil 7.1 & \hfil 7.4 & \hfil 8.0 & \hfil 6.5 & \hfil 4.8 & \hfil 3.0 & \hfil 1.3; 1.8 - 11.8 & \hfil 13.0 & \hfil 13.2 \\
                          & Vector         & \hfil 15.1  & \hfil 14.9 & \hfil 13.8 & \hfil 10.0 & \hfil 7.7 & \hfil 7.9 & \hfil 8.4 & \hfil 6.8 & \hfil 5.1 & \hfil 3.2 & \hfil 1.5; 2.0 - 13.2 & \hfil 14.6 & \hfil 14.8 \\
                        \hline\hline    
\end{tabular}
}
\caption{Lower bounds at 95.4\% CL on the masses of various thermal BSM particles in MeV. The bounds are given for a Majorana fermion, Dirac fermion, neutral scalar boson, complex scalar boson and vector boson.
The rows correspond to constraints using data from various sources as detailed in Section~\ref{sec:current_data_analyis}.
The columns correspond to the ratio of electrophilic to neutrinophilic particles. A `-' means that no bound is obtained at this confidence level and a `\# - \#' means that masses in this range are excluded.}
\label{tab:BRBounds}
\end{table}

\cleardoublepage

\cleardoublepage

\subsection{Implications for Lithium-7 and Helium-3}
\label{app:cosmo_imp_other}
\begin{figure}[!ht]
    \centering
    \includegraphics[width=0.46\textwidth]{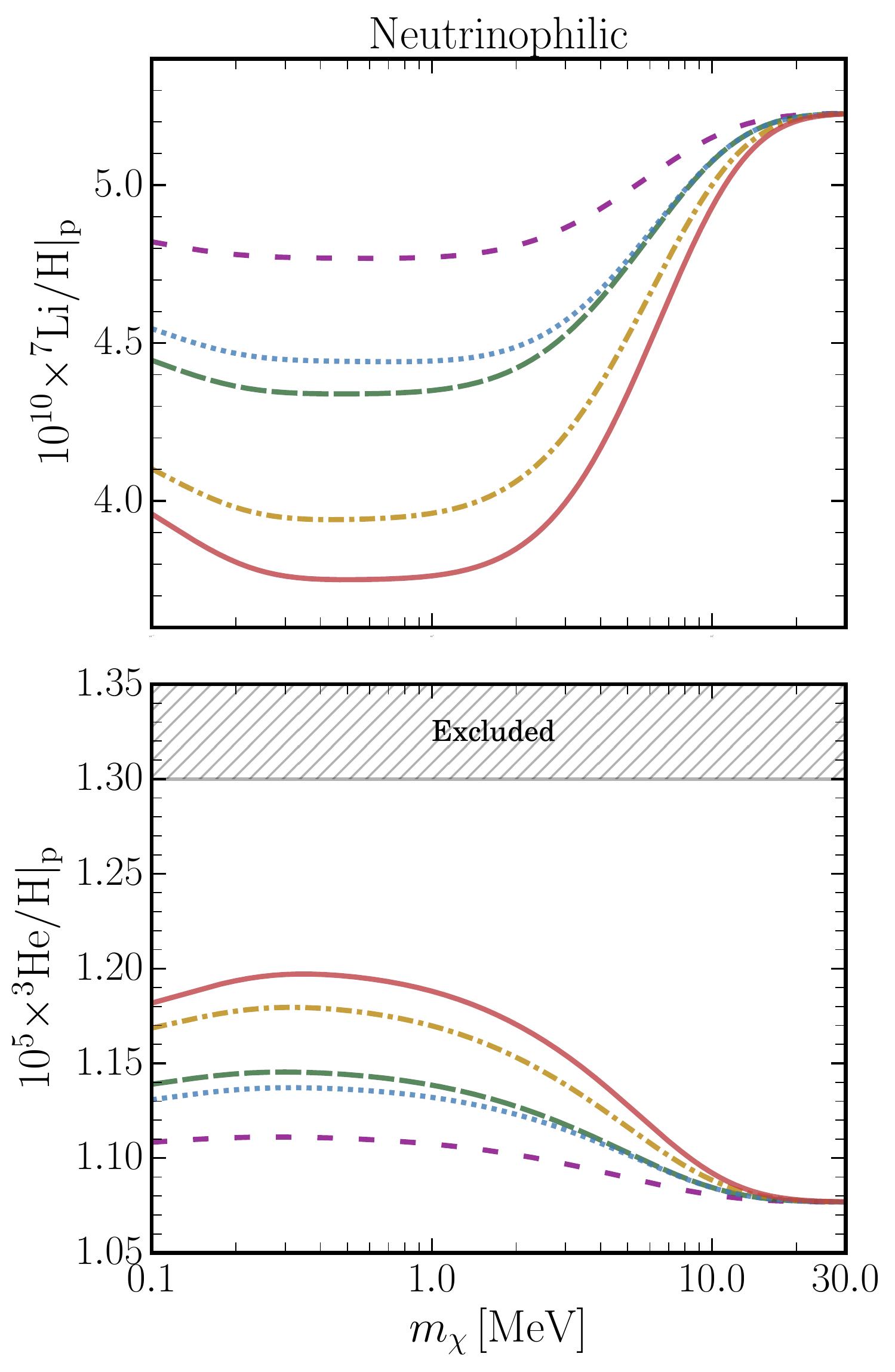} \qquad
    \includegraphics[width=0.45\textwidth]{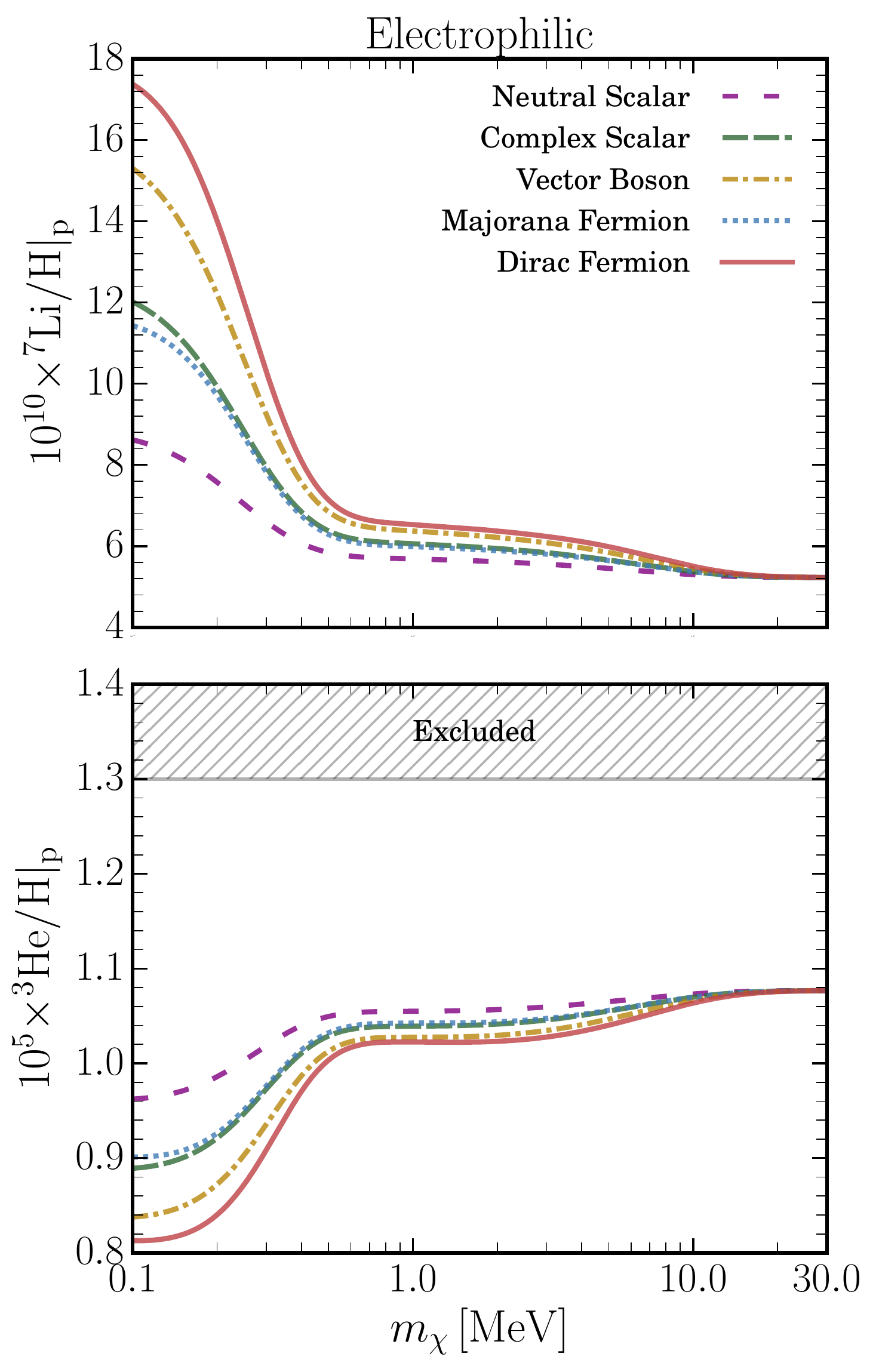}
    \caption{Cosmological impact of light BSM particles in thermal equilibrium with the SM plasma as a function of their mass $m_\chi$. The \textit{left/right panels} correspond to neutrinophilic/electrophilic particles. \textit{Upper panels:} The lithium-7 primordial abundance $^7\mathrm{Li}/\mathrm{H}|_\mathrm{P}$. Measurements of $^7\mathrm{Li}/\mathrm{H}|_\mathrm{P}$ are not shown for clarity, see e.g. \cite{pdg} for current measurements. \textit{Lower panels:} The helium-3 primordial abundance $^3\mathrm{He}/\mathrm{H}|_\mathrm{P}$. The grey contours correspond to an upper limit as reported by~\cite{Bania:2002yj}. The predictions are made with $\Omega_\mathrm{b} h^2 = 0.021875$ and $\tau_n = 879.5\,\text{s}$.}
    \label{fig:Cosmoimply_other}
\end{figure}

We show the evolution of the primordial $^7\mathrm{Li}/\mathrm{H}|_\mathrm{P}$ and $^3\mathrm{He}/\mathrm{H}|_\mathrm{P}$ abundances in Figure \ref{fig:Cosmoimply_other} as a function of the mass of a thermal BSM particle. We note that the upper panels do not include any confidence intervals, since it is well known that current measurements of the primordial lithium-7 are in disagreement with SM predictions using the baryon-to-photon ratio inferred from CMB observations \cite{pdg}. The excluded regions in the lower panels are based on observations of helium-3 in our galaxy \cite{Bania:2002yj}. Helium-3 can be both produced and destroyed in stars, which makes it difficult to precisely determine the time evolution of its primordial abundance \cite{VangioniFlam:2002sa}. Therefore, we have not included measurements of either lithium-7 or helium-3 in our analysis. Nevertheless, if the situation changes in the future, it will be straightforward to obtain bounds from Figure \ref{fig:Cosmoimply_other} and see how it improves the current BBN constraints.

\cleardoublepage

\subsection{CMB-S4 Forecast}
\label{app:CMBfisher}
In order to forecast the reach of CMB-S4 constraints, we first choose a fiducial cosmology with cosmological parameters equal to the Planck 2018 TTTEEE+lowE mean values as in Table 2 of \cite{Aghanim:2018eyx}, which are reproduced below in Table \ref{tab:FisherResults}.
The fiducial helium abundance is obtained by running \texttt{PRIMAT} within the Standard Model and fiducial cosmology.\\\\
\noindent To forecast the sensitivity of future CMB experiments, we employ the same procedure as used in the CMB-S4 Science Book \cite{Abazajian:2016yjj}. Assuming Gaussian statistics, the Fisher matrix for CMB experiments is given by
\begin{align}
    F_{ij} = \underset{X,Y}{\sum}\hspace{0.1 cm}\overset{\ell_\mathrm{max}}{\underset{\ell=\ell_\mathrm{min}}{\sum}}\frac{\partial\mathcal{C}_\ell^X}{\partial\theta_i}\left[\mathbf{C}_\ell^{XY}\right]^{-1}\frac{\partial\mathcal{C}_\ell^Y}{\partial\theta_i}\ ,
\end{align}
with indices $X = ab$, $Y = cd$ and $a,b,c,d\in\{T,E,B\}$. 
The covariance matrix $\mathbf{C}_\ell^{XY}$ for each multipole $\ell$ is defined as
\begin{align}
    \mathbf{C}_\ell^{abcd} =& \frac{1}{(2\ell+1)f_\mathrm{sky}}\left[\left(\mathcal{C}_\ell^{ac}+N_\ell^{ac}\right)\left(\mathcal{C}_\ell^{bd}+N_\ell^{bd}\right)
    + \left(\mathcal{C}_\ell^{ad}+N_\ell^{ad}\right)\left(\mathcal{C}_\ell^{bc}+N_\ell^{bc}\right)\right]\ ,
\end{align}
with $f_\mathrm{sky}$ the effective fraction of sky covered by the experiment, $\mathcal{C}_\ell^X$ the simulated CMB power spectra and $N_\ell^X$ (Gaussian) noise power spectra. The noise is approximated as
\begin{align}
    N_\ell^{aa} = (\Delta X)^2\exp\left(\frac{\ell(\ell+1)\theta^2_\mathrm{FWHM}}{8\ln2}\right)\ ,
\end{align}
where $\Delta X \in \{\Delta T,\Delta P\}$ and $N_\ell^{TE} = 0$.
We adopt a similar configuration as used in the CMB-S4 Science Book: lensed power spectra with $\ell_\mathrm{min} = 30$, $\{\ell_\mathrm{max}^{TT}, \ell_\mathrm{max}^{TE}\} = 3000$, $\{\ell_\mathrm{max}^{EE},\ell_\mathrm{max}^{BB}\} = 5000$, $f_\mathrm{sky} = 0.4$, $\theta_\mathrm{FWHM} = 1'$, $\Delta T = 1$ $\mu$K-arcmin and $\Delta P = \sqrt{2}$ $\mu$K-arcmin.\\\\
The \texttt{CLASS} code \cite{Blas:2011rf} is used to obtain the power spectra. The numerical derivatives are computed using the symmetric derivative $\mathcal{C}_\ell'(\theta) = \left[\mathcal{C}_\ell(\theta+\Delta\theta)-\mathcal{C}_\ell(\theta-\Delta\theta)\right]/(2\Delta\theta)$, with fiducial parameter $\theta$ and stepsize $\Delta\theta$. The stepsizes used are of order $\Delta\theta_i \sim \sigma(\theta_i)$, as to output a more reliable estimate of the confidence level \cite{Perotto:2006rj}. The CMB-S4 Fisher matrix is then added to the Planck 2018 low-$\ell$ TTTEEE+lowP+lowE Fisher matrix to obtain the combined constraints. The fiducial parameters and step sizes used in our computations, together with the forecasted sensitivities, are listed in Table \ref{tab:FisherResults}. We find good overall agreement with the forecasts performed in \cite{Abazajian:2016yjj} within $\Lambda$CDM. 

\begin{table}[!ht]
\begin{center}
{\def\arraystretch{1.35}
\begin{tabular}{p{2cm}|p{2.5cm}|p{2cm}|p{2cm}|p{3cm}}
\hline\hline
	\hfil \textbf{Parameter} &\hfil  \textbf{Fiducial Value} &\hfil  $\boldsymbol{\Delta\theta}$ &\hfil  \textbf{CMB-S4} &\hfil  \textbf{CMB-S4+Planck} \\ \hline\hline
   	\hfil $\Omega_\mathrm{b}h^2$ &\hfil  0.02236 & \hfil $3\times 10^{-5}$ &\hfil $4.9\times 10^{-5}$ &\hfil  $4.7\times 10^{-5}$\\ \hline
   	\hfil $\Omega_\mathrm{c}h^2$ &\hfil  0.1202 & \hfil $6\times 10^{-4}$ &\hfil $1.8\times 10^{-3}$ &\hfil  $1.3\times 10^{-3}$\\ \hline
   	\hfil $100\theta_\mathrm{s}$ &\hfil  1.04090 &\hfil  $2\times 10^{-4}$ &\hfil  $2.3\times 10^{-4}$&\hfil  $1.8\times 10^{-4}$\\ \hline
   	\hfil $\ln(10^{10}A_\mathrm{s})$ &\hfil  3.045 & \hfil $9.5\times 10^{-3}$  &\hfil  $1.2\times 10^{-2}$ &\hfil  $8.1\times 10^{-3}$\\ \hline
   	\hfil $n_\mathrm{s}$ &\hfil  0.9649 & \hfil  $2\times 10^{-3}$ &\hfil  $3.7\times 10^{-3}$ &\hfil  $2.9\times 10^{-3}$\\ \hline
   	\hfil $\tau$ & \hfil 0.0544 & \hfil $6\times 10^{-3}$ & \hfil $7.2\times 10^{-3}$ & \hfil $4.8\times 10^{-3}$\\ \hline
   	\hfil $N_\mathrm{eff}$ & \hfil 3.046 &$ \hfil 3\times 10^{-2}$ & \hfil $1.1\times 10^{-1}$ & \hfil $8.1\times 10^{-2}$\\ \hline
   	\hfil $Y_\mathrm{P}$ & \hfil 0.2472 & \hfil $4\times 10^{-3}$ & \hfil $6.1\times 10^{-3}$ & \hfil $4.3\times 10^{-3}$\\  \hline \hline

\end{tabular}
}
\end{center}
\vspace{-0.2cm}
\caption{Forecasted sensitivities of CMB-S4 and CMB-S4+Planck 2018 for the parameters of $\Lambda\mathrm{CDM}+N_\mathrm{eff}+Y_\mathrm{P}$. The column $\Delta\theta$ refers to the stepsizes used to compute the numerical derivatives.}
\label{tab:FisherResults}
\end{table}

\end{document}